\documentclass[12pt]{article}
\pdfoutput=1
\usepackage{authblk}
\newcommand{\SeriesTitle}{Persistent Misperception with Correctly Specified Models}
\newcommand{\PartITitle}{\SeriesTitle, Part~I: General Smooth Policy}

\newcommand{\ArticleTitle}{\PartITitle}
\newcommand{\AuthorOne}{Tetsuya Kaji}
\newcommand{\AuthorTwo}{Michihiro Kandori}
\newcommand{\AuthorThree}{Yuichi Yamamoto}
\newcommand{\AffiliationOne}{University of Chicago Booth School of Business}
\newcommand{\AffiliationTwo}{Department of Economics, University of Tokyo}
\newcommand{\AffiliationThree}{Department of Economics, University of Tokyo}

\author[1]{\AuthorOne}
\author[2]{\AuthorTwo}
\author[3]{\AuthorThree}
\affil[1]{\AffiliationOne}
\affil[2]{\AffiliationTwo}
\affil[3]{\AffiliationThree}
\title{\ArticleTitle%
\thanks{This work is supported by the Richard N.\ Rosett Faculty Fellowship at the University of Chicago Booth School of Business. Generative AI was used to assist with proofs and to polish the writing.}}
\date{\today}
\usepackage{amsmath,amssymb,amsthm,mathtools,mathrsfs,bm,bbm}
\usepackage{mleftright}
\mleftright
\usepackage{enumitem,graphicx,booktabs,longtable,array}
\usepackage[onehalfspacing,nodisplayskipstretch]{setspace}
\usepackage{microtype}
\usepackage[authoryear]{natbib}
\usepackage{xcolor}
\usepackage[colorlinks,citecolor=blue,urlcolor=blue,linkcolor=blue,hypertexnames=false]{hyperref}
\usepackage{cleveref}
\makeatletter
\newcommand*{\condtarget}[2]{%
  \begingroup\let\cond@savedHref\@currentHref
  \phantomsection\def\@currentlabel{#2}\label{#1}%
  \global\let\@currentHref\cond@savedHref\endgroup#2}
\makeatother
\usepackage{refcount}
\usepackage[justification=raggedright,singlelinecheck=false]{caption}
\usepackage{subcaption}
\usepackage[margin=1.25in]{geometry}
\allowdisplaybreaks[3]
\theoremstyle{plain}
\newtheorem{theorem}{Theorem}
\newtheorem{proposition}[theorem]{Proposition}
\newtheorem{lemma}[theorem]{Lemma}
\newtheorem{corollary}[theorem]{Corollary}

\newtheorem{auxlemma}[theorem]{Lemma}
\theoremstyle{definition}
\newtheorem{definition}[theorem]{Definition}
\newtheorem{assumption}[theorem]{Assumption}
\theoremstyle{remark}
\newtheorem{remark}[theorem]{Remark}
\usepackage{thmtools}
\newcommand{\exasymbol}{$\square$}
\declaretheoremstyle[
  spaceabove=6pt, spacebelow=6pt,
  headfont=\normalfont\bfseries,
  notefont=\mdseries, notebraces={(}{)},
  bodyfont=\normalfont, postheadspace=1em,
  qed=\exasymbol
]{exampleStyle}
\declaretheorem[style=exampleStyle,name=Example]{example}
\renewcommand\thmcontinues[1]{continued}
\crefname{example}{Example}{Examples}
\newcommand{\R}{\mathbb R}

\DeclareMathOperator{\Var}{Var}
\DeclareMathOperator{\sgn}{sgn}
\begin{document}
\pdfbookmark[0]{Article}{part1-article}
\maketitle
\begin{abstract}
Econometrics teaches that the slope of a regression \(y=\alpha+\beta x+\varepsilon\)
can be learned only if the regressor varies enough. What happens when the
regressor is chosen by the very agent who is trying to learn
\((\alpha,\beta)\)? We show that a correctly specified Bayesian decision maker
who chooses \(x_t\) as a function of its posterior generally fails to learn.
If the action is a smooth function of the posterior mean and covariance and
responds to slope revisions that leave the predicted outcome unchanged,
complete learning has probability zero at almost every such parameter, and incorrect learning occurs on every
set of regular parameters with exactly its prior probability. Near an incorrect
Berk--Nash equilibrium under the rule, concentrated priors keep the posterior
nearby with probability arbitrarily close to one, without any dogmatic belief,
so overconfidence and underconfidence arise endogenously. The results cover
myopic optimization, automated systems including delegation to AI, and rules of
thumb.
\end{abstract}

\section{Introduction}\label{sec:introduction}
A basic lesson of econometrics is that the slope of a linear regression
\[
 y_t=\alpha+\beta x_t+\varepsilon_t
\]
can be learned only if the regressor \(x_t\) exhibits enough variation.
Standard identification arguments take that variation as given. In many
economic environments, however, the regressor is itself a decision. A firm
chooses its advertising budget or its price, and a central bank chooses its
policy instrument, while using the resulting data to learn how outcomes
respond to those choices. Many economic decisions can be represented, at
least locally or approximately, by this structure. This raises a simple
question: what happens when \(x_t\) is chosen by the very agent who is
trying to learn \((\alpha,\beta)\)? Who generates the variation needed for
identification?

The difficulty is not the conventional endogeneity of a regressor. The
current action is chosen before the current shock is realized, so it is
independent of that shock. The difficulty is that the learner itself
determines whether enough identifying variation is generated in the first
place.

We show that, very generally, it is not. We consider a decision maker who
learns \((\alpha,\beta)\) by Bayesian updating, with a correctly specified
model, and who chooses \(x_t\) as a function of its posterior belief at date
\(t\). Under mild conditions on this function, the decision maker does not
learn the true parameter. Complete learning has prior probability zero on the
set of parameters at which the conditions hold, and, for almost every such
parameter, the posterior slope converges to an incorrect value with
probability one. The
failure is not caused by a misspecified model, by non-Bayesian updating,
or by an exogenously fixed action. It arises because the decision maker's own
rule suppresses the variation needed to distinguish the intercept from the
slope.

The conditions are simple. With a Gaussian prior and Gaussian noise, the
posterior is Gaussian, with mean \((a,b)\) and covariance matrix \(\Sigma\), and
the action is a function \(x=\psi(a,b,\Sigma)\). Let
\[
 h(a,b)=\psi(a,b,0)
\]
be the action the decision maker would take if it knew the parameter to be
\((a,b)\). Suppose that \condtarget{cond:intro-smooth}{(1)} \(\psi\) is smooth, i.e., continuously differentiable with locally Lipschitz derivatives,
near complete information, \(\Sigma=0\);
and \condtarget{cond:intro-transversality}{(2)} the \textit{transversality condition}
\[
 h_b\ne h\,h_a
\]
holds at the true parameter. Then, under the prior, complete learning has
probability zero on the set of parameters at which \ref{cond:intro-smooth} and \ref{cond:intro-transversality} hold, and for
almost every such true parameter the decision maker fails to learn it with
probability one. Condition \ref{cond:intro-transversality} has a transparent meaning. A
revision of beliefs that leaves the predicted outcome \(a+bx\) at the current
action \(x=h(a,b)\) unchanged satisfies \(da=-x\,db\), and it changes the action
by \(h_a\,da+h_b\,db=(h_b-h\,h_a)\,db\). So transversality says that the action
responds to slope revisions that the data at the current action cannot
detect.

Transversality also ensures that incorrect beliefs form a smooth family of
steady states. Given the true parameter \((\alpha,\beta)\), a belief \((a,b)\)
held with certainty is a \textit{Berk--Nash equilibrium belief} under the rule \citep[cf.][]{ep2016} when the action it prescribes, \(h(a,b)\), produces
no surprise on average:
\[
 \alpha+\beta h(a,b)=a+bh(a,b).
\]
Write \(f(a,b)=a-\alpha+(b-\beta)h(a,b)\), so that the equilibrium beliefs
are the zeros of \(f\). The derivative of \(f\) along the direction
\((-h,1)\), which leaves the predicted outcome at \(h\) unchanged, is
\[
 -h\,f_a+f_b=(b-\beta)(h_b-h\,h_a).
\]
At an incorrect belief (\(b\ne\beta\)), transversality therefore says that the
set of Berk--Nash equilibrium beliefs crosses the line of beliefs that
predict the same outcome at \(h\). In particular, the gradient of \(f\) does
not vanish, and by the implicit function theorem the equilibrium beliefs
near \((a,b)\) form a smooth curve.\footnote{The converse does not hold. A
rule that ignores beliefs, \(h\) constant, has a nonvanishing gradient
\(\nabla f=(1,h)\) but violates transversality.}

The leading example is a myopic decision maker who maximizes
\(E[y_t]-C(x_t)\) in each period, for a smooth cost \(C\) with \(C''>0\). An
advertiser who
weighs expected sales against the cost of the campaign is one. Its action
\(h(a,b)=(C')^{-1}(b)\) depends only on the posterior slope, so \(h_a=0\) and
\(h_b=1/C''(h)>0\), and transversality holds. Myopia is natural for automated
systems, including AI agents to which a recurring decision is delegated. They
maximize the per-period objective they were coded to use
\citep{aggarwal2019autobid,aggarwal2024survey,balseirogur2019}. Our results also apply
to rules that do not come from optimization, such as a budgeting rule of
thumb or a certainty-equivalent policy rule. By contrast, a rule that pins the
predicted outcome to a target violates transversality, and for such rules
complete learning can be likely.

Our results contrast with the mistaken self-assessment of \citet{hks2018}.
Their decision maker holds a dogmatic belief \(a=a^0\) about its own ability,
the intercept, and learns the slope. They show that the belief converges to a
Berk--Nash equilibrium with an incorrect slope. We show that convergence to
an incorrect Berk--Nash equilibrium does not require a dogmatic belief. Our
decision maker learns both coefficients with a correctly specified model and
a full-support prior, and yet, near an incorrect Berk--Nash equilibrium,
sufficiently concentrated priors keep the posterior near the incorrect belief
with probability arbitrarily close to one.

The behavioral types assumed by \citet{hks2018}, overconfidence
(\(a^0>\alpha\)) and underconfidence (\(a^0<\alpha\)), then arise endogenously
from the learning history. Decision makers with the same preferences, the
same model and the same prior end up overconfident or underconfident,
depending only on the outcomes they happened to observe. We illustrate this by
simulation in Section~\ref{sec:confidence}.

Our local result for rules that respond to beliefs also differs from local
stability results in finite-action environments \citep{fls2021}, where a
strict equilibrium action remains optimal under slightly perturbed beliefs, so
that the same action is chosen and the equilibrium belief continues to fit the
data. Our model has a counterpart of those results at equilibria where the rule
does not respond, such as a spending cap (Proposition~\ref{thm:corner-local}).
With a continuum of actions, a
small change in beliefs typically changes the action, and the resulting variation keeps
generating information that points toward the true parameter. The question is whether
the total amount of that information is large enough. One might expect a rule
that responds smoothly to beliefs to keep experimenting. The opposite
happens. A small belief revision produces a small change in the action, which
produces even less information, and near a regular limit this feedback is
contractive.

Rules that respond to beliefs also appear in the adaptive learning
literature, where convergence of least-squares learning is connected to the
stability of an associated differential equation \citep{marcetsargent1989}.
There the rules are evaluated by whether learning converges to a rational
expectations equilibrium; here the rule is held fixed, and the question is
whether Bayesian learning reaches the true parameter.

A companion paper \citep{kky2026partii} studies forward-looking decision makers,
who value the information generated by their actions; their optimal policy
need not be smooth, and the companion paper devises a way to handle it.

Section~\ref{sec:model} gives the model, decision rules, Berk--Nash equilibria
and transversality, and the examples. Section~\ref{sec:results} presents the
global and local results, the endogenous emergence of overconfidence and
underconfidence, and applications to myopic decision makers and to rules that
do not come from optimization. The \hyperref[app:posterior]{Appendix} contains
the proofs of the main results, and the \hyperref[supplement]{Supplement}
gives the remaining proofs and the verifications for the examples.

\section{Model}\label{sec:model}
Consider an infinite-horizon model where time is indexed by
\(t=1,2,\dots\). At date \(t\), a decision maker chooses an action
\(x_t\in X=[\underline x,\overline x]\), where \(\underline x\le\overline x\),
and observes an outcome
\[
 y_t=\alpha+\beta x_t+\varepsilon_t,
 \qquad \varepsilon_t\stackrel{\mathrm{iid}}{\sim}N(0,1).
\]
The parameter \(\vartheta=(\alpha,\beta)\in\R^2\) is unknown and fixed over
time. The decision maker has a nondegenerate Gaussian prior
\(\mu=N_2((a_0,b_0)^\top,\Sigma_0)\) about \(\vartheta\), where \((a_0,b_0)\)
is the mean and \(\Sigma_0\) is a positive-definite covariance matrix. The
parameter \(\vartheta\) and the noise sequence \((\varepsilon_t)_{t\ge1}\) are
mutually independent. A history at the end of date \(t\) consists of the past
actions and outcomes, \((x_s,y_s)_{s=1}^t\).

\subsection{Posterior beliefs}\label{sec:posterior}
The posterior belief after each history remains Gaussian, with mean
\((a_t,b_t)\) and covariance matrix \(\Sigma_t\). We parameterize its covariance
as
\begin{equation}\label{eq:sigma}
 \Sigma_t=
 \begin{pmatrix}
  A_t^{-1}+\bar x_t^2/V_t & -\bar x_t/V_t\\
  -\bar x_t/V_t & 1/V_t
 \end{pmatrix},
 \qquad A_t>0,\quad V_t>0.
\end{equation}
At \(t=0\), this representation accommodates every positive-definite
\(\Sigma_0\). Gaussian updating gives
\[
 A_t=A_0+t,\qquad
 \bar x_t=\frac{A_0\bar x_0+\sum_{j=1}^t x_j}{A_0+t}.
\]
The quantity \(A_t\) is an effective sample size, and \(\bar x_t\) is the
prior-augmented \textit{average action}. The initial value \(\bar x_0\) has
weight \(A_0\), and each subsequent action has weight one. These initial
quantities encode the prior covariance and need not represent actual past
observations. The quantity \(V_t\) measures \textit{slope precision}, since
\(V_t^{-1}=\Var_{\mu_t}(\beta)\).

It is convenient to describe the covariance by two variances,
\[
 u_t=A_t^{-1},\qquad w_t=V_t^{-1}.
\]
Here \(w_t\) is the posterior variance of the slope, and \(u_t\) is the
posterior variance of the mean outcome \(\alpha+\beta\bar x_t\) at the average
action, as a direct computation from \eqref{eq:sigma} shows. We call
\(s_t=(a_t,b_t,u_t,\bar x_t,w_t)\) the \textit{posterior state}. More generally,
a point \(s=(a,b,u,\bar x,w)\in\R^5\) is called a \textit{state}, and it is
\textit{feasible} if \(u>0\) and \(w>0\). By \eqref{eq:sigma}, a feasible state
and the pair consisting of the mean \((a,b)\) and a positive-definite
covariance matrix \(\Sigma\) determine each other,
with \(w=\Sigma_{22}\), \(\bar x=-\Sigma_{12}/\Sigma_{22}\) and
\(u=\Sigma_{11}-\Sigma_{12}^2/\Sigma_{22}\). We write \(\Sigma(u,\bar x,w)\)
for the matrix in \eqref{eq:sigma}. Complete information corresponds to
\(u=w=0\), where \(\Sigma=0\).

Let \(D_t=x_{t+1}-\bar x_t\) be the \textit{action gap}, which measures the
difference between today's action and the (augmented) average of past
actions. A standard argument shows that, with
\(\kappa_t=A_t/(A_t+1)\),\footnote{The argument is Bayes' rule for linear
Gaussian systems; see \citet[Theorem~4.4.1]{m2012}.}
\begin{equation}\label{eq:design}
 A_{t+1}=A_t+1,\qquad
 \bar x_{t+1}=\bar x_t+\frac{D_t}{A_t+1},\qquad
 V_{t+1}=V_t+\kappa_tD_t^2.
\end{equation}
The last equation shows that an action farther from the average adds more
slope precision. Choosing \(x_{t+1}=\bar x_t\) adds none, although the
observation still informs the mean outcome at that action. Equivalently,
accumulated slope precision is
\begin{equation}\label{eq:design-sum}
 V_t=V_0+A_0(\bar x_0-\bar x_t)^2+\sum_{j=1}^t(x_j-\bar x_t)^2,
\end{equation}
the (augmented) unnormalized variance of past actions. This is the precise
sense in which learning the slope requires variation in the regressor. The
posterior means are updated recursively by the prediction error
\(I_{t+1}=y_{t+1}-a_t-b_tx_{t+1}\):
\begin{equation}\label{eq:means}
 b_{t+1}=b_t+\frac{A_tD_t}{(A_t+1)V_{t+1}}I_{t+1},\qquad
 a_{t+1}=a_t+\frac{I_{t+1}}{A_t+1}-\bar x_{t+1}(b_{t+1}-b_t).
\end{equation}
So the slope estimate moves in proportion to the action gap. An action at the
average action leaves \(b_t\) unchanged and attributes the whole prediction
error to the intercept. The second formula says that the predicted mean
outcome at the new average action, \(a_{t+1}+b_{t+1}\bar x_{t+1}\), moves by
\(I_{t+1}/(A_t+1)\) plus the term \(b_t(\bar x_{t+1}-\bar x_t)\) from the move
of the average action. We write \(r_t=V_t/A_t\) for the \textit{precision
ratio}.

\subsection{Decision rules}\label{sec:policies}
The decision maker's behavior is described by a \textit{decision rule}, or
\textit{policy}, which maps each posterior to an action. Since the posterior
is determined by its mean \((a,b)\) and covariance \(\Sigma\), we may write a
rule as a function \(\Psi(a,b,\Sigma)\) of the mean and positive-definite covariance, or, through
\eqref{eq:sigma}, as a function \(\psi(s)=\Psi(a,b,\Sigma(u,\bar x,w))\) of the
feasible state \(s\); we use both notations. The action at
date \(t+1\) is
\[
 x_{t+1}=\psi(s_t).
\]
Thus the rule is \textit{Markov}, that is, it depends on the past history only through
the posterior, so any two histories that induce the same posterior lead to the
same action. We do not ask where the rule comes from. It may be the outcome of
optimization, a rule coded into an automated system, or a rule of
thumb.\footnote{Throughout, a decision rule is a Borel function, so that the
actions it induces are random variables.} Given a prior \(\mu\) and a rule \(\psi\), we write
\(P^{\mu,\psi}\) for the \textit{prior-predictive law}, under which
\(\vartheta\) is drawn from \(\mu\), and \(P^{\psi}_\vartheta\) for the
\textit{fixed-truth law} at the parameter \(\vartheta\).

The leading example is a myopic decision maker. Suppose that the decision maker receives a payoff \(U(x,y)\) each period. Given the posterior, the outcome of an action \(x\) is predicted to be
normal with mean \(a_t+b_tx\) and variance
\(1+(1,x)\Sigma_t(1,x)^\top=1+u_t+w_t(x-\bar x_t)^2\), so the expected payoff of
\(x\) is
\[
 g_t(x)=E\bigl[U\bigl(x,a_t+b_tx+\sqrt{1+u_t+w_t(x-\bar x_t)^2}\,Z\bigr)\bigr],\qquad Z\sim N(0,1).
\]
A \textit{myopic decision rule} \(\psi_0\) chooses an action that maximizes
\(g_t\) in each period. Its action may depend on the posterior covariance as
well as on the mean, because the payoff of an action may depend on how
precisely its outcome is predicted.

Our results concern whether the rule \(\psi(s)\) is a smooth function near \textit{complete-information states} where \(u=w=0\) (or equivalently \(\Sigma=0\)). While these complete-information states are not in the domain of the rule \(\psi\), we can discuss its smoothness by extending the domain of \(\psi\).  Specifically, let \(\mathcal N\subseteq\R^5\) be open and \(k\ge1\). A rule \(\psi\) admits a \(C^k\) extension to \(\mathcal N\) if there is a \(C^k\) function
\(\bar\psi:\mathcal N\to\R\) such that
\(\bar\psi(s)=\psi(s)\) at every feasible state \(s\in\mathcal N\). We call such \(\bar\psi\) an \textit{extension} of \(\psi\).  The derivatives of \(\bar\psi\) at points of \(\mathcal N\) with \(u\ge0\) and \(w\ge0\) are determined by \(\psi\), so they do not depend on the extension chosen.\footnote{The rule \(\psi(s)\) is not defined for \(s\) with \(u=0\) or \(w=0\). However, since we assume that the derivatives are continuous (i.e., the \(C^k\) property), the derivatives of \(\bar\psi\) for \(s\) with \(u\ge0\) and \(w\ge0\) are uniquely determined.} We write \(\bar\psi_a\), \(\bar\psi_b\), and so on for partial derivatives. Note that if \(\Psi\) admits a \(C^k\) extension to an open neighborhood of \((a,b,0)\) in the space of posterior means and symmetric covariance matrices, then the corresponding rule \(\psi\) also admits a \(C^k\) extension near \((a,b,0,\bar x,0)\), because \(\Sigma(u,\bar x,w)\) is polynomial in the state.

\subsection{Berk--Nash equilibria and transversality}\label{sec:equilibria}
\citet{ep2016} introduce the notion of a \textit{Berk--Nash equilibrium}, which
is a steady state of a (possibly misspecified) learning problem. A pair of an action and a belief is a Berk--Nash equilibrium if (i) the equilibrium action is optimal against the equilibrium belief, and (ii) the equilibrium belief concentrates on (a subset of) the parameters which best explain the data, in that these parameters minimize the Kullback--Leibler divergence (KL divergence) between the perceived distribution of outcomes and the true distribution given the equilibrium action. In many learning problems with a myopic decision maker, if the chosen action converges, it must be a Berk--Nash equilibrium action (e.g., \cite{fls2021} and \cite{epy2021}).

In our setup, the true parameter \(\vartheta=(\alpha,\beta)\) minimizes the KL divergence for every action $x$, as it always induces the correct prediction of the outcome distribution. However, given an action, there are many other parameters which explain the data equally well. Specifically, given the true parameter \(\vartheta=(\alpha,\beta)\) and an action $x$, any parameters  \((a,b)\) on the line $\{(a,b)| a+bx=\alpha+\beta x\}$ predict the outcome correctly, and hence minimize the KL divergence. In other words, given an action $x$, the set of KL minimizers forms the line \(a+bx=\alpha+\beta x\), and the decision maker cannot distinguish parameters on this line if the same action $x$ is chosen forever.
Based on this observation, we define a Berk--Nash equilibrium for a general rule $\psi$ as follows.

\begin{definition}[Berk--Nash equilibrium]\label{def:bn-psi}
Fix the true parameter \(\vartheta=(\alpha,\beta)\) and a decision rule
\(\psi\). A tuple $(a^e,b^e,x^e,w^e)$ with \(w^e\ge0\) is a
\textit{Berk--Nash equilibrium under \(\psi\)} if
\begin{equation}\label{eq:bn-psi}
 a^e+b^ex^e=\alpha+\beta x^e,
 \qquad
 \psi(s_n)\to x^e\ \text{for every sequence of feasible states}\ s_n\to z^e,
\end{equation}
where \(z^e=(a^e,b^e,0,x^e,w^e)\).  If \(w^e=0\), it is a \textit{complete-information Berk--Nash equilibrium under \(\psi\)}, and we denote it by  \((a^e,b^e,x^e)\).
\end{definition}

A tuple $z^e=(a^e,b^e,0,x^e,w^e)$ and $x^e$ in the definition above can be interpreted as an equilibrium belief and an equilibrium action, respectively. To see this, note that this equilibrium belief $z^e=(a^e,b^e,0,x^e,w^e)$ is a degenerate distribution, which concentrates on the line \(\{(a,b)|a+bx^e=a^e+b^e x^e\}\). The \textit{no-surprise condition} $ a^e+b^ex^e=\alpha+\beta x^e$ appearing in (\ref{eq:bn-psi}), which asserts that the posterior mean $(a^e,b^e)$ correctly predicts the average outcome, guarantees that this line coincides with the set of KL minimizers discussed earlier.
So the equilibrium belief $z^e$ indeed concentrates on the KL minimizers given the equilibrium action $x^e$. The second condition in the definition, $\psi(s_n)\to x^e$, simply asserts that any belief $s_n$ close to the equilibrium belief $z^e$ induces an action close to the equilibrium action $x^e$. Here we consider a limit of the actions $\psi(s_n)$, as the rule $\psi$ does not specify an action for the equilibrium belief $z^e$; the domain of the rule $\psi$ is the set of feasible beliefs, so the degenerate belief $z^e$ is ruled out. For the case  in which $w^e=0$, the equilibrium belief $z^e=(a^e,b^e,0,x^e,w^e)$ concentrates on the posterior mean $(a^e,b^e)$. Hence we call it a complete-information equilibrium.

For the special case of the myopic decision maker, a Berk--Nash equilibrium  defined above is exactly the same as that of \citet{ep2016}, provided that some regularity condition holds and an equilibrium belief is restricted to be normal. Note that under an equilibrium belief $z^e=(a^e,b^e,0,x^e,w^e)$ and an action $z$, the outcome is predicted to be normal with mean \(a^e+b^ez\) and variance \(1+w^e(z-x^e)^2\). Hence for the myopic case,  if the optimal action is unique for each belief,  a standard argument shows that the condition (\ref{eq:bn-psi}) is equivalent to
\begin{align}
 & x^e\in\arg\max_{z\in X}E\left[U\left(z,a^e+b^ez+\sqrt{1+w^e(z-x^e)^2}\,Z\right)\right], \label{eq:bne-ic}\\
 & a^e+b^ex^e=\alpha+\beta x^e. \label{eq:bne-ns}
\end{align}
The first condition is \textit{incentive compatibility}, which says that the equilibrium action \(x^e\) is optimal given the equilibrium belief. The second is
the no-surprise condition, which is the same as the one in the definition above. For complete-information Berk--Nash equilibrium, the equilibrium belief is concentrated on the mean \(( a^e,b^e)\), and hence incentive compatibility is further simplified as
\[
 x^e\in\arg\max_{z\in X}G(z,a^e+b^ez),
\]
where
\[
 G(x,m)=E[U(x,m+Z)]
\]
denotes the expected payoff from an action \(x\) when the mean outcome is \(m\).

As we will show through examples in the next subsection, given the true parameter $\vartheta=(\alpha,\beta)$, the true parameter itself is a (complete-information) Berk--Nash equilibrium belief under many rules including the myopic rule. However, in general, there are many other equilibria, and our question is whether such an ``incorrect'' equilibrium belief can arise as a long-run outcome. It turns out that the following \textit{transversality condition} is critical to answer this question. A rule \(\psi\) has a \textit{complete-information limit} on an open set \(\mathcal A\subseteq\R^2\) if there is a map \(h:\mathcal A\to X\) such that \(\psi(s_n)\to h(a,b)\) for every \((a,b)\in\mathcal A\), every \(\bar x\in X\), and every sequence of feasible states \(s_n\to(a,b,0,\bar x,0)\). We call \(h(a,b)\) the \textit{complete-information action}, that is, the action that the rule takes in the limit where the parameter \((a,b)\) is known. The definition requires it not to depend on the average past action, which is automatic if \(\Psi\) admits a continuous extension to \(\Sigma=0\), jointly in \((a,b,\Sigma)\); then \(h(a,b)\) is the value of this extension at \((a,b,0)\).

Consider a complete-information Berk–Nash equilibrium $(a^e,b^e,x^e)$ under $\psi$, and let $z^e=(a^e,b^e,0,x^e,0)$ be the corresponding equilibrium belief. Suppose that $\psi$ admits a $C^1$ extension $\bar\psi$ to an open neighborhood of $z^e$. Since $\bar\psi$ is continuous, the equilibrium condition implies $\bar\psi(z^e)=x^e$. The \textit{response coefficient} of a
complete-information equilibrium \((a^e,b^e,x^e)\) is
\[
 \gamma^e=\bar\psi_b(z^e)-x^e\bar\psi_a(z^e),
\]
which does not depend on the extension. Now, consider the special case in which the rule $\psi$ has a complete-information limit \(h\) on a neighborhood of \((a^e,b^e)\).\footnote{ This is not automatic even though we assume the existence of $C^1$ extension $\bar\psi$. Indeed, this $C^1$ function $\bar\psi$ may induce different limit actions for different $\bar x$, which is not allowed for the rule to have a complete-information limit.} Then the response coefficient above is rewritten as
\[
 \gamma^e=h_b(a^e,b^e)-h(a^e,b^e)\,h_a(a^e,b^e).
\]
The \textit{transversality condition} is \(\gamma^e\ne0\). Intuitively, it requires that the decision rule responds to changes in the posterior means along the line of KL minimizers associated with the equilibrium action. For the special case in which the rule depends on the belief only through the posterior slope $b_t$ (i.e., the rule takes the form $\psi(b)$), we have $\gamma^e= \psi^{\prime}(b^e)$, so the transversality reduces to $ \psi^{\prime}(b^e) \neq 0$. So in this special case, the transversality simply requires that the posterior slope has a first-order impact on actions.

\begin{remark}
\citet{epy2021} show that the frequency of actions can converge to a
mixed-action equilibrium even when the actions themselves do not. In our model we can restrict attention
to pure-strategy equilibria. For every nondegenerate mixed action, the minimizer of the
Kullback--Leibler divergence is unique (it is the true parameter
\(\vartheta\)), and a decision rule is deterministic.
\end{remark}

\subsection{Examples}\label{sec:examples}

For the myopic case, the true parameter forms a complete-information Berk--Nash equilibrium belief whenever
\(z\mapsto G(z,\alpha+\beta z)\) has a maximizer on \(X\), for example when \(G\)
is continuous. However, there are typically many others. We now present three
economic examples and show what the set of Berk--Nash equilibrium beliefs
looks like. Let \(\Pi_X(z)=\min\{\overline x,\max\{\underline x,z\}\}\) denote
the projection onto the action interval \(X\). In the examples,
\(\underline x<\overline x\).

\begin{figure}[t]
\centering
\begin{subfigure}[t]{0.31\textwidth}
\centering
\includegraphics{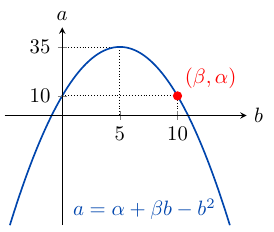}
\caption{Advertising, \(\alpha=\beta=10\).}
\label{figure:BerkNash-Ad}
\end{subfigure}
\
\begin{subfigure}[t]{0.30\textwidth}
\centering
\includegraphics{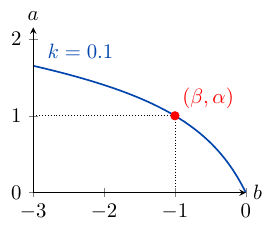}
\caption{Pricing, \(k=0.1\).}
\label{figure:BerkNash-Monopoly}
\end{subfigure}
\
\begin{subfigure}[t]{0.32\textwidth}
\centering
\includegraphics{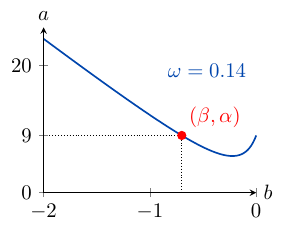}
\caption{Inflation, \((\alpha,\beta,y^*,\omega)=(9,-0.7,0,0.14)\).}
\label{figure:BerkNash-Inflation}
\end{subfigure}
\caption{Berk--Nash equilibrium beliefs \((a,b)\) in the three examples.}
\label{figure:BerkNash}
\end{figure}

\begin{example}[name=Advertising]\label{ex:advertising}
A firm chooses a (signed) campaign reallocation \(x\), observes a resulting
sales performance \(y=\alpha+\beta x+\varepsilon\), and receives the payoff
\(U(x,y)=y-C(x)\), where \(C\) is a smooth cost of reallocation with
\(C''>0\).
Reallocation can move effort in either direction around a neutral allocation.
The expected payoff \(a_t+b_tx-C(x)\) does not depend on the posterior
covariance, and the myopically optimal action maximizes it over \(X\). When
the maximizer is interior, it is \(h(a,b)=(C')^{-1}(b)\), so \(h_a=0\),
\(h_b=1/C''(h)\), and the response coefficient is
\(\gamma^e=1/C''(x^e)>0\). For the quadratic cost \(C(x)=x^2/2\), simple algebra
shows that the myopic rule is \(\psi_0(s_t)=\Pi_X(b_t)\), which we use below.
Automated allocation motivates this decision-feedback problem
\citep{aggarwal2019autobid,aggarwal2024survey}. Learning with advertising
budgets is studied in a strategic auction environment by
\citet{balseirogur2019}; our model abstracts from competition among
advertisers and focuses on how a firm's decisions affect its own learning
about advertising returns.

With the quadratic cost, the equilibrium conditions show that every posterior
mean \((a,b)\) satisfying
\[
 a=\alpha+\beta b-b^2,\qquad b\in\operatorname{int}X,
\]
is a Berk--Nash equilibrium belief, with action \(x=b\). So there is a continuum
of Berk--Nash equilibrium beliefs, and for every incorrect intercept \(a\) that
is not too large relative to \(\alpha\) there is a corresponding equilibrium
belief. Figure~\ref{figure:BerkNash-Ad} describes the set of
Berk--Nash equilibrium beliefs when \(\alpha=\beta=10\).
\end{example}

\begin{example}[name=Pricing]\label{ex:pricing}
A seller posts a price \(x\in X\Subset(0,\infty)\),\footnote{We write
\(A\Subset B\) when the closure of \(A\) is compact and contained in the
interior of \(B\).} observes a noisy demand index
\(y=\alpha+\beta x+\varepsilon\), and receives the payoff \(U(x,y)=(x-k)y\),
where \(k\ge0\) is a constant marginal cost. Since the prior is nondegenerate
Gaussian, the true slope \(\beta\) may be positive, even though the
economically relevant case is \(\beta<0\); and since the noise is Gaussian, the
demand index may be negative. Bayesian pricing and demand learning provide the
model precedent \citep{hkz2012,kz2014}.

When the posterior slope is negative, \(b_t<0\), simple algebra shows that the
myopically optimal price is \(\psi_0(s_t)=\Pi_X(k/2-a_t/(2b_t))\). As in the
advertising example, the model admits a continuum of Berk--Nash equilibrium
beliefs. For example, suppose that the true parameter is \(\vartheta=(1,-1)\).
Then every \((a,b)\) with \(b<0\) satisfying
\[
 a=\frac{b[2-k(1+b)]}{b-1},\qquad
 \frac k2-\frac a{2b}\in\operatorname{int}X,
\]
is a Berk--Nash equilibrium belief. Figure~\ref{figure:BerkNash-Monopoly}
describes the set of Berk--Nash equilibrium beliefs with \(b<0\) for the cost
\(k=0.1\), ignoring the constraint that the equilibrium price lies in
\(\operatorname{int}X\).
\end{example}

\begin{example}[name=Inflation]\label{ex:inflation}
A central bank chooses a policy instrument \(x\in X\) and
observes an inflation rate \(y=\alpha+\beta x+\varepsilon\). Its payoff takes the
loss-function form \(U(x,y)=-[(y-y^*)^2+\omega x^2]\), where \(y^*\) is the
target inflation rate and \(\omega>0\) measures the cost of using the policy
instrument. Simple algebra shows that the myopically optimal action is
\[
 \Psi_0(a,b,\Sigma)
 =\Pi_X\left(\frac{b(y^*-a)-\Sigma_{12}}{b^2+\Sigma_{22}+\omega}\right)
 =\Pi_X\left(\frac{b(y^*-a)+w\bar x}{b^2+w+\omega}\right).
\]
The covariance terms reflect caution. When the slope is uncertain, an action
far from past actions makes inflation less predictable, so the bank stays
closer to the average past action. The complete-information action is
\(h(a,b)=\Pi_X(b(y^*-a)/(b^2+\omega))\). At a degenerate belief the expected
loss \((a+bx-y^*)^2+1+\omega x^2\) is strictly convex in \(x\). Thus, for
\(\beta b+\omega\ne0\), the set of Berk--Nash equilibrium beliefs is
characterized by
\[
 a=\frac{(\alpha-y^*)b^2+\beta y^*b+\alpha\omega}{\beta b+\omega},
 \qquad
 \frac{b(y^*-a)}{b^2+\omega}\in\operatorname{int}X,
\]
again a continuum. Figure~\ref{figure:BerkNash-Inflation} describes the set of
Berk--Nash equilibrium beliefs with \(b<0\) for the calibration
\((\alpha,\beta,y^*,\omega)=(9,-0.7,0,0.14)\) of
\citet{wieland2000monetary}, ignoring the constraint that the equilibrium
action lies in \(\operatorname{int}X\).
\end{example}

\subsection{Complete learning}\label{sec:complete-learning}
This paper concerns whether \textit{complete learning} occurs, in the sense that
the posterior eventually concentrates on the true parameter. The following
lemma shows that this happens precisely when slope precision \(V_t\) diverges to infinity, except on an event of probability zero. Since \eqref{eq:design-sum} shows that \(V_t\) is determined by the variance of past actions, this implies that complete learning occurs only when \textit{the chosen actions have a sufficiently large variance}. Let \(V_\infty=\lim_{t\to\infty}V_t\), which exists because \(V_t\) is nondecreasing. The proof is in
Appendix~\ref{app:posterior}.

\begin{lemma}[Complete learning and action variance]\label{lem:complete-learning}
For every decision rule \(\psi\), under the prior-predictive law
\(P^{\mu,\psi}\), there is an event of probability one on which all of the
following hold:
\begin{enumerate}[label=\arabic*.,ref=\arabic*]
\item The posterior means \((a_t,b_t)\) converge to finite limits\label{lem:complete-learning:1}
\((a_\infty,b_\infty)\).
\item If \(V_t\to\infty\), then \((a_\infty,b_\infty)=(\alpha,\beta)\) and\label{lem:complete-learning:2}
each entry of the covariance matrix \(\Sigma_t\) converges to zero.
\item If \(V_\infty<\infty\), then \(b_\infty\ne\beta\), the posterior distribution of the slope\label{lem:complete-learning:3}
converges weakly to \(N(b_\infty,V_\infty^{-1})\), the average action
\(\bar x_t\) and the action \(x_t\) converge to a common finite limit
\(\bar x_\infty\), and \(a_\infty+b_\infty\bar x_\infty=\alpha+\beta\bar x_\infty\).
\end{enumerate}
\end{lemma}

The lemma shows that when the slope precision $V_{t}$ goes to infinity, correct learning occurs and the posterior converges to the point-mass belief on the true parameter. On the other hand, when the slope precision $V_{\infty}$ is bounded, the decision maker fails to learn the state and the posterior mean $(a_t,b_t)$ converges to a point which differs from the truth. In the latter case, the limiting posterior mean $(a_{\infty},b_{\infty})$ satisfies the no-surprise condition for the limiting action \(\bar x_\infty\), so that the decision maker correctly predicts the outcome distribution. For the special case in which the rule \(\psi\) has a limit as feasible states approach \(z_\infty=(a_\infty,b_\infty,0,\bar x_\infty,V_\infty^{-1})\), this implies that the limit outcome \((a_\infty,b_\infty,\bar x_\infty,V_\infty^{-1})\) is almost surely an incorrect Berk–Nash equilibrium under \(\psi\).\footnote{ This follows from the fact that $s_t \to z_{\infty}$ and $\psi(s_t)=x_{t+1} \to \bar x_{\infty}$. }

A high-level intuition behind this lemma is as follows. In our model, the
decision maker needs to learn two parameters \((\alpha,\beta)\) from a
one-dimensional outcome \(y\). Accordingly, complete learning requires a
sufficiently large variation of the chosen actions;  indeed, if the
same action \(x\) is chosen every period, then the decision maker cannot
distinguish \((\tilde\alpha,\tilde\beta)\) and
\((\tilde\alpha',\tilde\beta')\) with
\(\tilde\alpha+\tilde\beta x=\tilde\alpha'+\tilde\beta'x\), as they yield the
same mean outcome given this action \(x\). The question is how to measure
the variation of actions, and the proof of the lemma shows that \(V_t\)  is the
right one.  The result here may be reminiscent of the classical
consistency result of the ordinary least-squares estimator, but the critical
difference is that the action \(x_t\) is endogenously chosen in our
model.\footnote{\citet{lr1981} consider a similar problem in a linear
regression model and give an example in which the least-squares slope
estimate is inconsistent although the unnormalized variance of the regressors
diverges. Their example relies on the assumption that the decision maker's
action each period can depend directly on the past noise terms. Here the
decision maker's action depends only on past actions and outcomes, which do
not reveal the noise terms unless the parameter is known, so there is no
contradiction.}

Lemma~\ref{lem:complete-learning} does not need a bounded action set. It
holds for every action set \(X\subseteq\R\), such as \(\R\) or
\([0,\infty)\), and every decision rule with values in that
set.\footnote{Here \(X\) is a Borel set, so that, as for decision rules, the
actions are random variables.} A key is that the average action cannot move without adding slope precision, which rules out divergence of the average action $\bar x_t$ when $V_{\infty} <\infty$.

\section{Main Results}\label{sec:results}

\subsection{Generic failure of complete learning}\label{sec:global}
We first ask whether a decision maker following a decision rule \(\psi\) can learn completely at all. The answer depends on how the rule behaves near complete information, and the following regularity condition is critical. Let \(\mathbb S\) denote the set of symmetric \(2\times2\) matrices.

\begin{definition}[Regular parameters]\label{def:regular}
A parameter \(\vartheta=(\alpha,\beta)\) is \textit{interior regular} for \(\psi\) if \(\Psi\) admits a \(C^1\) extension \(\widetilde\Psi\), with locally Lipschitz first derivatives, to an open neighborhood of \((\vartheta,0)\) in \(\R^2\times\mathbb S\), and the complete-information action \(h(a,b)=\widetilde\Psi(a,b,0)\) satisfies
\[
 h_b(\vartheta)\ne h(\vartheta)\,h_a(\vartheta).
\]
A parameter \(\vartheta=(\alpha,\beta)\) is \textit{locally constant} for
\(\psi\) if there are an action \(e\in X\) and \(r>0\) such that \(\psi\) has a
complete-information limit on
\(\{(a,b):\lvert a-\alpha\rvert<r,\ \lvert b-\beta\rvert<r\}\) and
\(\psi(s)=e\) at every feasible state \(s=(a,b,u,\bar x,w)\) with
\(\lvert a-\alpha\rvert<r\), \(\lvert b-\beta\rvert<r\), \(u<r\), \(w<r\), and
\(\lvert\bar x-e\rvert<r\). Let \(\Theta^T_\psi\) and
\(\Theta^C_\psi\) be the sets of interior regular and locally constant
parameters, and \(\Theta_\psi=\Theta^T_\psi\cup\Theta^C_\psi\); a parameter in
\(\Theta_\psi\) is \textit{regular} for \(\psi\).
\end{definition}

In words, an interior regular parameter is one at which the rule is a smooth
function of the mean and the covariance matrix near complete information and
satisfies the transversality condition. Such a rule has a
complete-information limit on a neighborhood of \(\vartheta\), equal to
\(\Psi(\cdot,0)\) there by continuity of \(\Psi\) (step~\ref{cond:I1} of the proof of
Corollary~\ref{cor:covariance}, which uses only the representation).
So \(h\) does not depend on the choice of \(\Psi\), and
\(\gamma(\vartheta)=h_b(\vartheta)-h(\vartheta)h_a(\vartheta)\) is the response
coefficient at the true parameter. A locally constant parameter is typically
one at which the rule saturates at an endpoint of \(X\), so there the action does
not move at all once the decision maker is close to complete information and
the average action is close to that endpoint. The complete-information limit
cannot be dropped. Without it, a rule could choose \(e\) near \(\bar x=e\) but,
away from \(e\), alternate between \(\bar x+\delta\) and \(\bar x-\delta\)
according to the parity of the sample size, which keeps \(\bar x\) away from
\(e\) while the slope precision grows without bound. Both
sets are open (Appendix~\ref{app:global}), so \(\Theta_\psi\) is a Borel set.

Let \(\mathcal W=\{V_\infty<\infty,\ b_t\to b_\infty\ne\beta\}\) be the event on
which the decision maker ends up with an incorrect posterior slope, and let
\(\mathcal L\) be the event on which \((a_t,b_t)\) converges to a finite limit
\((a_\infty,b_\infty)\), \(\bar x_t\) and \(x_t\) converge to a common finite
limit \(\bar x_\infty\), and
\(a_\infty+b_\infty\bar x_\infty=\alpha+\beta\bar x_\infty\). The
following theorem shows that the event \(\mathcal W\) occurs with probability one for
almost all regular parameters, and that the actions then settle at a
limit at which the decision maker is not surprised. The proof is in Appendix~\ref{app:global},
which proves the same conclusions on a set of parameters at least as large, defined
through smoothness of the rule in the posterior state
(Theorem~\ref{thm:smooth-global}).

\begin{theorem}[Generic failure of complete learning]\label{thm:global}
For every nondegenerate Gaussian prior \(\mu\),
\[
 P^{\mu,\psi}(V_t\to\infty,\ \vartheta\in\Theta_\psi)=0.
\]
Consequently, for every Borel set \(B\subseteq\Theta_\psi\),
\[
 P^{\mu,\psi}(\mathcal W,\ \vartheta\in B)=\mu(B),
\]
and \(P^{\psi}_\vartheta(\mathcal W)=1\) for Lebesgue-almost every
\(\vartheta\in\Theta_\psi\). The last two conclusions remain true with
\(\mathcal W\) replaced by \(\mathcal W\cap\mathcal L\).
\end{theorem}

The theorem implies that
\[
 \mu(\Theta_\psi)>0\ \Longrightarrow\ P^{\mu,\psi}(\mathcal W)>0,
 \qquad
 \mu(\Theta_\psi)=1\ \Longrightarrow\ P^{\mu,\psi}(\mathcal W)=1.
\]
So when regularity holds almost everywhere under the prior, the decision rule
leads to incorrect learning almost surely.  In particular, from Lemma \ref{lem:complete-learning},  the limiting outcome is an incorrect Berk--Nash equilibrium, provided that the rule \(\psi\) has a limit as feasible states approach \(z_\infty=(a_\infty,b_\infty,0,\bar x_\infty,V_\infty^{-1})\).

We prove this theorem by contradiction. Pick an interior regular parameter. Suppose, contrary to the theorem, that complete learning occurs. Then the posterior converges to the true parameter and the action converges to the complete-information action \(h(\vartheta)\). Near complete information the rule is smooth and responds to slope revisions. So the noise keeps moving the posterior means, and the posterior means keep moving the action. In the proof, we show that a variation in actions generated by this mechanism is bounded in that $V_{\infty}<\infty$. This contradicts Lemma \ref{lem:complete-learning}, which says that, almost surely, complete learning requires $V_{\infty} = \infty$.

So the key step in the proof is to show that the variation in actions is so small that $V_{\infty} < \infty$. To illustrate the idea, suppose that the rule takes the form \(x_{t+1}=\psi(b_t)\), so that
the action depends on the posterior only through the posterior slope \(b_t\); the myopic advertising rule \(\psi(b)=b\) is an instance. When
the posterior is concentrated on the truth and \(A_t\) is large, we have \(V_{t+1}/V_t\approx 1\), \(A_t/(A_t+1)\approx 1\), and $I_{t+1}=\varepsilon_{t+1} +o(1)$. Hence the recursion for the posterior slope $b_t$ appearing in (\ref{eq:means}) is approximately
\[
 b_{t+1}-b_t\approx\frac{D_t}{V_t}\varepsilon_{t+1}.
\]
So the size of the slope updating is proportional to the action gap \(D_t=x_{t+1}-\bar x_t\). Since the next action depends on \(b_{t+1}\),
\[
 x_{t+2}-x_{t+1}
 \approx\psi'(b_t)(b_{t+1}-b_t)
 \approx\psi'(b_t)\frac{D_t}{V_t}\varepsilon_{t+1}.
\]
Since the average action changes slowly for large \(A_t\), this gives
\[
 D_{t+1}\approx\left(1+\frac{\psi'(b_t)}{V_t}\varepsilon_{t+1}\right)D_t.
\]
Thus the action gap follows an approximately \textit{multiplicative process}. In particular, since transversality ensures $\psi'(b_t) \neq 0$, the multiplier $1+\frac{\psi'(b_t)}{V_t}\varepsilon_{t+1}$ is stochastic.

This stochastic multiplicative structure creates a contraction force on the action gap $D_t$. While the mean of the
multiplier \(1+\psi'(b_t)\varepsilon_{t+1}/V_t\) is approximately one, the expected logarithm of its absolute value is negative, as
\(E[\log\lvert1+c\varepsilon\rvert]<0\) for every small nonzero \(c\). Hence the action gap $D_t$ tends to shrink. The economic intuition is simple. A small action gap generates less informative data, which moves the posterior slope less, which in turn moves the next action less. Hence, once the action gap becomes small, it is hard for
it to become large again.

An important complication is that this contraction force becomes weaker as \(V_t\) increases. Indeed, when $V_t$ goes to infinity, even the expected  logarithm, \(E[ \log |1+\psi'(b_t)\varepsilon_{t+1}/V_t|]\), approaches zero. This suggests two possibilities: If the action gap shrinks quickly, little information accumulates, \(V_t\) remains bounded, and the contraction continues. On the other hand, if it shrinks slowly, more information accumulates, \(V_t\) grows, which weakens the contraction further. In the proof, we show that the first possibility prevails, so that slope precision $V_t$ cannot diverge to infinity at a regular parameter.

Transversality is what makes this contraction force work. If the transversality condition fails so that \(\psi'=0\), the stochastic component of the multiplier vanishes, so that the contraction force disappears;  accordingly, the action gap may not shrink sufficiently quickly. This may sound paradoxical, because naively, it is natural to expect that a rule whose action does not respond to beliefs at the margin, \(\psi'=0\), should produce less variation in actions. A key is that when the action is responsive to the belief, there is a chance that the action moves to the direction with which the action gap $D_t$ shrinks. Once the gap becomes small, subsequent belief revisions become smaller, making it difficult for the action gap to grow again. This is why responsiveness of actions \textit{helps} to stabilize the evolution of the action gap $D_t$. Transversality is indeed critical for the theorem above; in Section \ref{sec:rules}, we present an example in which transversality fails so that correct learning occurs with  probability close to one.

The same argument extends to a general rule \(\psi(a,b,u,\bar x,w)\). Tedious algebra shows that
\[
D_{t+1}\approx
\left(1+\frac{\gamma_t}{V_t}
\varepsilon_{t+1}\right)D_t
+\frac{\psi_a}{A_t}\varepsilon_{t+1}
+\frac{\psi_{\bar x}-1}{A_t+1}D_t
-\frac{\psi_u}{A_t^2}
-\frac{\psi_w}{V_t^2}D_t^2, \]
where $\gamma_t=
\psi_b(s_t)-\bar x_t\psi_a(s_t)$ is the response coefficient evaluated at the belief $s_t$. Very roughly, transversality, $\gamma^e\neq 0$, guarantees that the first term on the right-hand side generates the same contraction force as in the special case above. In the proof, we show that the other terms in the right-hand side do not undermine this contraction mechanism, so that the action gap $D_t$ indeed converges to zero quickly.

The theorem has a stark implication when \(\mu(\Theta_\psi)=1\). At the outset,
the decision maker knows the prior, the correctly specified model, and the rule
it follows. Under these beliefs, it assigns probability one to eventually
holding an incorrect belief about the slope. Thus the failure of complete
learning is not an unforeseen consequence of the rule---the decision maker knows
ex ante that its own decisions will not reveal the true parameter.

The conclusion also does not require the true parameter to be drawn from the
decision maker's prior. Since \(P^\psi_\vartheta(\mathcal W)=1\) at
Lebesgue-almost every \(\vartheta\in\Theta_\psi\), Fubini's theorem shows
that, if the true parameter is drawn from a distribution with a density
while the decision maker starts from \(\mu\), incorrect learning occurs on
every Borel set \(B\subseteq\Theta_\psi\) with exactly the probability that
this distribution assigns to \(B\).

Theorem~\ref{thm:global} shows that if the rule \(\psi\) is regular at
almost all parameters $\vartheta=(\alpha,\beta)$, then incorrect learning occurs almost surely. We next consider how common such a rule is, and show that the set of such rules contains a residual set in a space of rules that are \(C^2\) near complete information. In topology, a property which holds on a residual set is typically thought of as a generic property; for example, the set of irrational numbers is residual in $\mathbb R$. In this sense, for a generic choice of the rule $\psi$, complete learning occurs with zero probability, whatever the prior. Also, we show that a somewhat similar (but slightly weaker) result holds for an open and dense set of rules, i.e., we show that the probability that complete learning occurs is arbitrarily close to zero for such rules.\footnote{ An open and dense set is a stronger notion of genericity than a residual set. Indeed, any open and dense set is residual in any topological space, while the converse is not true. }

Suppose \(\underline x<\overline x\), and fix an open set
\(U\subseteq\R^2\times\mathbb S\) with \(\R^2\times\{0\}\subseteq U\). Let
\(\mathcal P(U)\) be the set of Borel functions
\(\Psi:\R^2\times\mathbb S\to X\) that are \(C^2\) on \(U\) with bounded
derivatives of order at most two there. Each \(\Psi\) defines the rule
\(\psi(s)=\Psi(a,b,\Sigma(u,\bar x,w))\), its complete-information action
\(h(a,b)=\Psi(a,b,0)\), and its response coefficient \(\gamma=h_b-h\,h_a\) on
\(\R^2\), and we measure the distance between two such functions by
\[
 d(\Psi,\Psi')=\sup_{\R^2\times\mathbb S}\lvert\Psi-\Psi'\rvert
 +\sum_{1\le\lvert\kappa\rvert\le2}\ \sup_U\lvert\partial^\kappa\Psi-\partial^\kappa\Psi'\rvert,
\]
where \(\kappa\) runs over multi-indices in the five coordinates of
\(\R^2\times\mathbb S\).

\begin{corollary}[Generic rules]\label{cor:generic}
Suppose \(\underline x<\overline x\), and fix an open set
\(U\supseteq\R^2\times\{0\}\).
\begin{enumerate}[label=\textup{(\roman*)},ref=(\roman*)]
\item The set \(\mathcal R\) of \(\Psi\in\mathcal P(U)\) whose response\label{cor:generic:i}
coefficient vanishes only on a Lebesgue-null subset of \(\R^2\) is a dense
\(G_\delta\), hence residual, subset of \(\mathcal P(U)\) with respect to
\(d\). For every \(\Psi\in\mathcal R\) and every nondegenerate Gaussian prior
\(\mu\), \(P^{\mu,\psi}(V_t\to\infty)=0\), and
\(P^{\psi}_\vartheta(\mathcal W)=1\) for Lebesgue-almost every
\(\vartheta\in\R^2\).
\item For every nondegenerate Gaussian prior \(\mu\) and every\label{cor:generic:ii}
\(\varepsilon>0\), the set of \(\Psi\in\mathcal P(U)\) under which
\(P^{\mu,\psi}(V_t\to\infty)<\varepsilon\) contains an open and dense subset of
\(\mathcal P(U)\) with respect to \(d\).
\end{enumerate}
\end{corollary}

The residual set \(\mathcal R\) does not depend on the prior. The open and
dense set in~\ref{cor:generic:ii} consists of the rules whose response
coefficient vanishes only on a set of prior probability less than
\(\varepsilon\). Asking for transversality at every parameter of a
disk instead would not work. That set of
rules is open but not dense, because a response coefficient that changes sign
on a disk keeps changing sign under every small perturbation of the rule. The
proof is in Appendix~\ref{app:global}.

\subsection{Convergence to incorrect Berk--Nash equilibria}\label{sec:local}
The previous theorem does not say where the posterior goes. We next ask
whether it can settle at a particular incorrect Berk--Nash equilibrium. The existence of incorrect equilibria does not imply that they arise as long-run outcomes of Bayesian learning. Even if actions converge to an equilibrium action, small continuing adjustments may provide enough information to identify the true parameter. The following theorem shows that an incorrect Berk--Nash equilibrium can nevertheless persist.

To be more precise, pick a complete-information Berk--Nash equilibrium belief $(a^e,b^e,0,x^e,0)$ which is approximated by a ``near-complete-information'' Berk--Nash equilibrium belief $(a_w,b_w,0,x^e,w)$ for $w>0$. The first part of the following theorem shows that such approximation is indeed possible under transversality and smoothness of the rule $\psi$. Then the second part of the theorem shows that if the initial prior is sufficiently close to that near-complete-information equilibrium belief, then with probability at least $1-\varepsilon$, the posterior belief remains in a neighborhood of that equilibrium, which is in a neighborhood of the original complete-information equilibrium, and eventually converges.\footnote{ Here it is convenient to consider an equilibrium with $w>0$, because in the limit we are interested in, we have $V_{\infty}<\infty$ so that $w_t$ does not converge to zero (although it is close to zero).  } The proof is in Appendix~\ref{app:local}.

\begin{theorem}[Local persistence]\label{thm:smooth-local}
Fix the true parameter \(\vartheta=(\alpha,\beta)\) and a complete-information
Berk--Nash equilibrium \((a^e,b^e,x^e)\) under \(\psi\). Suppose that
\(\psi\) admits a \(C^1\) extension \(\bar\psi\), with locally Lipschitz
first derivatives, to an open neighborhood \(\mathcal N\) of
\(z^e=(a^e,b^e,0,x^e,0)\), and that the response coefficient satisfies \(\gamma^e\ne0\).
Then there are \(\bar w>0\) and \(K\ge0\) such that, for every
\(w\in[0,\bar w)\), there is \((a_w,b_w)\) with
\(\lvert a_w-a^e\rvert+\lvert b_w-b^e\rvert\le Kw\) such that
\((a_w,b_w,0,x^e,w)\in\mathcal N\) and \((a_w,b_w,x^e,w)\) is a Berk--Nash
equilibrium under \(\psi\). Also, there is \(\bar r>0\) such that, for every \(r\in(0,\bar r)\), there is
\(\overline w_r\in(0,\bar w)\) such that for every
\(w^e\in(0,\overline w_r]\) and every \(\varepsilon\in(0,1)\), there is \(r_0>0\)
such that, for every nondegenerate Gaussian prior whose state
\(s_0=(a_0,b_0,u_0,\bar x_0,w_0)\) is \(r_0\)-close to the state of the
Berk--Nash equilibrium \((a_{w^e},b_{w^e},x^e,w^e)\), that is,
\[
 \lvert a_0-a_{w^e}\rvert+\lvert b_0-b_{w^e}\rvert+u_0+\lvert\bar x_0-x^e\rvert
 +\lvert w_0-w^e\rvert<r_0,
\]
the following hold simultaneously with \(P_\vartheta^{\psi}\)-probability at
least \(1-\varepsilon\):
\begin{enumerate}[label=\textup{(\roman*)},ref=(\roman*)]
\item For every \(t\ge0\),\label{thm:smooth-local:i}
\(\frac{w^e}{2}\le w_t\le 2w^e\), and \((a_t,b_t,\bar x_t)\) is \(r\)-close to
the complete-information Berk--Nash equilibrium \((a^e,b^e,x^e)\), that is,
\(\lvert a_t-a^e\rvert+\lvert b_t-b^e\rvert+\lvert\bar x_t-x^e\rvert<r\).
\item \(V_\infty<\infty\); the posterior slope converges to a finite limit\label{thm:smooth-local:ii}
\(b_\infty\); the average action \(\bar x_t\) and the action \(x_t\) converge to
a common finite limit \(\bar x_\infty\); and
\(a_t\to\alpha+\beta\bar x_\infty-b_\infty\bar x_\infty\). If \(b^e\ne\beta\), then
\(b_\infty\ne\beta\).
\end{enumerate}
\end{theorem}

When the equilibrium belief is incorrect, the limiting slope is incorrect,
even when \(r\) exceeds \(\lvert b^e-\beta\rvert\); by
\ref{thm:smooth-local:i}, it also lies within \(r\) of \(b^e\).

This result shows that the result of \cite{hks2018}, who consider  Bayesian learning by a decision maker with dogmatic overconfidence on her capability, extends to the case in which a decision maker does not have such a dogmatic belief, provided that the initial belief is concentrated near the equilibrium. Consider a decision maker who chooses effort $x$ and observes an output $y$ each period. The output distribution depends on the chosen effort $x$, as well as the decision maker's capability $\alpha$ and an unknown state $\beta$, as in our model. Crucially, \cite{hks2018} assume that the decision maker has overconfidence about her own capability, in that she incorrectly believes that her capability is $A>\alpha$, and updates only the belief about the unknown state $\beta$. They show that the posterior belief almost surely converges to a Berk--Nash equilibrium.

A natural question is whether a similar result obtains if the decision maker's belief about her capability is not dogmatic. Indeed, if she observes a large amount of data which contradicts her view, she may start to doubt her capability and update the belief. Our local persistence theorem shows that the answer is affirmative. A point is that a Berk--Nash equilibrium in the overconfidence model is still a Berk--Nash equilibrium in our environment. To be precise, pick the true parameters $\vartheta=(\alpha,\beta)$, and let $b^*(A)$ denote the Berk--Nash equilibrium when the decision maker incorrectly thinks that her capability is $A>\alpha$. Clearly, this pair $(A,b^*(A))$ forms a complete-information Berk--Nash equilibrium with an appropriate choice of $x^e$. Hence, provided that \(\psi\) admits a \(C^1\) extension with locally Lipschitz first derivatives near this equilibrium and \(\gamma^e\ne0\) there, the theorem above ensures that if the initial prior is chosen from (a subset of) a neighborhood of the degenerate belief on the Berk--Nash equilibrium $(A,b^*(A))$ of the overconfidence model, then with probability close to one, the posterior stays in that neighborhood and eventually converges;  so the limiting outcome is very similar to that of the overconfidence model. In other words, even though we allow the decision maker to revise the belief in a flexible way, her overconfidence about the intercept parameter $\alpha$ may persist as long as the initial belief is highly concentrated on the associated Berk--Nash equilibrium.

The mechanism is the multiplicative contraction of
Section~\ref{sec:global}, now operating near an incorrect equilibrium.
Observations generated by two distinct actions identify both coefficients, so
the variation in actions near the equilibrium keeps generating information
that points toward the true parameter. This is the key difference from strict
equilibria in finite-action environments \citep{fls2021}, where the same action
is chosen as long as beliefs stay nearby. Near our equilibrium, the variation
does not stop, but it shrinks fast enough that the total information it
generates is finite, and the posterior never accumulates enough information to
leave the incorrect belief. Transversality is essential for this, for the
reason given in Section~\ref{sec:global}. If \(\gamma^e=0\), the
multiplicative term disappears. Section~\ref{sec:rules} shows that rules that pin the predicted
outcome to a target, which violate transversality identically, can lead to
complete learning with probability close to one.

Transversality concerns a rule whose action responds to beliefs. At the other
extreme, the rule does not respond at all near the equilibrium. We say that
\(\psi\) is \textit{constant near} a complete-information Berk--Nash
equilibrium \((a^e,b^e,x^e)\) under \(\psi\) if, for some \(\rho>0\),
\(\psi(s)=x^e\) at every feasible state \(s=(a,b,u,\bar x,w)\) with
\(\lvert a-a^e\rvert,\lvert b-b^e\rvert,u,w,\lvert\bar x-x^e\rvert<\rho\).
Unlike local constancy of a parameter (Definition~\ref{def:regular}), this
asks nothing of the rule away from \(\bar x=x^e\). It is the typical case at
an endpoint of \(X\), where the rule saturates. In the advertising example,
the myopic rule \(\Pi_X(b)\) is constant near every equilibrium with
\(b^e>\overline x\). A firm that believes the return to advertising to exceed
the spending cap spends at the cap, and data at the cap do not reveal the
slope. The following proposition needs neither transversality nor
smoothness. The proof is in Appendix~\ref{app:local}.

\begin{proposition}[Local persistence where the rule is constant]\label{thm:corner-local}
Fix the true parameter \(\vartheta=(\alpha,\beta)\) and a
complete-information Berk--Nash equilibrium \((a^e,b^e,x^e)\) under \(\psi\)
such that \(\psi\) is constant near it.
There is \(\bar r>0\) such that, for every \(r\in(0,\bar r)\), there is
\(\overline V_r<\infty\) such that for every \(V^e\ge\overline V_r\) and every
\(\varepsilon\in(0,1)\), there exist \(\overline A<\infty\) and \(r_0>0\) such
that, for every nondegenerate Gaussian prior \(\mu\) with
\[
 A_0>\overline A,
 \qquad
 \lvert a_0-a^e\rvert+\lvert b_0-b^e\rvert+\lvert V_0-V^e\rvert<r_0,
 \qquad
 A_0\lvert\bar x_0-x^e\rvert\bigl(\lvert\bar x_0-x^e\rvert+\lvert e_0\rvert\bigr)<r_0,
\]
where \(e_0=\alpha+\beta x^e-a_0-b_0x^e\), the following hold simultaneously
with \(P_\vartheta^{\psi}\)-probability at least \(1-\varepsilon\):
\begin{enumerate}[label=\textup{(\roman*)},ref=(\roman*)]
\item For every \(t\ge0\),\label{thm:corner-local:i}
\(x_{t+1}=x^e\),
\[
 \lvert a_t-a^e\rvert+\lvert b_t-b^e\rvert+\lvert\bar x_t-x^e\rvert<r,
 \qquad
 \frac{V^e}{2}\le V_t\le2V^e.
\]
\item \(V_\infty<\infty\); the posterior slope converges to a finite limit\label{thm:corner-local:ii}
\(b_\infty\); \(\bar x_t\to x^e\); and
\(a_t\to\alpha+\beta x^e-b_\infty x^e\). If \(b^e\ne\beta\), then
\(b_\infty\ne\beta\).
\end{enumerate}
These priors form a nonempty open set. Moreover, almost surely on the event
that \(x_{t+1}=x^e\) for every \(t\ge0\),
\[
 V_\infty=V_0+A_0d^2,
 \qquad
 b_\infty=b_0+\frac{A_0de_0}{V_0+A_0d^2},
 \qquad
 d=x^e-\bar x_0.
\]
\end{proposition}

Near an equilibrium where the rule is constant, the data come from the single action
\(x^e\). They reveal the mean outcome at \(x^e\) but nothing about the slope,
so the only slope information is what the prior already carries. The prior
treats its average action \(\bar x_0\) as if it had been tried \(A_0\) times,
so the gap \(d\) between \(\bar x_0\) and \(x^e\) acts as variation in past
actions. As the average action moves to \(x^e\), this variation adds slope
precision \(A_0d^2\), and the prediction error \(e_0\) at \(x^e\) moves the
slope by \(A_0de_0/(V_0+A_0d^2)\). The posterior can stay near the
equilibrium only if both are small, and the last condition on the prior makes
them small. At an interior equilibrium under transversality,
Theorem~\ref{thm:smooth-local} needs no such condition. There the rule moves
the action with the beliefs, and the multiplicative contraction of
Section~\ref{sec:global} shrinks the action gap at a rate that does not depend
on \(A_0\). Where the rule is constant, the action stays at \(x^e\),
and the gap \(x^e-\bar x_t=A_0d/A_t\) closes only through averaging, which is
slow when \(A_0\) is large. The condition cannot be replaced by
\(\sqrt{A_0}\lvert\bar x_0-x^e\rvert<r_0\). Take a prior with
\(a_0=a^e-r_0/2\), \(b_0=b^e\), \(V_0=V^e\) and \(d=r_0/(2\sqrt{A_0})\). It
satisfies the first two conditions and
\(\sqrt{A_0}\lvert\bar x_0-x^e\rvert<r_0\), and it has \(e_0=r_0/2\). The
formula gives
\(b_\infty-b^e=\sqrt{A_0}\,r_0^2/\bigl(4(V^e+r_0^2/4)\bigr)\) on the event that
the action stays at \(x^e\). This exceeds \(r\) when \(A_0\) is large, so
conclusion~\ref{thm:corner-local:i} then has probability zero.

Theorems~\ref{thm:global} and~\ref{thm:smooth-local} and
Proposition~\ref{thm:corner-local} describe different aspects of the same problem.
Theorem~\ref{thm:smooth-local} and Proposition~\ref{thm:corner-local} show that a
particular incorrect Berk--Nash equilibrium can persist from nearby
concentrated priors, with probability arbitrarily close to one, the first
when the rule responds to beliefs and the second when it does not.
Theorem~\ref{thm:global} does not identify the limiting belief;
instead, it rules out complete learning on the regular parameter set. Neither
kind of result implies the other. Together, they show both how a specific
incorrect belief can be sustained and why learning under a smooth rule need
not reach the true parameter.

\subsection{Endogenous overconfidence and underconfidence}\label{sec:confidence}
At an incorrect Berk--Nash equilibrium belief \((a,b)\) with action \(x\), the
no-surprise condition gives \(a-\alpha=(\beta-b)x\). So, when the
equilibrium action is nonzero, an incorrect slope comes with an incorrect
intercept. In the language of
\citet{hks2018}, the decision maker is \textit{overconfident} if it
overestimates the intercept, \(a>\alpha\), and \textit{underconfident} if
\(a<\alpha\). In the advertising example with \(x=b>0\), the decision maker is
overconfident exactly when it underestimates the return to advertising,
\(0<b<\beta\). It attributes too much of its sales to its own baseline.

\begin{figure}[t]
\centering
\begin{subfigure}[b]{0.305\textwidth}
\centering
\includegraphics[page=1]{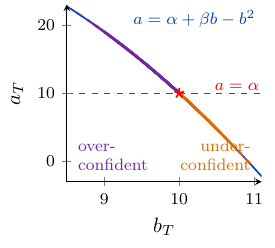}
\caption{Terminal beliefs.}
\label{figure:confidence:beliefs}
\end{subfigure}
\
\begin{subfigure}[b]{0.32\textwidth}
\centering
\includegraphics[page=2]{tikz/fig_myopic_advertising_confidence.pdf}
\caption{Terminal slope.}
\label{figure:confidence:slope}
\end{subfigure}
\
\begin{subfigure}[b]{0.315\textwidth}
\centering
\includegraphics[page=3]{tikz/fig_myopic_advertising_confidence.pdf}
\caption{Intercept error.}
\label{figure:confidence:intercept}
\end{subfigure}
  \caption{Endogenous overconfidence and underconfidence of a myopic
  advertiser. True parameter \(\alpha=\beta=10\), \(X=[0,20]\), rule
  \(x_{t+1}=\Pi_X(b_t)\), and a common Gaussian prior with mean \((10,10)\) and
  precision equal to one observation at \(x=9.5\) plus slope precision
  \(0.1\); 2,000 paths of length \(T=5{,}000\) differ only in their shocks.
  (a)~Terminal posterior means \((b_T,a_T)\) and the Berk--Nash curve
  \(a=\alpha+\beta b-b^2\); the star is the true parameter and the dashed line is
  \(a=\alpha\). (b)~Histogram of \(b_T\), by whether \(a_T>\alpha\)
  (overconfident) or \(a_T<\alpha\) (underconfident). (c)~\(a_t-\alpha\) for the
  first five overconfident and first five underconfident paths (logarithmic
  time axis). Finite-horizon simulation; frequencies are for this fixed true
  parameter and prior.}
  \label{figure:confidence}
\end{figure}

Figure~\ref{figure:confidence} illustrates this by simulation. It follows
2,000 advertisers that use the myopic rule \(x_{t+1}=\Pi_X(b_t)\) of
Example~\ref{ex:advertising}, with the true parameter \(\alpha=\beta=10\) and
\(X=[0,20]\). All advertisers have the same payoff, the same correctly
specified model and the same prior, whose mean is the true parameter; they
differ only in the shocks they observe. After 5,000 periods, their posterior
means lie very close to the curve of Berk--Nash equilibrium beliefs, and their
slope estimates and slope precision have essentially stopped moving. In this
simulation, 56 percent of the paths end overconfident and 44 percent end
underconfident. Which belief a path ends near by the final date depends on
its history, and the same prior produces both behavioral types.

\subsection{Myopic decision makers}\label{sec:applications}
We now verify the conditions for a myopic decision maker with a general
payoff. Myopia is a natural model of many decisions. For an automated system,
it is a design choice rather than an assumption about the patience of the firm.
The system maximizes the per-period objective it was coded to use. Delegating a
recurring decision to an AI agent that is instructed to optimize a per-period
objective, such as automated bidding or pricing, is a recent form of such
myopic automation. Myopia also describes successive managers who inherit the
historical record, and the managerial myopia documented by
\citet{mizikjacobson2007} and \citet{mizik2010}.

We maintain the following assumption on the payoff.

\begin{assumption}\label{ass:payoff}
\begin{enumerate}[label=\textup{(U\arabic*)},ref=\textup{(U\arabic*)}]
\item The action set \(X=[\underline x,\overline x]\) of\label{ass:payoff:U1}
Section~\ref{sec:model} has \(\underline x<\overline x\).
\item For every \(y\in\R\), \(x\mapsto U(x,y)\) is three times continuously\label{ass:payoff:U2}
differentiable on \(X\), with one-sided derivatives at the endpoints, and for
every \(x\in X\), \(y\mapsto U(x,y)\) is Borel.
\item There are constants \(C_U\) and \(c_U\) such that\label{ass:payoff:U3}
\(\lvert\partial_x^jU(x,y)\rvert\le C_Ue^{c_U\lvert y\rvert}\) for every
\(x\in X\), \(y\in\R\) and \(j\le3\).
\end{enumerate}
\end{assumption}

Condition \ref{ass:payoff:U3} ensures that the expected payoffs below are finite and can be
differentiated under the expectation, which the first- and second-order
conditions of the analysis use. The payoff needs to be smooth in the action
only. Because the outcome is normal given the belief, the expected payoff is
smooth in the predicted mean and variance even when the payoff is not smooth
in the outcome. So the assumption covers every polynomial payoff, CARA
utility \(-e^{-\rho(y-C(x))}\), and kinked payoffs such as the profit
\((x-k)\max\{y,0\}\) of a seller whose demand is truncated at zero and the
capacity-constrained payoff \(\min\{y,\kappa\}-C(x)\), with a \(C^3\) cost
\(C\) in the CARA and capacity cases. Write
\(g^{\mathrm{full}}(x;\vartheta)=G(x,\alpha+\beta x)\) for the expected payoff
when the parameter is known. A complete-information Berk--Nash equilibrium
\((a^e,b^e,x^e)\) is
\textit{interior regular} if \condtarget{cond:R1}{(R1)} \(x^e\in\operatorname{int}X\) is the unique
maximizer of \(z\mapsto G(z,a^e+b^ez)\) over \(X\), \condtarget{cond:R2}{(R2)} this function has a
strictly negative second derivative \(Q^e\) at \(x^e\), and \condtarget{cond:R3}{(R3)}
\(M^e=G_m(x^e,a^e+b^ex^e)\ne0\). Conditions \ref{cond:R1} and \ref{cond:R2} require the action to
be interior, uniquely optimal, and to satisfy the second-order condition.
Condition \ref{cond:R3} requires the mean outcome to be locally payoff-relevant,
that is, a marginal change in the expected outcome changes the expected payoff
when the action is held fixed. At the parameter level, let \(\Theta^I\) be the set of
parameters \(\vartheta\) at which \(g^{\mathrm{full}}(\cdot;\vartheta)\) has a
unique interior maximizer \(x^\star(\vartheta)\) with
\[
 Q(\vartheta)=\partial_{xx}g^{\mathrm{full}}(x^\star(\vartheta);\vartheta)<0,
 \qquad
 M(\vartheta)=G_m\bigl(x^\star(\vartheta),\alpha+\beta x^\star(\vartheta)\bigr)\ne0.
\]
So \(\Theta^I\) is the set of parameters \(\vartheta\) for which
\((\alpha,\beta,x^\star(\vartheta))\) is an interior regular Berk--Nash
equilibrium when the true parameter is \(\vartheta\). Let \(\Theta^B\) be the
set of parameters at which \(g^{\mathrm{full}}(\cdot;\vartheta)\) has a unique
maximizer at the upper endpoint \(\overline x\) with
\(g^{\mathrm{full}}_x(\overline x;\vartheta)>0\), or at the lower endpoint
\(\underline x\) with \(g^{\mathrm{full}}_x(\underline x;\vartheta)<0\), where
the derivatives at the endpoints are one-sided. Let
\(\Theta^R=\Theta^I\cup\Theta^B\). The excluded parameters are those at which
the complete-information optimum is not unique or fails one of these local
regularity conditions.

The following corollary shows that a myopic rule satisfies the hypotheses of
Theorems~\ref{thm:global} and~\ref{thm:smooth-local}. The proof is in Supplement~\ref{supp:myopic}.

\begin{corollary}[Myopic decision rules]\label{cor:myopic}
Suppose Assumption~\ref{ass:payoff} holds, and let \(\psi_0\) be a Borel
myopic decision rule.
\begin{enumerate}[label=\textup{(\roman*)},ref=(\roman*)]
\item Every interior regular Berk--Nash equilibrium \((a^e,b^e,x^e)\) is a\label{cor:myopic:i}
complete-information Berk--Nash equilibrium under \(\psi_0\); \(\psi_0\) admits a
\(C^1\) extension, with locally Lipschitz first derivatives, to an open
neighborhood of \((a^e,b^e,0,x^e,0)\), with response coefficient \(\gamma^e=-M^e/Q^e\ne0\);
and \((a^e,b^e,x^e,w)\) is a Berk--Nash equilibrium under \(\psi_0\) for
every sufficiently small \(w>0\). So the conclusions of
Theorem~\ref{thm:smooth-local} hold at it, with priors whose state is close to
\((a^e,b^e,0,x^e,w^e)\) for every sufficiently small \(w^e>0\).
\item \(\Theta^I\subseteq\Theta^T_{\psi_0}\), with complete-information\label{cor:myopic:ii}
action the complete-information optimum \(x^\star(\vartheta)\) and
\(\gamma(\vartheta)=-M(\vartheta)/Q(\vartheta)\), and
\(\Theta^B\subseteq\Theta^C_{\psi_0}\). Consequently, for every nondegenerate
Gaussian prior \(\mu\), \(P^{\mu,\psi_0}(V_t\to\infty,\ \vartheta\in\Theta^R)=0\);
\(P^{\mu,\psi_0}(\mathcal W,\ \vartheta\in B)=\mu(B)\) for every Borel
\(B\subseteq\Theta^R\); and \(P^{\psi_0}_\vartheta(\mathcal W)=1\) for
Lebesgue-almost every \(\vartheta\in\Theta^R\).
\item Conversely, if \((a^e,b^e,x^e)\) is a complete-information Berk--Nash\label{cor:myopic:iii}
equilibrium under \(\psi_0\), then it is a complete-information Berk--Nash
equilibrium, that is, \(x^e\) maximizes
\(z\mapsto G(z,a^e+b^ez)\) on \(X\).
\end{enumerate}
\end{corollary}

Part~\ref{cor:myopic:iii} shows that the complete-information equilibria under
the myopic rule are complete-information Berk--Nash equilibria in the usual
sense, and part~\ref{cor:myopic:i} gives the converse under
regularity. The formula \(\gamma^e=-M^e/Q^e\) shows that, at a regular
equilibrium, transversality holds exactly because the predicted outcome
matters at the margin. In the introduction's example \(U=y-C(x)\), \(M^e=1\)
and \(Q^e=-C''(x^e)\), so \(\gamma^e=1/C''(x^e)\) as computed there.
Supplement~\ref{supp:examples} verifies the hypotheses in the three examples,
and we summarize the conclusions. In each example, at every interior
regular equilibrium \((a^e,b^e,x^e)\), the tuple \((a^e,b^e,x^e,w)\) is a
Berk--Nash equilibrium under the rule for every sufficiently small \(w>0\)
(part~\ref{cor:myopic:i} and Lemma~\ref{Ex-lem:projected}\ref{Ex-lem:projected:b}),
so the priors of Theorem~\ref{thm:smooth-local} are those whose state is close
to \((a^e,b^e,0,x^e,w^e)\).

\begin{example}[name=Advertising,continues=ex:advertising]
With the quadratic cost, every \(b^e\in\operatorname{int}X\) gives an interior
regular Berk--Nash equilibrium, with response coefficient \(\gamma^e=1\).
Hence Theorem~\ref{thm:smooth-local} applies to every such equilibrium,
so sufficiently concentrated priors in its neighborhood remain nearby with
probability arbitrarily close to one. The complete-information action is
\(\Pi_X(\beta)\). Every parameter is regular except when \(\beta\) equals one of
the two endpoints of \(X\). Parameters with \(\beta\in\operatorname{int}X\) are
interior regular, and those with \(\beta\notin X\) are locally constant,
because the rule saturates at an endpoint. Hence the regular set has prior
probability one under every nondegenerate Gaussian prior, and
Theorem~\ref{thm:global} implies that the myopic firm learns the return
to advertising incorrectly with probability one.
\end{example}

\begin{example}[name=Pricing,continues=ex:pricing]
At a complete-information Berk--Nash equilibrium with \(b^e<0\), \(b^e\ne-\beta\) and interior price
\(x^e\), the response coefficient is
\[
 \gamma^e=\frac{\alpha+\beta k}{2b^e(b^e+\beta)},
\]
which is nonzero exactly when true expected demand at marginal cost is nonzero,
\(\alpha+\beta k\ne0\). The predicted demand at the equilibrium price equals the
true mean demand there, and the rule responds to a slope revision along the
mean-preserving direction only when that demand is nonzero. Under this
condition, Theorem~\ref{thm:smooth-local} applies. For the global result, every
parameter is regular except on a Lebesgue-null set of knife-edge cases: a
negative slope with a boundary price at which the derivative of the
complete-information profit vanishes, an interior price equal to marginal
cost, a positive slope with tied endpoint profits, and a zero slope with zero
intercept. Hence incorrect learning occurs with prior-predictive probability
one. This holds in particular on the downward-sloping parameters \(\beta<0\) of
the economic model, that is, incorrect learning occurs on every Borel set of regular
parameters with \(\beta<0\) with exactly its prior probability.
\end{example}

\begin{example}[name=Inflation,continues=ex:inflation]
At a complete-information Berk--Nash equilibrium with slope \(b^e\), interior action and
\(\omega+b^e\beta\ne0\), the response coefficient is
\[
 \gamma^e=\frac{\omega(y^*-\alpha)}{(\omega+b^e\beta)[(b^e)^2+\omega]},
\]
which is nonzero exactly when the true intercept differs from the target,
\(\alpha\ne y^*\). When \(\alpha=y^*\), inflation is on target on average
without intervention, the equilibrium action is zero, and the equilibrium
intercept is \(y^*\). A bank that believes the intercept to be \(y^*\) leaves
the instrument at zero whatever it believes about its effect, so its action
does not respond to slope revisions. So
Theorem~\ref{thm:smooth-local} applies at every such equilibrium when
\(\alpha\ne y^*\). The complete-information payoff is strictly concave, and
for the same reason every interior optimum with \(\alpha\ne y^*\) is regular;
the irregular parameters lie among those with \(\alpha=y^*\) and those with a
boundary optimum at which the derivative of the complete-information payoff
vanishes. These form a null set, so Theorem~\ref{thm:global} gives
prior-predictive probability-one incorrect learning. No sign restriction on
\(X\) is needed, so the instrument may be a deviation from a neutral setting
that takes either sign.
\citet{wieland2000monetary,wieland2000learning} studies this learning problem
and distinguishes the myopic central bank from one that experiments; our
results show that the myopic bank, whose caution keeps policy close to its past
average, fails to learn the effect of policy almost surely.
\end{example}

\subsection{Rules that do not come from optimization}\label{sec:rules}
Theorems~\ref{thm:global} and~\ref{thm:smooth-local} do not ask how the
rule was chosen. We give two examples of rules that need not come from
optimization, and a contrast that shows why transversality is essential. Both
rules depend on the posterior mean only. So they choose the equilibrium action
at every slope precision, which centers the priors of
Theorem~\ref{thm:smooth-local} at the complete-information equilibrium, and
Definition~\ref{def:regular} applies directly in the case where the
projection onto \(X\) is inactive; the parameters at which the rule saturates
are locally constant.

\begin{example}[name=Advertising rule of thumb]\label{ex:rule-of-thumb}
Consider the firm of Example~\ref{ex:advertising}, and suppose that its
budgeting system follows the stylized rule of thumb ``spend a base amount \(\ell\) plus
\(\tau\) times the estimated return'', capped by the campaign limits
\citep[for budgeting heuristics, see][]{bigne1995,kolsarici2020anatomy}:
\[
 \psi_{\ell,\tau}(s)=\Pi_X(\ell+\tau b),\qquad \tau\ne0.
\]
For \(\tau>0\), the rule is also the myopic optimum for the payoff
\(y-(x-\ell)^2/(2\tau)\), but the theorems do not use that. For \(\tau<0\), it is
not the myopic optimum of any payoff \(y-C(x)\) with a convex cost \(C\). For
every \(b^e\) with \(x^e=\ell+\tau b^e\in\operatorname{int}X\), the triple
\((a^e,b^e,x^e)\) with \(a^e=\alpha+(\beta-b^e)x^e\) is a complete-information
Berk--Nash equilibrium under the rule with response coefficient \(\gamma^e=\tau\ne0\), so
Theorem~\ref{thm:smooth-local} applies. Every parameter is regular except on
the two lines \(\ell+\tau\beta\in\{\underline x,\overline x\}\), so
Theorem~\ref{thm:global} implies that incorrect learning occurs with
probability one. A rule that ignored the estimated return (\(\tau=0\)) would be
constant. The global conclusion would still hold, and, although no equilibrium
would satisfy transversality, every equilibrium would persist by
Proposition~\ref{thm:corner-local} instead of Theorem~\ref{thm:smooth-local}.
\end{example}

\begin{example}[name=Certainty-equivalent policy]\label{ex:certainty-equivalent}
Consider the central bank of Example~\ref{ex:inflation}, and suppose that it
plugs the posterior means into the complete-information optimum, ignoring
parameter uncertainty \citep{pekelmantse1980,wieland2000monetary}.
It then uses the certainty-equivalent rule
\[
 \psi^{CE}(s)=\Pi_X\left(\frac{b(y^*-a)}{b^2+\omega}\right).
\]
Before projection, this classic adaptive rule differs from the myopic optimum
of Example~\ref{ex:inflation} at every posterior with
\(b(y^*-a)/(b^2+\omega)\ne\bar x\). The myopic rule moves only part of the way
from \(\bar x\) toward \(b(y^*-a)/(b^2+\omega)\), so the certainty-equivalent
rule is more aggressive. Its complete-information action, its Berk--Nash
equilibria under the rule and its response coefficient coincide with those of
the myopic rule. Hence Theorems~\ref{thm:global} and~\ref{thm:smooth-local}
apply to it with the same conditions and the same conclusions. The failure of learning in the inflation example
therefore does not come from caution.
\end{example}

\begin{remark}[Transversality cannot be dropped]\label{rem:targeting}
Consider a rule that sets the predicted mean
outcome at a target, \(a+bx=y^*\), so that \(h(a,b)=(y^*-a)/b\) in the case
where the action is interior. Then \(h_a=-1/b\) and \(h_b=-h/b\), so
\(h_b-h\,h_a=0\) identically, that is, a slope revision that leaves the predicted
outcome unchanged leaves the targeted action unchanged. None of
Theorems~\ref{thm:global} and~\ref{thm:smooth-local} and
Proposition~\ref{thm:corner-local} applies in that case. The myopic rule of the targeting payoff \(U(x,y)=-y^2\), whose
complete-information action is the target-hitting rule \(-a/b\), shows what
can happen instead.\footnote{For this payoff, \(G_m=0\) at every interior
complete-information optimum with \(b\ne0\); when \(b=0\), the objective is
constant in the action, so uniqueness and strict curvature fail. Hence no
equilibrium is interior regular, and the interior-regular conclusions of
Corollary~\ref{cor:myopic} do not apply; the response coefficient is computed
directly above, and it is zero.} An incorrect Berk--Nash equilibrium need not
persist. At the true parameter \((0,1)\), the posterior stays near the
incorrect equilibrium belief \((0,2)\) with finite slope precision with
probability zero. And for a prior concentrated around \((0,1)\), complete
learning occurs with prior-predictive probability close to one. Almost surely
on the complete-learning event of Supplement~\ref{supp:targeting}, the action
gaps vanish, but not fast enough to be square summable, so the small
continuing adjustments toward the target keep identifying the slope.
Supplement~\ref{supp:targeting} gives the precise statements and their
proofs. So, for a rule that responds to beliefs, the responsiveness of the
action to undetectable slope revisions is what keeps the decision maker from
learning the true parameter.
\end{remark}

\section{Conclusion}\label{sec:conclusion}
Econometrics teaches that learning the slope of a regression requires variation
in the regressor. When the regressor is chosen by the learner itself, the
learner's own decision rule determines how much variation is generated. We have
shown that a smooth rule that satisfies transversality generates too little---a
small belief revision produces a small change in the action, which produces
even less information. As a result, a correctly specified Bayesian decision
maker does not learn the true parameter, it can settle near an incorrect
Berk--Nash equilibrium without any dogmatic belief, and overconfidence and
underconfidence arise endogenously from its learning history. These
conclusions concern the rule, not the reason for it, so they cover myopic
optimization, automated systems including delegation to AI, and rules of
thumb.

The analysis is confined to a stable, nonstrategic linear environment with
conjugate Gaussian learning. Relevant information arrives through outcomes
generated by the chosen action; outside data, mandated experiments,
competitors' actions, or naturally occurring variation could change the
conclusion. Decision makers who value information may experiment. Their
optimal policy need not be smooth, and the companion paper
\citep{kky2026partii} devises a way to handle it and shows that the
conclusions of this paper survive forward-looking incentives.

\appendix
\numberwithin{equation}{section}
\numberwithin{theorem}{section}
\renewcommand{\theproposition}{\thetheorem}
\renewcommand{\thelemma}{\thetheorem}
\renewcommand{\thecorollary}{\thetheorem}
\renewcommand{\theclaim}{\thetheorem}
\renewcommand{\thedefinition}{\thetheorem}
\renewcommand{\theassumption}{\thetheorem}
\renewcommand{\theauxproposition}{\thetheorem}
\renewcommand{\theauxlemma}{\thetheorem}
\renewcommand{\theremark}{\thetheorem}

\section{Posterior foundations}\label{app:posterior}

Throughout the appendices, \(\psi\) is a Borel decision rule, and the prior
is the nondegenerate Gaussian law \(\mu=N_2((a_0,b_0)^\top,\Sigma_0)\).  The
path space is
\(\Omega\coloneqq(X\times\R)^{\mathbb N}\), with coordinates
\((x_t,y_t)_{t\ge1}\), and \(\mathcal Q\) is the law of the i.i.d.\
\(N(0,1)\) sequence \((\varepsilon_t)_{t\ge1}\) on \(\R^{\mathbb N}\).  For
\(\vartheta'=(\alpha',\beta')\in\R^2\) and \(e=(e_t)_{t\ge1}\in\R^{\mathbb N}\),
let \(\Phi(\vartheta',e)\in\Omega\) be the path generated from the prior
state \(s_0\) by
\[
 x_t=\psi(s_{t-1}),\qquad y_t=\alpha'+\beta'x_t+e_t,
\]
with \(s_t\) the posterior update of \(s_{t-1}\) by \((x_t,y_t)\).  The
fixed-truth law is
\(P^\psi_{\vartheta'}\coloneqq\mathcal Q\circ\Phi(\vartheta',\cdot)^{-1}\),
and the prior-predictive law \(P^{\mu,\psi}\) is the joint law of
\((\vartheta,\Phi(\vartheta,\varepsilon))\) on \(\R^2\times\Omega\), where
\(\vartheta\sim\mu\) is independent of \(\varepsilon\).  For
\(E\subseteq\R^2\times\Omega\), write
\(E_{\vartheta'}\coloneqq\{\omega:(\vartheta',\omega)\in E\}\) for its
section.  Put
\[
 \mathcal F_t\coloneqq\sigma(y_1,\ldots,y_t),\qquad
 \mathcal F_\infty\coloneqq\sigma\left(\bigcup_t\mathcal F_t\right),
\]
with \(\mathcal F_0\) trivial.  Write
\(\xi_t\coloneqq\alpha+\beta\bar x_t-a_t-b_t\bar x_t\) for the error of the
fitted mean outcome at the average action, \(\kappa_t=A_t/(A_t+1)\), and
\(E_\infty\coloneqq\{V_\infty=\infty\}\) for the event that slope precision
diverges, where \(V_\infty=\lim_tV_t\) exists because \(V_t\) is
nondecreasing by \eqref{eq:design}.  On
the null set where the posterior means fail to converge,
\((a_\infty,b_\infty)\) is defined arbitrarily, as an
\(\mathcal F_\infty\)-measurable pair.  In
Lemmas~\ref{lem:posterior-interface} and~\ref{lem:complete-learning}, \(X\)
may be an arbitrary Borel set \(X\subseteq\R\), bounded or not, and \(\psi\) a
Borel decision rule with values in \(X\). Their proofs below use no bound on
the actions.

\begin{lemma}[Posterior interface]\label{lem:posterior-interface}
For every \(t\), there is a Borel map \(R_t\) with
\(s_t=R_t(y_1,\ldots,y_t)\) on the image of \(\Phi\), hence
\(P^{\mu,\psi}\)-almost surely; thus \(s_t\) and \(x_{t+1}=\psi(s_t)\) are
\(\mathcal F_t\)-measurable up to \(P^{\mu,\psi}\)-null sets.  The following
hold.
\begin{enumerate}[label=(\alph*)]
\item \emph{Recursions.}  The recursions \eqref{eq:design} and\label{lem:posterior-interface:a}
\eqref{eq:means} hold pathwise, and so does
\(A_{t+1}\xi_{t+1}=A_t\xi_t-\varepsilon_{t+1}\).  Under \(P^{\mu,\psi}\),
for every \(t\), the conditional law of \(\vartheta\) given
\(\mathcal F_t\) is the Gaussian posterior
\(N((a_t,b_t)^\top,\Sigma_t)\), with
\[
 \Sigma_t=\begin{pmatrix}u_t+\bar x_t^2w_t&-\bar x_tw_t\\
 -\bar x_tw_t&w_t\end{pmatrix}.
\]
\item \emph{Prediction errors.}  Under \(P^{\mu,\psi}\), conditional on\label{lem:posterior-interface:b}
\(\mathcal F_t\),
\[
 I_{t+1}\sim N\left(0,\sigma_t^2\right),
 \qquad
 \sigma_t^2\coloneqq1+u_t+w_tD_t^2.
\]
The standardized error \(\eta_{t+1}\coloneqq I_{t+1}/\sigma_t\) is
\(N(0,1)\), independent of \(\mathcal F_t\), and
\(\mathcal F_{t+1}\)-measurable up to \(P^{\mu,\psi}\)-null sets, so
\((\eta_t)_{t\ge1}\) is i.i.d.\
\(N(0,1)\).  Moreover \(\kappa_t\sigma_t^2V_t=V_{t+1}\) for every \(t\).
\item \emph{Limits and dichotomy.}  \(P^{\mu,\psi}\)-almost surely,\label{lem:posterior-interface:c}
\((a_t,b_t)\) converges to a finite limit \((a_\infty,b_\infty)\) and
\(\xi_t\to0\).  A regular conditional law of \(\beta\) given
\(\mathcal F_\infty\) is \(N(b_\infty,V_\infty^{-1})\) at almost every path
of \(\{V_\infty<\infty\}\) and the point mass at \(b_\infty\) at almost
every path of \(E_\infty\).  Almost surely on \(E_\infty\),
\[
 (a_t,b_t,u_t,w_t)\to(\alpha,\beta,0,0)
 \quad\text{and}\quad
 \Sigma_t\to0.
\]
Finally,
\[
 P^{\mu,\psi}(V_\infty<\infty,\ b_\infty=\beta)=0.
\]
\end{enumerate}
\end{lemma}

\begin{lemma}[Fixed-truth transfer]\label{lem:disintegration}
The map \(\Phi\) is Borel, and:
\begin{enumerate}[label=(\roman*)]
\item \(\vartheta\mapsto P^\psi_\vartheta\) is a probability kernel from\label{lem:disintegration:i}
\(\R^2\) to \(\Omega\).  For every joint Borel event
\(E\subseteq\R^2\times\Omega\), the map
\(\vartheta\mapsto P^\psi_\vartheta(E_\vartheta)\) is Borel and
\[
 P^{\mu,\psi}(E)=\int P^\psi_\vartheta(E_\vartheta)\,\mu(d\vartheta).
\]
\item If \(P^{\mu,\psi}(E)=0\), then \(P^\psi_\vartheta(E_\vartheta)=0\) for\label{lem:disintegration:ii}
\(\mu\)-almost every \(\vartheta\); if \(P^{\mu,\psi}(E)=1\), then
\(P^\psi_\vartheta(E_\vartheta)=1\) for \(\mu\)-almost every \(\vartheta\).
\item A Borel subset of \(\R^2\) is \(\mu\)-null if and only if it is\label{lem:disintegration:iii}
Lebesgue-null.  Hence ``\(\mu\)-almost every'' in \ref{lem:disintegration:ii} may be replaced by
``Lebesgue-almost every''.
\end{enumerate}
\end{lemma}

Events such as \(E_\infty\) and \(\mathcal W\) are joint Borel events,
since each posterior quantity is a Borel function of the path and
\(\mathcal W\) also involves \(\beta\). So is the event \(\mathcal L\) of Section~\ref{sec:global}, since the
set on which a sequence of Borel functions converges is Borel and the limit is
Borel there.

For the standard probabilistic inputs we use \citet{m2012}, Theorem~4.4.1
(p.~119) and Section~7.6.1 (p.~232), for Bayes' rule for linear Gaussian
systems; \citet{d2019}, Theorems~4.4.6 and~4.6.8 (pp.~205 and~214), for
martingale convergence and L\'evy's upward theorem, and Theorem~2.4.1
(p.~65) for the strong law of large numbers; and \citet{k2021},
Lemma~1.28(ii) and Theorem~1.29(ii) (p.~25), for the measurability of
sections and Fubini's theorem.

\begin{proof}[Proof of Lemma~\ref{lem:posterior-interface}]
\emph{Measurability.}  The posterior update
\(s_{t+1}=F(s_t,x_{t+1},y_{t+1})\) defined by
\eqref{eq:design}--\eqref{eq:means} is continuous and maps feasible states
to feasible states.  On the image of \(\Phi\), \(x_j=\psi(s_{j-1})\) for
every \(j\), so by induction \(s_t=R_t(y_1,\ldots,y_t)\) there for a Borel
map \(R_t\).  Off the image of \(\Phi\) the \(x\)-coordinates are not
functions of the \(y\)-coordinates, so this identity, and the
\(\mathcal F_t\)-measurability of \(s_t\) and \(x_{t+1}=\psi(s_t)\), hold
only almost surely. The Borel event \(\{s_t=R_t(y_1,\ldots,y_t)\}\) contains
the image of \(\Phi\), and \(P^{\mu,\psi}\) is the law of
\((\vartheta,\Phi(\vartheta,\varepsilon))\).  Read on the space of
\((\vartheta,\varepsilon)\),
\(\mathcal F_t\subseteq\sigma(\vartheta,\varepsilon_1,\ldots,\varepsilon_t)\).
Hence \(\varepsilon_{t+1}\) is independent of
\(\sigma(\vartheta)\vee\mathcal F_t\).

\emph{Part~\ref{lem:posterior-interface:a}.}  Write \(\theta=(\alpha,\beta)^\top\),
\(\theta_t=(a_t,b_t)^\top\) and \(z_{t+1}=(1,x_{t+1})^\top\).  The covariance
\(\Sigma_t\) has inverse
\begin{equation}\label{S1-eq:precision}
 \Sigma_t^{-1}
 =\begin{pmatrix}A_t&A_t\bar x_t\\ A_t\bar x_t&A_t\bar x_t^2+V_t\end{pmatrix}.
\end{equation}
Suppose that the conditional law of \(\vartheta\) given \(\mathcal F_t\) is
\(N(\theta_t,\Sigma_t)\), as it is at \(t=0\).  Since \(x_{t+1}\) is
\(\mathcal F_t\)-measurable up to \(P^{\mu,\psi}\)-null sets and \(y_{t+1}=z_{t+1}^\top\theta+\varepsilon_{t+1}\)
with \(\varepsilon_{t+1}\sim N(0,1)\) independent of
\(\sigma(\vartheta)\vee\mathcal F_t\), Bayes' rule for linear Gaussian
systems gives the posterior \(N(\theta_{t+1},\Sigma_{t+1})\) given
\(\mathcal F_{t+1}\), with
\[
 \Sigma_{t+1}^{-1}=\Sigma_t^{-1}+z_{t+1}z_{t+1}^\top,
 \qquad
 \theta_{t+1}-\theta_t=\Sigma_{t+1}z_{t+1}I_{t+1}.
\]
Reading the entries of \(\Sigma_{t+1}^{-1}\) in the parametrization
\eqref{S1-eq:precision} gives \(A_{t+1}=A_t+1\),
\(A_{t+1}\bar x_{t+1}=A_t\bar x_t+x_{t+1}\) and
\(V_{t+1}=V_t+\kappa_tD_t^2\), which is \eqref{eq:design}; the mean update
is \eqref{eq:means}.  These identities are algebraic in the realized
\((x_{t+1},y_{t+1})\), so they hold on every path.  Finally, a direct
computation from \eqref{eq:means}, using
\(\bar x_{t+1}-x_{t+1}/(A_t+1)=\kappa_t\bar x_t\) and
\(I_{t+1}=(\alpha-a_t)+(\beta-b_t)x_{t+1}+\varepsilon_{t+1}\), gives
\begin{equation}\label{S1-eq:xi}
 A_{t+1}\xi_{t+1}=A_t\xi_t-\varepsilon_{t+1}.
\end{equation}

\emph{Part~\ref{lem:posterior-interface:b}.}  Conditional on \(\mathcal F_t\),
\(I_{t+1}=z_{t+1}^\top(\theta-\theta_t)+\varepsilon_{t+1}\) is Gaussian with
mean zero and variance
\(1+z_{t+1}^\top\Sigma_tz_{t+1}=1+u_t+w_tD_t^2\).  The conditional law of
\(\eta_{t+1}\) given \(\mathcal F_t\) is therefore \(N(0,1)\), which does
not depend on \(\mathcal F_t\); so \(\eta_{t+1}\) is independent of
\(\mathcal F_t\), and, since \(\eta_1,\ldots,\eta_t\) are
\(\mathcal F_t\)-measurable up to \(P^{\mu,\psi}\)-null sets, the sequence is
i.i.d.  The identity
\(\kappa_t\sigma_t^2V_t=V_t+\kappa_tD_t^2=V_{t+1}\) is direct.

\emph{Part~\ref{lem:posterior-interface:c}.}  By part~\ref{lem:posterior-interface:a}, \(a_t=E[\alpha\mid\mathcal F_t]\) and
\(b_t=E[\beta\mid\mathcal F_t]\) are \(L^2\)-bounded martingales, hence
converge almost surely \citep[Theorem~4.4.6]{d2019}.  By part~\ref{lem:posterior-interface:a},
\(E[e^{ir\beta}\mid\mathcal F_t]=\exp(irb_t-r^2w_t/2)\), and \(w_t\)
decreases to \(V_\infty^{-1}\), which is zero on \(E_\infty\).  L\'evy's
upward theorem \citep[Theorem~4.6.8]{d2019} gives, almost surely for every
rational \(r\) and hence, by continuity, for every \(r\),
\[
 E[e^{ir\beta}\mid\mathcal F_\infty]=\exp\left(irb_\infty-r^2V_\infty^{-1}/2\right).
\]
This identifies the regular conditional law of \(\beta\) given
\(\mathcal F_\infty\).  Since \(\{V_\infty<\infty\}\) and \(b_\infty\) are
\(\mathcal F_\infty\)-measurable and a Gaussian law with positive variance
has no atoms,
\(P^{\mu,\psi}(V_\infty<\infty,\ \beta=b_\infty)=0\); likewise
\(\beta=b_\infty\) almost surely on \(E_\infty\).  Iterating
\eqref{S1-eq:xi} gives
\(\xi_t=(A_0\xi_0-\sum_{j\le t}\varepsilon_j)/(A_0+t)\to0\) almost surely by
the strong law of large numbers.

It remains to identify the intercept on \(E_\infty\).  For every
\(c\in\R\), a direct computation from \eqref{eq:design} gives
\[
 V_{t+1}+A_{t+1}(\bar x_{t+1}-c)^2=V_t+A_t(\bar x_t-c)^2+(x_{t+1}-c)^2,
\]
so \(V_t+A_t(\bar x_t-c)^2\) is nondecreasing in \(t\) on every path.  Taking
\(c=\bar x_t\) and comparing dates \(T\le t\) gives
\(A_T(\bar x_t-\bar x_T)^2\le V_t\), hence
\(\bar x_t^2w_t\le2\bar x_T^2w_t+2/A_T\).  On \(E_\infty\), letting
\(t\to\infty\) and then \(T\to\infty\) gives \(\bar x_t^2w_t\to0\), whatever
the actions.  Since \(u_t\to0\) on every path and \(w_t\to0\) on
\(E_\infty\), every entry of \(\Sigma_t\), namely \(u_t+\bar x_t^2w_t\),
\(-\bar x_tw_t\) and \(w_t\), tends to zero on \(E_\infty\), the middle one
because \((\bar x_tw_t)^2=(\bar x_t^2w_t)w_t\).  By part~\ref{lem:posterior-interface:a},
\(E[e^{ir\alpha}\mid\mathcal F_t]=\exp\bigl(ira_t-r^2(u_t+\bar x_t^2w_t)/2\bigr)\),
and L\'evy's upward theorem, as for \(\beta\), gives
\(E[e^{ir\alpha}\mid\mathcal F_\infty]=e^{ira_\infty}\) almost surely on
\(E_\infty\) for every \(r\).  So the regular conditional law of \(\alpha\)
given \(\mathcal F_\infty\) is the point mass at \(a_\infty\) there, and
\(\alpha=a_\infty\) almost surely on \(E_\infty\).
\end{proof}

\begin{proof}[Proof of Lemma~\ref{lem:disintegration}]
By the induction in the proof of Lemma~\ref{lem:posterior-interface}, the
date-\(t\) coordinate of \(\Phi(\vartheta',e)\) is a Borel function of
\((\vartheta',e_1,\ldots,e_t)\), so \(\Phi\) is Borel and the realized path is
\(\Phi(\vartheta,\varepsilon)\).  Part~\ref{lem:disintegration:i} then follows from the measurability
of sections and Fubini's theorem, since \(\vartheta\sim\mu\) is independent of
\(\varepsilon\sim\mathcal Q\); part~\ref{lem:disintegration:ii} because a nonnegative function with
zero integral vanishes almost everywhere; and part~\ref{lem:disintegration:iii} because \(\mu\) has
a strictly positive Lebesgue density.
\end{proof}

\begin{proof}[Proof of Lemma~\ref{lem:complete-learning}]
Work under \(P^{\mu,\psi}\).  Let \(\Omega_0\) denote the intersection of
the probability-one events of Lemma~\ref{lem:posterior-interface}\ref{lem:posterior-interface:c}. On
\(\Omega_0\), \((a_t,b_t)\to(a_\infty,b_\infty)\) and, if \(V_t\to\infty\),
\((a_t,b_t,u_t,w_t)\to(\alpha,\beta,0,0)\) and \(\Sigma_t\to0\).  This gives
statements~\ref{lem:complete-learning:1} and~\ref{lem:complete-learning:2} on \(\Omega_0\).

For statement~\ref{lem:complete-learning:3}, by Lemma~\ref{lem:posterior-interface}\ref{lem:posterior-interface:a}, the posterior
law of \(\beta\) given \(\mathcal F_t\) is the second marginal of
\(N((a_t,b_t)^\top,\Sigma_t)\), namely \(N(b_t,w_t)\).  By
\eqref{eq:design}, \(V_t\) is nondecreasing, so \(V_t\uparrow V_\infty\in[V_0,\infty]\).
On \(\Omega_0\cap\{V_\infty<\infty\}\), \(b_t\to b_\infty\) and
\(w_t\to V_\infty^{-1}\), so for every \(r\in\R\) the characteristic
functions \(\exp(irb_t-r^2w_t/2)\) converge to
\(\exp(irb_\infty-r^2V_\infty^{-1}/2)\).  L\'evy's continuity theorem gives
weak convergence of the posterior law of \(\beta\) to
\(N(b_\infty,V_\infty^{-1})\).

For the actions, fix a path in \(\Omega_0\cap\{V_\infty<\infty\}\). By the
proof of Lemma~\ref{lem:posterior-interface},
\(A_T(\bar x_t-\bar x_T)^2\le V_t\le V_\infty\) whenever \(T\le t\), and
\(A_T=A_0+T\to\infty\), so \((\bar x_t)\) is a Cauchy sequence and converges to a
finite limit \(\bar x_\infty\). By \eqref{eq:design},
\(\kappa_tD_t^2=V_{t+1}-V_t\to0\), and \(\kappa_t\ge A_0/(A_0+1)\), so
\(D_t=x_{t+1}-\bar x_t\to0\) and \(x_{t+1}\to\bar x_\infty\). Finally,
\(\xi_t=\alpha+\beta\bar x_t-a_t-b_t\bar x_t\to0\) on \(\Omega_0\), and letting
\(t\to\infty\) gives \(a_\infty+b_\infty\bar x_\infty=\alpha+\beta\bar x_\infty\).

By the last assertion of
Lemma~\ref{lem:posterior-interface}\ref{lem:posterior-interface:c}, the event
\(\{V_\infty<\infty,\ b_\infty=\beta\}\) is null. Removing it from
\(\Omega_0\) leaves an event of probability one on which all three statements hold.
\end{proof}

\section{Proof of local persistence}\label{app:local}

Throughout this appendix the truth \(\vartheta=(\alpha,\beta)\) is fixed.
The proof applies the trapping lemma of the Supplement, in the form for one
slope precision (Lemma~\ref{lem:trapping} and Remark~\ref{B-rem:single}).
That lemma is stated in the coordinates
\((a,b,u,\bar x,V)\), with \(V=1/w\) the slope precision, in the
generality used by the companion paper; a decision rule that coincides with
a smooth map \(\varphi\) of these coordinates on the relevant states
satisfies its hypotheses (Section~\ref{supp:general-policies}).

\begin{proof}[Proof of Theorem~\ref{thm:smooth-local}]
Let \(\bar\psi\) be the \(C^1\) extension of \(\psi\) on the open
neighborhood \(\mathcal N\) of \(z^e=(a^e,b^e,0,x^e,0)\) in the theorem, and put
\(\lambda^0(s)\coloneqq\bar\psi_b(s)-\bar x\bar\psi_a(s)\) on \(\mathcal N\), so that
\(\lambda^0(z^e)=\gamma^e\).  The proof finds the steady states \(b_w\), and
then, at each large \(V^e\), applies the trapping lemma at the target
\((a_w,b_w,x^e)\) with \(w=1/V^e\) and transfers its conclusions back to
\((a^e,b^e)\).

\emph{Choice of the box.}  Since \(\gamma^e\ne0\) and \(\lambda^0\) is
continuous on \(\mathcal N\), choose \(r_*,u_*,w_*\in(0,1)\) such that the
closed box
\[
 \mathcal K_Q\coloneqq\{(a,b,u,\bar x,w):
 \lvert a-a^e\rvert,\lvert b-b^e\rvert,\lvert\bar x-x^e\rvert\le r_*,\
 \lvert u\rvert\le u_*,\ \lvert w\rvert\le w_*\}
\]
lies in \(\mathcal N\) and \(\lvert\lambda^0-\gamma^e\rvert<\lvert\gamma^e\rvert/2\)
on \(\mathcal K_Q\).  In particular, \(\lambda^0\) has the sign of
\(\gamma^e\) and \(\lvert\lambda^0\rvert>\lvert\gamma^e\rvert/2\) there.

\emph{Steady states.}  Put \(\eta\coloneqq r_*/(1+\lvert x^e\rvert)\),
\(p(b,w)\coloneqq(\alpha+(\beta-b)x^e,b,0,x^e,w)\) and
\(F(b,w)\coloneqq\bar\psi(p(b,w))-x^e\).  Since
\(a^e=\alpha+(\beta-b^e)x^e\), the first coordinate of \(p(b,w)\) differs from
\(a^e\) by \((b^e-b)x^e\), so \(p(b,w)\in\mathcal K_Q\subseteq\mathcal N\)
whenever \(\lvert b-b^e\rvert\le\eta\) and \(0\le w\le w_*\).  Each such
\(p(b,w)\) satisfies the no-surprise condition, and, being a limit of feasible
states in \(\mathcal N\), on which \(\bar\psi\) is continuous, it satisfies
\(\psi(s_n)\to\bar\psi(p(b,w))\) along every sequence of feasible states
\(s_n\to p(b,w)\).  So
\((\alpha+(\beta-b)x^e,b,x^e,w)\) is a Berk--Nash equilibrium under \(\psi\)
exactly when \(F(b,w)=0\).  The average
action at \(p(b,w)\) is \(x^e\), so the chain rule gives
\[
 \partial_bF(b,w)=\bar\psi_b(p(b,w))-x^e\bar\psi_a(p(b,w))=\lambda^0(p(b,w)).
\]
Hence \(b\mapsto\gamma^eF(b,w)\) is increasing on
\([b^e-\eta,b^e+\eta]\), with slope at least \((\gamma^e)^2/2\).  Since
\((a^e,b^e,x^e)\) is a Berk--Nash equilibrium under \(\psi\) and
\(p(b^e,0)=z^e\), \(F(b^e,0)=0\), and, by the mean value theorem,
\(\lvert F(b^e,w)\rvert\le Lw\) with \(L\coloneqq\max_{\mathcal K_Q}\lvert\bar\psi_w\rvert\).
Put \(K\coloneqq2L/\lvert\gamma^e\rvert\) and choose \(\bar w\in(0,w_*]\) with
\(K\bar w<\eta\).  For \(w\in[0,\bar w)\),
\begin{align*}
 \gamma^eF(b^e+Kw,w)&\ge\gamma^eF(b^e,w)+\lvert\gamma^e\rvert Lw\ge0,\\
 \gamma^eF(b^e-Kw,w)&\le\gamma^eF(b^e,w)-\lvert\gamma^e\rvert Lw\le0,
\end{align*}
so the intermediate value theorem gives a zero \(b_w\) of \(F(\cdot,w)\) with
\(\lvert b_w-b^e\rvert\le Kw<\eta\), and strict monotonicity makes it the
only zero with \(\lvert b-b^e\rvert<\eta\).  If, in addition, \((a^e,b^e,x^e,w)\) is a Berk--Nash
equilibrium under \(\psi\) for every sufficiently small \(w>0\), then \(b^e\)
is such a zero, since \(p(b^e,w)=(a^e,b^e,0,x^e,w)\), so \(b_w=b^e\) and
\((a_w,b_w)=(a^e,b^e)\); this is used in
Corollary~\ref{cor:myopic}\ref{cor:myopic:i} and in Section~\ref{sec:applications}.  Since \(a_w-a^e=(b^e-b_w)x^e\),
\begin{equation}\label{eq:steady-shift}
 \lvert a_w-a^e\rvert+\lvert b_w-b^e\rvert\le(1+\lvert x^e\rvert)Kw
 \qquad(0\le w<\bar w).
\end{equation}
This proves the first assertion of the theorem, with \((1+\lvert x^e\rvert)K\)
as its constant.

\emph{The map \(\varphi\) and the collars.}  Put
\(\varphi(a,b,u,\bar x,V)\coloneqq\bar\psi(a,b,u,\bar x,1/V)\) when \(V>0\) and
\((a,b,u,\bar x,1/V)\in\mathcal N\), and \(\varphi\coloneqq0\) otherwise; \(\varphi\)
is Borel, being continuous on the open set where the first case applies and
constant on its complement.  Put \(V_*\coloneqq4/w_*\).  For \(V^e\ge V_*\) with
\((1+\lvert x^e\rvert)K/V^e<r_*/2\) and \(w=1/V^e<\bar w\), let
\(\mathcal C_w(V^e)\) be the collar of Lemma~\ref{lem:trapping} around the
target \((a_w,b_w,x^e)\), with radius \(r_*/2\) and the bound \(u_*\).  By
\eqref{eq:steady-shift}, every point of \(\mathcal C_w(V^e)\) has
\(\lvert a-a^e\rvert+\lvert b-b^e\rvert+\lvert\bar x-x^e\rvert<r_*\),
\(\lvert u\rvert<u_*\) and \(1/V\in[1/(2V^e),2/V^e]\subseteq(0,w_*/2]\), so
its image under \(V\mapsto1/V\) lies in \(\mathcal K_Q\).

\emph{Hypotheses \textup{\ref{lem:trapping:a}}--\textup{\ref{lem:trapping:c}}.}  \ref{lem:trapping:a} On the compact convex set
\(\mathcal K_Q\), \(\bar\psi\) and its first derivatives are bounded, and the
first derivatives, being locally Lipschitz, are Lipschitz.  On
\(V\ge V_*/2\ge2\), the map \(V\mapsto1/V\) and its first derivative are
bounded and Lipschitz, and \((a,b,u,\bar x,V)\mapsto(a,b,u,\bar x,1/V)\) is
\(1\)-Lipschitz.  By the chain rule, \(\varphi\) therefore has a \(C^{1,1}\)
norm at most a constant \(C_*\) on every collar \(\mathcal C_w(V^e)\) above,
with \(C_*\) independent of \(V^e\).  \ref{lem:trapping:b} The target satisfies
\(a_w+b_wx^e=\alpha+\beta x^e\), and
\(\varphi(a_w,b_w,0,x^e,V^e)=\bar\psi(a_w,b_w,0,x^e,w)=x^e\) by the
definition of \(b_w\).  \ref{lem:trapping:c} \(\varphi_a\) and \(\varphi_b\) are
\(\bar\psi_a\) and \(\bar\psi_b\) evaluated at \(w=1/V\), so
\(\lvert\varphi_b-\bar x\varphi_a\rvert\ge\lambda^0_*\coloneqq\lvert\gamma^e\rvert/2\)
on every collar.

\emph{Agreement with \(\varphi\).}  At every feasible state
\(s\in\mathcal C_w(V^e)\), the point \((a,b,u,\bar x,1/V)\) is a feasible
point of \(\mathcal K_Q\subseteq\mathcal N\), so
\(\psi(s)=\bar\psi(a,b,u,\bar x,1/V)=\varphi(s)\).  The realized action is
\(x_{t+1}=\psi(s_t)\), so \(x_{t+1}=\varphi(s_t)\) whenever
\(s_t\in\mathcal C_w(V^e)\), so the condition of Definition~\ref{B-def:tracking}
holds on \(\mathcal C_w(V^e)\) with zero error, for every initial prior.  The
events of conclusions \ref{thm:smooth-local:i} and \ref{thm:smooth-local:ii} are determined by the actions and the
outcomes, so their probabilities are those under \(P^\psi_\vartheta\)
(Section~\ref{supp:general-policies}).

\emph{Radii.}  Let \(\widehat V\) be the threshold of
Remark~\ref{B-rem:single} for \(x^e\), the radius \(r_*/2\), \(u_*\),
\(C_*\) and \(\lambda^0_*\); it does not depend on \(V^e\) or on the target.
Put \(\bar r\coloneqq r_*/2\), fix \(r\in(0,\bar r)\), and put
\(\Delta\coloneqq\lvert b^e-\beta\rvert\), \(\rho\coloneqq\min\{r,\Delta/4\}\) if
\(\Delta>0\) and \(\rho\coloneqq r\) if \(\Delta=0\).  Put
\[
 \overline V_r\coloneqq\max\bigl\{\widehat V,\ V_*,\ 2/\bar w,\
 2(1+\lvert x^e\rvert)K/\rho\bigr\}.
\]
For \(V^e\ge\overline V_r\) and \(w=1/V^e\), we have \(w<\bar w\) and, by
\eqref{eq:steady-shift}, \(\lvert a_w-a^e\rvert+\lvert b_w-b^e\rvert\le\rho/2<r_*/2\),
so the collar above is defined.  Put \(\Delta_w\coloneqq\lvert b_w-\beta\rvert\),
\(r'\coloneqq\min\{\rho/2,\Delta_w/4\}\) if \(\Delta_w>0\) and \(r'\coloneqq\rho/2\) if
\(\Delta_w=0\).  Then \(r'\in(0,r_*/2)\), and \(r'<\Delta_w/2\) when
\(\Delta_w>0\).  Remark~\ref{B-rem:single} at this \(V^e\), the target
\((a_w,b_w,x^e)\), the radius \(r'\) and \(\varepsilon\) gives
\(\overline A\) and \(r_0'\) such that, for every prior with
\(A_0>\overline A\) and
\(\lvert a_0-a_w\rvert+\lvert b_0-b_w\rvert+\lvert\bar x_0-x^e\rvert+\lvert V_0-V^e\rvert<r_0'\),
its conclusions \ref{lem:trapping:i} and \ref{lem:trapping:ii} about
\((a_w,b_w)\) hold with \(P^\psi_\vartheta\)-probability at least
\(1-\varepsilon\).  On that event,
\(\lvert a_t-a^e\rvert+\lvert b_t-b^e\rvert+\lvert\bar x_t-x^e\rvert<r'+\rho/2\le\rho\le r\),
which is \ref{thm:smooth-local:i}, and the limits in \ref{thm:smooth-local:ii} are the same.  When \(\Delta>0\),
\(\Delta_w\ge\Delta-\rho/2>0\), so
\(\lvert b_\infty-\beta\rvert\ge\Delta_w-r'\ge\Delta-\rho\ge3\Delta/4>0\),
which gives \(b_\infty\ne\beta\) in \ref{thm:smooth-local:ii}.  The bounds
\(V^e/2\le V_t\le2V^e\) of the trapping lemma are \(w^e/2\le w_t\le2w^e\) with
\(w^e=1/V^e\).

\emph{The prior condition in the state.}  Put
\(\overline w_r\coloneqq1/\overline V_r\), which lies in \((0,\bar w)\) because
\(\overline V_r\ge2/\bar w\).  For \(w^e\in(0,\overline w_r]\), apply the above
with \(V^e=1/w^e\ge\overline V_r\), take \(\overline A>0\) without loss of
generality, and put
\[
 r_0\coloneqq\min\bigl\{1/\overline A,\ w^e/2,\ r_0'/2,\ r_0'(w^e)^2/4\bigr\}.
\]
Let the state of a prior satisfy the condition of the theorem with this
\(r_0\).  Then \(u_0<1/\overline A\), so \(A_0>\overline A\).  Also
\(\lvert w_0-w^e\rvert<w^e/2\), so \(w_0>w^e/2\) and
\[
 \lvert V_0-V^e\rvert=\frac{\lvert w_0-w^e\rvert}{w_0w^e}
 <\frac{2r_0}{(w^e)^2}\le\frac{r_0'}2 .
\]
Together with
\(\lvert a_0-a_{w^e}\rvert+\lvert b_0-b_{w^e}\rvert+\lvert\bar x_0-x^e\rvert<r_0\le r_0'/2\),
the prior satisfies the conditions of the trapping lemma above, and the
conclusions follow.
\end{proof}

\begin{proof}[Proof of Proposition~\ref{thm:corner-local}]
\emph{Thresholds.}  Let \(\rho\) be as in the definition of a rule constant
near the equilibrium.  Put \(\bar r\coloneqq1\), fix \(r\in(0,\bar r)\), and
put \(m\coloneqq\min\{r,\rho\}\) if \(b^e=\beta\) and
\(m\coloneqq\min\{r,\rho,\lvert b^e-\beta\rvert/4\}\) otherwise.  Put
\(\overline V_r\coloneqq2+4/\rho\), and fix \(V^e\ge\overline V_r\) and
\(\varepsilon\in(0,1)\).  The thresholds \(r_0\le\min\{1,V^e/4\}\) and
\(\overline A\ge\max\{1,1/\rho\}\) are chosen in the bounds below; they depend
only on \(m\), \(\rho\), \(x^e\), \(V^e\) and \(\varepsilon\).  Fix a prior
that satisfies the three conditions of the theorem.
Write \(c=a+bx^e\) for the mean outcome at \(x^e\) under \((a,b)\),
\(c^e=\alpha+\beta x^e=a^e+b^ex^e\), \(c_0=a_0+b_0x^e\), so that
\(e_0=c^e-c_0\), and \(d=x^e-\bar x_0\).  Let
\(S_t=\sum_{j\le t}\varepsilon_j\); under \(P^\psi_\vartheta\) the
\(\varepsilon_j\) are i.i.d.\ \(N(0,1)\) whatever the actions.

\emph{The posterior while the action is \(x^e\).}  Suppose that
\(x_1=\dots=x_t=x^e\).  By \eqref{eq:design}, \(A_t=A_0+t\),
\(x^e-\bar x_t=A_0d/A_t\) and
\(V_t=V_0+A_0^2d^2\sum_{j<t}(A_jA_{j+1})^{-1}=V_0+A_0d^2(1-A_0/A_t)\).  By
\eqref{eq:sigma}, the prior precision matrix is
\(\left(\begin{smallmatrix}A_0&A_0\bar x_0\\A_0\bar x_0&A_0\bar x_0^2+V_0\end{smallmatrix}\right)\),
so, in the coordinates \((c,b)\), the log posterior density is, up to a
constant,
\[
 -\tfrac12\Bigl[A_0\bigl(c-c_0-(b-b_0)d\bigr)^2+V_0(b-b_0)^2
 +\textstyle\sum_{j\le t}(y_j-c)^2\Bigr],
\]
with \(y_j=c^e+\varepsilon_j\).  The posterior mean \((c_t,b_t)\) maximizes
it.  Maximizing over \(b\) first gives
\(b-b_0=A_0d(c-c_0)/(V_0+A_0d^2)\) and leaves
\(-\frac12[A'(c-c_0)^2+\sum_{j\le t}(y_j-c)^2]\) with
\(A'\coloneqq A_0V_0/(V_0+A_0d^2)\).  Hence
\[
 c_t-c_0=\frac{te_0+S_t}{A'+t},
 \qquad
 b_t-b_0=\frac{A_0d\,(c_t-c_0)}{V_0+A_0d^2},
 \qquad
 a_t=c_t-b_tx^e.
\]

\emph{Limits.}  If \(x_{t+1}=x^e\) for every \(t\), then
\(S_t/t\to0\) almost surely, so \(c_t\to c^e\), \(V_t\to V_0+A_0d^2\),
\(b_t\to b_0+A_0de_0/(V_0+A_0d^2)\), \(\bar x_t\to x^e\) and
\(a_t\to c^e-b_\infty x^e\).  These are the limits in
\ref{thm:corner-local:ii} and the exact formulas.

\emph{A maximal bound.}  For \(\lambda>0\), Kolmogorov's maximal
inequality \citep[Section~2.5]{d2019}, applied on the block \(t\le A'\), where
\(A'+t\ge A'\), and on the blocks \(2^{k-1}A'<t\le2^kA'\), \(k\ge1\), where
\(A'+t>2^{k-1}A'\), gives
\[
 P^\psi_\vartheta\Bigl(\sup_{t\ge1}\frac{\lvert S_t\rvert}{A'+t}\ge\lambda\Bigr)
 \le\frac{1}{\lambda^2A'}\Bigl(1+\sum_{k\ge1}\frac{2^kA'}{4^{k-1}A'}\Bigr)
 =\frac{5}{\lambda^2A'}.
\]
Let \(\Omega_\lambda\) be the complement of this event, with
\(\lambda\coloneqq\sqrt{5/(\varepsilon A')}\), so that
\(P^\psi_\vartheta(\Omega_\lambda)\ge1-\varepsilon\).

\emph{Bounds.}  By the conditions on the prior,
\(A_0d^2<r_0\), \(A_0\lvert d\rvert\lvert e_0\rvert<r_0\),
\(\lvert d\rvert<(r_0/A_0)^{1/2}\le r_0^{1/2}\),
\(\lvert e_0\rvert<(1+\lvert x^e\rvert)r_0\) and
\(V_0>V^e-r_0\ge1\), so \(A'\ge A_0/2\).  If
\(x_1=\dots=x_t=x^e\), then on \(\Omega_\lambda\), by the posterior formulas,
\(\lvert c_t-c^e\rvert=\lvert S_t-A'e_0\rvert/(A'+t)\le\lambda+\lvert e_0\rvert\),
\[
 \lvert b_t-b_0\rvert\le\frac{A_0\lvert d\rvert\bigl(\lvert e_0\rvert+\lambda\bigr)}{V_0}
 \le r_0+\Bigl(\frac{10\,r_0}{\varepsilon}\Bigr)^{1/2},
\]
where \(A_0\lvert d\rvert\lambda\le(10A_0d^2/\varepsilon)^{1/2}\) because
\(A_0/A'\le2\), and \(\lvert\bar x_t-x^e\rvert\le\lvert d\rvert\),
\(V_0\le V_t\le V_0+A_0d^2\), \(u_t\le1/A_0\) and \(w_t\le1/V_0\).  Since
\(a_t-a^e=(c_t-c^e)-(b_t-b^e)x^e\), the quantity
\(\lvert a_t-a^e\rvert+\lvert b_t-b^e\rvert+\lvert\bar x_t-x^e\rvert\) is at
most a multiple, depending only on \(x^e\), of
\(r_0+(r_0/\varepsilon)^{1/2}+(\varepsilon\overline A)^{-1/2}\), using
\(\lambda\le(10/(\varepsilon A_0))^{1/2}\).  Choose \(r_0\) small and
\(\overline A\) large so that this bound is below \(m\).  Then also
\(V^e/2\le V_t\le2V^e\), since \(\lvert V_t-V^e\rvert<2r_0\le V^e/2\);
\(u_t<1/\overline A\le\rho\); and \(w_t<2/V^e<\rho\).

\emph{Induction.}  On \(\Omega_\lambda\), the prior state \(s_0\)
satisfies these bounds with \(t=0\), so it lies in the box on which
\(\psi=x^e\), and \(x_1=x^e\).  If \(x_1=\dots=x_t=x^e\), the bounds put
\(s_t\) in the same box, so \(x_{t+1}=\psi(s_t)=x^e\).  Hence, on
\(\Omega_\lambda\), \(x_{t+1}=x^e\) for every \(t\), and the bounds give
\ref{thm:corner-local:i}, since \(m\le r\).  The limits give
\ref{thm:corner-local:ii} almost surely on \(\Omega_\lambda\).  If
\(b^e\ne\beta\), then \(\lvert b_\infty-b^e\rvert\le m\le\lvert b^e-\beta\rvert/4\), so
\(b_\infty\ne\beta\).  The three conditions on the prior are strict
inequalities between continuous functions of
\((A_0,\bar x_0,V_0,a_0,b_0)\), so they define an open set, which contains
every prior with \((a_0,b_0,\bar x_0,V_0)=(a^e,b^e,x^e,V^e)\) and
\(A_0>\overline A\).
\end{proof}

\section{Proof of generic failure}\label{app:global}

This appendix proves Theorem~\ref{thm:global}. It first states a more general
result, in which interior regularity is defined through smoothness of the rule
in the posterior state rather than in the mean and the covariance matrix, and
derives Theorem~\ref{thm:global} from it.

\subsection{Regularity in the posterior state}\label{app:global-state}

\begin{definition}[Interior regularity in the posterior state]\label{def:regular-psi}
A parameter \(\vartheta=(\alpha,\beta)\) is \textit{interior regular in the
state} for \(\psi\) if:
\begin{enumerate}
\item[\condtarget{cond:I1}{(I1)}] \(\psi\) has a complete-information limit \(h\) on an open
neighborhood \(\mathcal A\) of \(\vartheta\); write
\(x^\star(\vartheta)=h(\vartheta)\) and
\(s^\star(\vartheta)=(\alpha,\beta,0,x^\star(\vartheta),0)\);
\item[\condtarget{cond:I2}{(I2)}] \(\psi\) admits a \(C^1\) extension \(\bar\psi\), with locally Lipschitz
first derivatives, to an open neighborhood \(\mathcal O\) of
\(s^\star(\vartheta)\);
\item[\condtarget{cond:I3}{(I3)}] \(\gamma(\vartheta)=\bar\psi_b(s^\star(\vartheta))
-x^\star(\vartheta)\bar\psi_a(s^\star(\vartheta))\ne0\).
\end{enumerate}
Let \(\Theta^I_\psi\) be the set of such parameters, let \(\Theta^C_\psi\) be
the set of locally constant parameters of Definition~\ref{def:regular}, and
put \(\Theta^S_\psi=\Theta^I_\psi\cup\Theta^C_\psi\).
\end{definition}

\begin{theorem}[Generic failure, state form]\label{thm:smooth-global}
For every nondegenerate Gaussian prior \(\mu\),
\[
 P^{\mu,\psi}(V_t\to\infty,\ \vartheta\in\Theta^S_\psi)=0.
\]
Consequently, for every Borel set \(B\subseteq\Theta^S_\psi\),
\[
 P^{\mu,\psi}(\mathcal W,\ \vartheta\in B)=\mu(B),
\]
and \(P^{\psi}_\vartheta(\mathcal W)=1\) for Lebesgue-almost every
\(\vartheta\in\Theta^S_\psi\).
\end{theorem}

\begin{corollary}[Smooth rules of the mean and covariance]\label{cor:covariance}
The set \(\Theta^T_\psi\) of Definition~\ref{def:regular} is open,
\(\Theta^T_\psi\subseteq\Theta^I_\psi\), and \(x^\star=h\) and
\(\gamma=h_b-h\,h_a\) on \(\Theta^T_\psi\). Consequently, the conclusions of
Theorem~\ref{thm:smooth-global} hold with \(\Theta^S_\psi\) replaced by
\(\Theta^T_\psi\).
\end{corollary}

Theorem~\ref{thm:smooth-global} is proved in Section~\ref{app:global-proof}
and Corollary~\ref{cor:covariance} in Section~\ref{app:covariance}.

\begin{proof}[Proof of Theorem~\ref{thm:global}]
By Corollary~\ref{cor:covariance},
\(\Theta_\psi=\Theta^T_\psi\cup\Theta^C_\psi\subseteq\Theta^S_\psi\). The
set \(\Theta^T_\psi\) is open by the corollary and \(\Theta^C_\psi\) by the
proof of Theorem~\ref{thm:smooth-global}, so \(\Theta_\psi\) is open, hence
Borel. Theorem~\ref{thm:smooth-global}, restricted to \(\Theta_\psi\) by
monotonicity, gives the three conclusions.

For \(\mathcal W\cap\mathcal L\), let
\(E=\{V_\infty<\infty\}\setminus\mathcal L\), a joint Borel event. By
statements~\ref{lem:complete-learning:1} and~\ref{lem:complete-learning:3} of
Lemma~\ref{lem:complete-learning}, \(P^{\mu,\psi}(E)=0\). Since
\(\mathcal W\subseteq\{V_\infty<\infty\}\), this gives, for every Borel set \(B\subseteq\Theta_\psi\),
\(P^{\mu,\psi}(\mathcal W\cap\mathcal L,\ \vartheta\in B)
=P^{\mu,\psi}(\mathcal W,\ \vartheta\in B)=\mu(B)\). By
Lemma~\ref{lem:disintegration}\ref{lem:disintegration:ii}--\ref{lem:disintegration:iii},
\(P^\psi_\vartheta(E_\vartheta)=0\) for Lebesgue-almost every \(\vartheta\),
so \(P^\psi_\vartheta(\mathcal W\cap\mathcal L)=P^\psi_\vartheta(\mathcal W)=1\)
for Lebesgue-almost every \(\vartheta\in\Theta_\psi\).
\end{proof}

\begin{proof}[Proof of Corollary~\ref{cor:generic}]
For \(\Psi\in\mathcal P(U)\), let \(h_\Psi(a,b)=\Psi(a,b,0)\),
\(\gamma_\Psi=\partial_bh_\Psi-h_\Psi\,\partial_ah_\Psi\), and
\(Z_\Psi=\{\vartheta\in\R^2:\gamma_\Psi(\vartheta)=0\}\). These are defined on
all of \(\R^2\) because \(U\supseteq\R^2\times\{0\}\), and \(Z_\Psi\) is closed.
A nondegenerate Gaussian prior and Lebesgue measure have the same null sets.

\emph{The bound.} At every \(\vartheta\notin Z_\Psi\), the restriction
of \(\Psi\) to the neighborhood \(U\) of \((\vartheta,0)\) is a \(C^1\) extension
with locally Lipschitz first derivatives, and \(h_b\ne h\,h_a\) at
\(\vartheta\). So \(\vartheta\) is interior regular
(Definition~\ref{def:regular}), and Theorem~\ref{thm:global} gives
\(P^{\mu,\psi}(V_t\to\infty,\ \vartheta\notin Z_\Psi)=0\) and
\(P^\psi_\vartheta(\mathcal W)=1\) for Lebesgue-almost every
\(\vartheta\notin Z_\Psi\). Since the marginal law of \(\vartheta\) is \(\mu\),
\[
 P^{\mu,\psi}(V_t\to\infty)\le\mu(Z_\Psi).
\]
For \(\Psi\in\mathcal R\), \(Z_\Psi\) is null, which gives the conclusions
of~\ref{cor:generic:i} for every nondegenerate Gaussian \(\mu\).

\emph{Openness.} Fix a nondegenerate Gaussian \(\mu\) and \(\eta>0\), and let
\(\mathcal G_\eta=\{\Psi:\mu(Z_\Psi)<\eta\}\). If \(d(\Psi_n,\Psi)\to0\), then
\(h_{\Psi_n}\) and its first partial derivatives converge uniformly on
\(\R^2\). Since \(h_\Psi\) takes values in \(X\) and \(\partial_ah_\Psi\) is
bounded, \(\gamma_{\Psi_n}\to\gamma_\Psi\) uniformly on \(\R^2\). For every
\(\delta>0\),
\(Z_{\Psi_n}\subseteq\{\vartheta:\lvert\gamma_\Psi(\vartheta)\rvert\le\delta\}\)
for large \(n\). These closed sets decrease to \(Z_\Psi\) as \(\delta\downarrow0\),
so their \(\mu\)-measures decrease to \(\mu(Z_\Psi)\). Hence
\(\mu(Z_\Psi)<\eta\) implies \(\mu(Z_{\Psi_n})<\eta\) for large \(n\), and
\(\mathcal G_\eta\) is open.

\emph{Density.} Fix \(\Psi\) and \(\rho>0\). Let \(m\) be the midpoint of
\(X\), and let \(g(b)=\arctan b\), which is bounded with
\(g'(b)=1/(1+b^2)>0\) and bounded derivatives. For small \(\tau>0\) and
\(\lambda=(\lambda_0,\lambda_2)\) in a small square \(\Lambda\),
\[
 \Psi^\lambda=(1-\tau)\Psi+\tau m+\lambda_0+\lambda_2\,g(b)
\]
takes values in \(X\), lies in \(\mathcal P(U)\), and satisfies
\(d(\Psi,\Psi^\lambda)<\rho\). Its complete-information action is
\(h+\lambda_0+\lambda_2g(b)\) with \(h=(1-\tau)h_\Psi+\tau m\), so
\[
 \gamma_{\Psi^\lambda}=\gamma_{\Psi^0}-\lambda_0\,h_a+\lambda_2\,\bigl(g'(b)-g(b)\,h_a\bigr).
\]
At each \(\vartheta\in\R^2\), this is an affine function of \(\lambda\) with a
nonzero gradient, since \(g'(b)-g(b)h_a=g'(b)>0\) whenever \(h_a=0\). So the
\(\lambda\in\Lambda\) with \(\gamma_{\Psi^\lambda}(\vartheta)=0\) lie on a line and
form a Lebesgue-null set. For a nondegenerate Gaussian \(\mu\), Tonelli's
theorem gives \(\int_\Lambda\mu(Z_{\Psi^\lambda})\,d\lambda=0\), so
\(\mu(Z_{\Psi^\lambda})=0\) for some \(\lambda\in\Lambda\), and this
\(\Psi^\lambda\) lies in \(\mathcal R\).

\emph{Conclusion.} Fix a nondegenerate Gaussian \(\mu\). Then
\(\mathcal R=\bigcap_{n\ge1}\mathcal G_{1/n}\) is a countable intersection of
open sets, and it is dense, so it is a dense \(G_\delta\). Each
\(\mathcal G_{1/n}\) contains \(\mathcal R\), so it is open and dense, and the
complement of \(\mathcal R\) is meagre. This proves~\ref{cor:generic:i}. For
\ref{cor:generic:ii}, \(\mathcal G_\varepsilon\) is open, contains the dense
set \(\mathcal R\), and satisfies \(P^{\mu,\psi}(V_t\to\infty)<\varepsilon\) by
the bound.
\end{proof}

\subsection{Pathwise facts}\label{app:global-pathwise}

Throughout the rest of this appendix \(\psi\) is a decision rule and \(\mu\) a
nondegenerate Gaussian prior.
Probabilities refer to \(P^{\mu,\psi}\) unless the truth is subscripted.  Put
\[
 {X_{\mathrm{avg}}}\coloneqq \operatorname{co}(\{{\bar x}_0\}\cup X),
 \qquad
 {C_D}\coloneqq \sup\{\lvert x-{\bar x}\rvert :x\in X,\ {\bar x}\in {X_{\mathrm{avg}}}\}<\infty.
\]
The following facts are pathwise and hold for every policy with actions in
\(X\).  By \eqref{eq:design}, the average action satisfies
\begin{equation}\label{C-average-action}
 A_t\bar x_t=A_0\bar x_0+\sum_{j=0}^{t-1}x_{j+1},
\end{equation}
and \(A_t=A_0+t\). Thus \(\bar x_t\) is a convex combination of \(\bar x_0\),
with weight \(A_0/A_t\), and of points of \(X\). Hence
\({\bar x}_t\in {X_{\mathrm{avg}}}\) at every date, so \(\lvert D_t\rvert \le {C_D}\) and
\begin{equation}\label{C-xbar-bound}
 \lvert\bar x_t\rvert\le C_{\bar x}\coloneqq\max\{\lvert\bar x_0\rvert,\lvert\underline x\rvert,\lvert\overline x\rvert\}
 \qquad(t\ge0);
\end{equation}
moreover, on every path,
\begin{equation}\label{C-xbar-dist}
 \operatorname{dist}(\bar x_t,X)
 \le\frac{A_0}{A_0+t}\operatorname{dist}(\bar x_0,X)\longrightarrow0
 \qquad(t\to\infty),
\end{equation}
so every limit point of \((\bar x_t)\) lies in \(X\).  The gap bound gives
\(V_t\le V_0+C_D^2t\), so \(r_t\coloneqq V_t/A_t\le C_r\) for a finite
constant \(C_r\).

Lemma~\ref{lem:posterior-interface}\ref{lem:posterior-interface:c} gives
\begin{equation}
 (a_t,b_t,u_t,w_t)\to(\alpha,\beta,0,0)
 \quad\text{almost surely on }E_\infty,
 \qquad
 P^{\mu,\psi}(V_\infty<\infty,\ \beta=b_\infty)=0.
 \label{C-collapse-limit}
\end{equation}

\subsection{The exclusion package for a regular rule}
\label{app:global-package}

The coherent exclusion package (Definition~\ref{def:package}) and the
exclusion theorem (Theorem~\ref{thm:exclusion}) are stated in
Section~\ref{supp:exclusion} of the Supplement.  For the decision rule
\(\psi\), the smooth reference map of the package is a smooth extension of
\(\psi\) itself, read in the coordinates \((a,b,u,\bar x,\ell_V)\)
(Lemma~\ref{lem:smooth-package}), so the action equals the reference action and clause
\ref{package:approximation} holds with equality.  The
events in the package and in the conclusion of the theorem are determined by
\(\vartheta\), the actions and the outcomes, so their probabilities are those
under \(P^{\mu,\psi}\) and \(P^\psi_\vartheta\)
(Section~\ref{supp:general-policies}).

The locally constant case of Theorem~\ref{thm:smooth-global} uses the
following elementary fact.

\begin{auxlemma}[Eventual constancy starves slope precision]
\label{lem:eventual-constant}
If \(x_{t+1}=e\) for every \(t\ge T\), where \(T<\infty\) and \(e\) may be
path dependent, then
\[
 A_tD_t=A_T(e-\bar x_T)\qquad(t\ge T).
\]
Consequently \(\limsup_tA_t\lvert D_t\rvert<\infty\) and
\(\sum_{t\ge T}\kappa_tD_t^2<\infty\).
\end{auxlemma}

\begin{proof}
Equation~\eqref{C-average-action} gives
\(A_t\bar x_t=A_T\bar x_T+(A_t-A_T)e\) for \(t\ge T\). Since
\(D_t=e-\bar x_t\), the displayed identity follows. Thus
\(\lvert D_t\rvert\le C/A_t\) after \(T\), for the finite random constant
\(C=A_T\lvert e-\bar x_T\rvert\), and the square sum is bounded by
\(C^2\sum_{t\ge T}A_t^{-2}\).
\end{proof}

The complete-information limit \(h\) of Section~\ref{sec:equilibria} on an
open set \(\mathcal A\subseteq\R^2\) is unique and continuous on
\(\mathcal A\). For continuity, let \(\vartheta_n\to\vartheta\) in
\(\mathcal A\) and fix \(\bar x\in X\). The definition at \(\vartheta_n\) gives
a feasible state \(s_n\) within \(1/n\) of \((\vartheta_n,0,\bar x,0)\) with
\(\lvert\psi(s_n)-h(\vartheta_n)\rvert<1/n\); since \(s_n\to(\vartheta,0,\bar x,0)\), the
definition at \(\vartheta\) gives \(\psi(s_n)\to h(\vartheta)\), hence
\(h(\vartheta_n)\to h(\vartheta)\).

\begin{lemma}[Openness and the exclusion package]
\label{lem:smooth-package}
The set \(\Theta^I_\psi\) is open. For every nondegenerate Gaussian prior
\(\mu\), there is an at most countable, possibly empty, cover of \(\Theta^I_\psi\) by open rectangles
\((O_j)_{j\in J}\), \(J\subseteq\mathbb N\), such that \(\mu\) and the rule \(\psi\) admit a
coherent exclusion package on \(\Theta^I_\psi\) \textup{(}Definition~\textup{\ref{def:package})} with \(\eta=1\), the
cover \(\Theta^I_\psi=\bigcup_{j\in J}O_j\), and the smooth reference map of member \(j\)
equal to \(\bar\psi_j\circ\iota\), where
\(\iota(a,b,u,\bar x,\ell_V)=(a,b,u,\bar x,e^{\ell_V})\) and \(\bar\psi_j\) is
a \(C^1\) extension of \(\psi\) with locally Lipschitz first derivatives.
\end{lemma}

\begin{proof}
Fix \(\vartheta_0=(\alpha_0,\beta_0)\in\Theta^I_\psi\) with the data
\(\mathcal A\), \(h\), \(\mathcal O\) and \(\bar\psi\) of \ref{cond:I1}--\ref{cond:I2}, and
put \(x_0^\star\coloneqq x^\star(\vartheta_0)\) and
\(s_0^\star\coloneqq s^\star(\vartheta_0)\).

\emph{Step 1: the extension at complete information.} Approaching a point
\((a,b,0,\bar x,0)\in\mathcal O\) with \((a,b)\in\mathcal A\) and
\(\bar x\in X\) by the feasible states \((a,b,1/n,\bar x,1/n)\), continuity of
\(\bar\psi\) and the definition of \(h\) give
\begin{equation}
 \bar\psi(a,b,0,\bar x,0)=h(a,b).
 \label{C-ci-identity}
\end{equation}
With \(\bar x=x_0^\star\), \(h\) agrees near \(\vartheta_0\) with a \(C^1\) map
whose gradient \((\bar\psi_a,\bar\psi_b)(s_0^\star)\) is nonzero, since
\(\gamma(\vartheta_0)\ne0\); so \(h\) has no local extremum at
\(\vartheta_0\), and since \(h\) takes values in \(X\),
\(x_0^\star\in\operatorname{int}X\). Choose \(\epsilon_0\in(0,1/2)\) such
that the closed cube \(K\coloneqq\{s:\lVert s-s_0^\star\rVert_\infty\le\epsilon_0\}\) lies in
\(\mathcal O\), with projections on \((a,b)\) in \(\mathcal A\) and on
\(\bar x\) in \(X\). By \eqref{C-ci-identity}, on \(K\cap\{u=w=0\}\),
\(\bar\psi\) does not depend on \(\bar x\), so \(\bar\psi_{\bar x}=0\) there,
\(h\) is \(C^1\) near \(\vartheta_0\) with \((h_a,h_b)=(\bar\psi_a,\bar\psi_b)\),
and \(\bar\psi(s^\star(\vartheta))=x^\star(\vartheta)\) whenever
\(s^\star(\vartheta)\in K\).

\emph{Step 2: a neighborhood of truths.} By continuity of \(h\), choose an
open square \(B^0\) about \(\vartheta_0\), with closure in \(\mathcal A\), such
that \(\lvert\alpha-\alpha_0\rvert,\lvert\beta-\beta_0\rvert,\lvert x^\star(\vartheta)-x_0^\star\rvert<\epsilon_0/2\)
for every \(\vartheta\in\overline{B^0}\). Then
\(\Gamma^0\coloneqq\{s^\star(\vartheta):\vartheta\in\overline{B^0}\}\) is a compact
subset of the interior of \(K\), and by Step 1
\(\gamma(\vartheta)=\bar\psi_b(s^\star(\vartheta))
-x^\star(\vartheta)\bar\psi_a(s^\star(\vartheta))\) is continuous on
\(\overline{B^0}\) and nonzero at \(\vartheta_0\). Shrinking \(B^0\), assume
\(\lvert\gamma\rvert\ge2\lambda>0\) on \(\overline{B^0}\). Every \(\vartheta\in B^0\)
then satisfies \ref{cond:I1}--\ref{cond:I3} with the same \(\mathcal A\) and \(\mathcal O\), so
\(B^0\subseteq\Theta^I_\psi\), and \(\Theta^I_\psi\) is open.

\emph{Step 3: transversality and flatness on the cube.} Put
\(\bar x_+(s)\coloneqq\bar x+\frac{u}{1+u}(\bar\psi(s)-\bar x)\), so that
\(\bar x_+\circ\iota\) is the quantity \(\bar x_m^{+,\varphi}\) of
\ref{package:regularity} for the smooth reference map
\(\varphi=\bar\psi\circ\iota\) chosen below. On \(\Gamma^0\), \(\bar x_+=x^\star(\vartheta)\)
and \(\bar\psi_b-\bar x_+\bar\psi_a=\gamma(\vartheta)\), so, by continuity,
there is an open neighborhood \(\mathcal M\) of \(\Gamma^0\) with
\(\overline{\mathcal M}\subseteq\operatorname{int}K\) on which
\begin{equation}
 \lvert\bar\psi_b-\bar x_+\bar\psi_a\rvert\ge\lambda.
 \label{C-transversality-margin}
\end{equation}
On the compact convex set \(K\), the first derivatives of \(\bar\psi\) are
bounded and Lipschitz; let \(C_1\) be a common bound for their absolute values
and their Lipschitz constants in the maximum norm. For
\(s=(a,b,u,\bar x,w)\in K\) with \(u,w\ge0\), the point
\((a,b,0,\bar x,0)\) lies in \(K\), and \(\bar\psi_{\bar x}\) vanishes there
by Step 1; hence
\begin{equation}
 \lvert\bar\psi_{\bar x}(s)\rvert
 =\lvert\bar\psi_{\bar x}(s)-\bar\psi_{\bar x}(a,b,0,\bar x,0)\rvert
 \le C_1\max\{u,w\}\le C_1(u+w).
 \label{C-flat}
\end{equation}

\emph{Clause \textup{\ref{package:geometry}}.} Choose \(\epsilon\in(0,\epsilon_0)\) such that
the cube \(\{s:\lVert s-s_0^\star\rVert_\infty\le\epsilon\}\) lies in \(\mathcal M\),
and shrink \(B^0\) so that \(\lvert\alpha-\alpha_0\rvert,\lvert\beta-\beta_0\rvert,
\lvert x^\star(\vartheta)-x_0^\star\rvert<\epsilon/2\) on \(\overline{B^0}\). Let
\(\mathcal B^{\mathrm{in}}_{ab\bar x}\) and \(\mathcal B^{\mathrm{mid}}_{ab\bar x}\)
be the open cubes of radii \(\epsilon/2\) and \(\epsilon\) about
\((\alpha_0,\beta_0,x_0^\star)\), and put
\(\mathcal B^{\mathrm{in}}\coloneqq\mathcal B^{\mathrm{in}}_{ab\bar x}\times[0,\epsilon/2)\)
and \(\mathcal B^{\mathrm{mid}}\coloneqq\mathcal B^{\mathrm{mid}}_{ab\bar x}\times[0,\epsilon)\)
in \((a,b,\bar x,u)\), with \(\ell^{\mathrm{in}}\coloneqq\log(\epsilon/2)\) and
\(\ell^{\mathrm{out}}\coloneqq\log\epsilon\). The sets
\(\mathcal S^{\mathrm{tail}}=\mathcal B^{\mathrm{mid}}\times(-\infty,\ell^{\mathrm{out}})\)
and \(\mathcal K^0=\overline{\mathcal B^{\mathrm{in}}}\times(-\infty,\ell^{\mathrm{in}}]\)
of \ref{package:geometry},
read in \((a,b,u,\bar x,w)\) with \(w=e^{\ell_V}\), lie in the cube of radius
\(\epsilon\) about \(s_0^\star\), hence in \(\mathcal M\subseteq K\); and
\((\alpha,\beta,x^\star(\vartheta))\in\mathcal B^{\mathrm{in}}_{ab\bar x}\) for
every \(\vartheta\in\overline{B^0}\).

\emph{Clause \textup{\ref{package:regularity}}.} Take the smooth reference map
\(\varphi\coloneqq\bar\psi\circ\iota\) on \(\mathcal S^{\mathrm{tail}}\), that is,
\(\bar\psi\) read in \((a,b,u,\bar x,\ell_V)\) through \(w=e^{\ell_V}\). The
coordinate map \(\iota\) is \(1\)-Lipschitz on \(\mathcal S^{\mathrm{tail}}\), since
\(e^{\ell_V}<\epsilon<1\) there. The first derivatives of \(\varphi\) in
\(a,b,u,\bar x\) are those of \(\bar\psi\) composed with \(\iota\), hence
bounded by \(C_1\) and \(C_1\)-Lipschitz; and
\(\partial_{\ell_V}\varphi=e^{\ell_V}\,\bar\psi_w\circ\iota\) is the product of
two functions bounded by \(1\) and \(C_1\) and Lipschitz with constants
\(1\) and \(C_1\), hence bounded by \(C_1\) and \(2C_1\)-Lipschitz. Thus
\(\varphi\) has a finite \(C^{1,1}\) norm on \(\mathcal S^{\mathrm{tail}}\),
which is the \(C^{1,\eta}\) clause with \(\eta=1\); in particular
\(\varphi_a\) and \(\varphi_b\) have a common Lipschitz constant in
\(\ell_V\) on \(\mathcal K^0\). The transversality clause is
\eqref{C-transversality-margin} with \(\lambda_m\coloneqq\lambda\), since
\(\mathcal K^0\subseteq\mathcal M\). For the covariance clause, \eqref{C-flat} and
\(\lvert\partial_{\ell_V}\varphi\rvert=w\lvert\bar\psi_w\rvert\le C_1w\) give, on \(\mathcal K^0\),
\[
 \lvert\varphi_{\bar x}\rvert+\lvert\partial_{\ell_V}\varphi\rvert
 \le 2C_1(u+e^{\ell_V}).
\]

\emph{Clause \textup{\ref{package:corridor}}.} Choose the four thresholds of the corridor
\(\mathcal K_m\subseteq\mathcal K^0\) so that they satisfy the finitely many inequalities
that Section~\ref{S2-section-5} requires (for \(\eta=1\), read at one fixed \(\eta'\in(6/7,1)\)); they
depend only on \(\eta\), \(X\), \(\mu\), the data of \ref{package:geometry}--\ref{package:regularity} and the
constant \(C_{2,m}\coloneqq1\) of \ref{package:approximation}, and are fixed before any path is observed.

\emph{Clause \textup{\ref{package:approximation}}.} Every state of \(\mathcal K_m\) is feasible and lies in
\(\mathcal O\), so \(\psi=\bar\psi=\varphi\) there. The realized action is
\(x_{t+1}=\psi(s_t)\), so at
every date with \(s_t\in\mathcal K_m\), \(x_{t+1}=\psi(s_t)=\varphi(s_t)\); hence
\eqref{eq:tracking} holds with \(C_{2,m}=1\).

\emph{The cover.} For each \(\vartheta_0\in\Theta^I_\psi\), select an open
rectangle with rational vertices that contains \(\vartheta_0\) and has closure
in the final \(B^0\) of \(\vartheta_0\). Only countably many such rectangles
occur, possibly none; enumerate them as \((O_j)_{j\in J}\) with \(J\subseteq\mathbb N\), fix for each one representative
\(\vartheta_0\), and retain its data. Then \(\Theta^I_\psi=\bigcup_{j\in J}O_j\),
each \(O_j\subseteq\Theta^I_\psi\), and \ref{package:geometry}--\ref{package:approximation} hold for member \(j\) with
truth set \(\Theta_j\coloneqq O_j\), \(\bar\psi_j\coloneqq\bar\psi\), and the smooth reference map \(\bar\psi_j\circ\iota\);
the data are deterministic and truth-independent.

\emph{Clause \textup{\ref{package:anchoring}}.} Fix \(j\) with its representative data. On
\(E_\infty\cap\{\vartheta\in O_j\}\), outside a \(P^{\mu,\psi}\)-null set,
\eqref{C-collapse-limit} gives \((a_t,b_t,u_t,w_t)\to(\alpha,\beta,0,0)\), and
\(\lvert\bar x_t\rvert\le C_{\bar x}\) by \eqref{C-xbar-bound}. On such a path,
\(x_{t+1}=\psi(s_t)\to x^\star(\vartheta)\). Otherwise there are
\(\delta>0\) and a subsequence along which
\(\lvert\psi(s_t)-x^\star(\vartheta)\rvert\ge\delta\); passing to a further subsequence
with \(\bar x_t\to\hat x\), where \(\hat x\in X\) by \eqref{C-xbar-dist}, the
feasible states
\(s_t\) converge to \((\alpha,\beta,0,\hat x,0)\) with
\((\alpha,\beta)\in O_j\subseteq\mathcal A\), and the definition of the
complete-information limit gives
\(\psi(s_t)\to h(\vartheta)=x^\star(\vartheta)\), a
contradiction.
Lemma~\ref{lem:anchored-collapse}
states that, for a set \(B\) and a map \(x^\star:B\to X\), if
\(x_{t+1}\to x^\star(\vartheta)\) almost surely on
\(E_\infty\cap\{\vartheta\in B\}\), then, almost surely on that event,
\((a_t,b_t,u_t,V_t^{-1})\to(\alpha,\beta,0,0)\),
\(\bar x_t\to x^\star(\vartheta)\), \(D_t\to0\) and \(V_t/A_t\to0\). It
applies with \(B=O_j\). Hence \(s_t\to s^\star(\vartheta)\),
\(D_t\to0\) and \(u_t/w_t=V_t/A_t\to0\) there. Since
\((\alpha,\beta,x^\star(\vartheta))\) lies in the open cube
\(\mathcal B^{\mathrm{in}}_{ab\bar x}\), and
\(\lvert\varphi(s_t)-\bar x_t\rvert=\lvert x_{t+1}-\bar x_t\rvert=\lvert D_t\rvert\) once \(s_t\in\mathcal O\),
the state eventually satisfies every defining inequality of the corridor
\(\mathcal K_m\) and remains there. This is \ref{package:anchoring} for member \(j\).
\end{proof}

\subsection{Proof of Theorem~\ref{thm:smooth-global}}
\label{app:global-proof}

\begin{proof}[Proof of Theorem~\ref{thm:smooth-global}]
\emph{The set \(\Theta^S_\psi\).} \(\Theta^C_\psi\) is open. If a truth is
locally constant with radius \(r\), every truth within \(r/2\) of it in each
coordinate is locally constant with radius \(r/2\).
Lemma~\ref{lem:smooth-package} proves that \(\Theta^I_\psi\) is open. Hence
\(\Theta^S_\psi\) is open, and in particular Borel.

\emph{Interior-regular truths.} Lemma~\ref{lem:smooth-package} gives a
coherent exclusion package on the open set \(\Theta^I_\psi\) for \(\mu\) and
\(\psi\). Theorem~\ref{thm:exclusion}, applied to \(\psi\) with
\(\Theta_0=\Theta^I_\psi\), gives
\(P^{\mu,\psi}(V_t\to\infty,\ \vartheta\in\Theta^I_\psi)=0\).

\emph{Locally constant truths.} Let \(\omega\) be a path in
\(E_\infty\cap\{\vartheta\in\Theta^C_\psi\}\) on which
\((a_t,b_t,u_t,w_t)\to(\alpha,\beta,0,0)\); by \eqref{C-collapse-limit}, the
remaining paths of that event form a null set. Take \(e\) and \(r\) as in
Definition~\ref{def:regular} for this \(\vartheta\), and let \(h\) be the
complete-information limit. Along feasible states converging to
\((a,b,0,e,0)\) with \(\lvert a-\alpha\rvert<r\) and \(\lvert b-\beta\rvert<r\),
the rule eventually chooses \(e\), so \(h=e\) there. As in the proof of Lemma~\ref{lem:smooth-package}
(clause~\ref{package:anchoring}), every
subsequence of \((\bar x_t)\) has a further subsequence converging to a point
of \(X\) by \eqref{C-xbar-bound} and \eqref{C-xbar-dist}, so
\(x_{t+1}=\psi(s_t)\to h(\vartheta)=e\). Then \(\bar x_t\to e\), since
\(\bar x_t\) is a weighted average of \(\bar x_0\) and \(x_1,\dots,x_t\) in
which the weight of \(\bar x_0\) vanishes. Hence there is \(T<\infty\) such
that \(\lvert a_t-\alpha\rvert,\lvert b_t-\beta\rvert,u_t,w_t<r\) and
\(\lvert\bar x_t-e\rvert<r\) for \(t\ge T\), so \(x_{t+1}=\psi(s_t)=e\) for
\(t\ge T\). Lemma~\ref{lem:eventual-constant}
gives \(\sum_{t\ge T}\kappa_tD_t^2<\infty\). By
\eqref{eq:design}, \(V_\infty=V_T+\sum_{t\ge T}\kappa_tD_t^2<\infty\),
contradicting \(\omega\in E_\infty\). Hence
\(P^{\mu,\psi}(V_t\to\infty,\ \vartheta\in\Theta^C_\psi)=0\).

Combining the interior-regular and locally constant truths gives the first
display of Theorem~\ref{thm:smooth-global}. For a Borel
\(B\subseteq\Theta^S_\psi\), this display gives
\(P^{\mu,\psi}(V_\infty<\infty,\ \vartheta\in B)=\mu(B)\), since the
marginal law of \(\vartheta\) is \(\mu\). On \(\{V_\infty<\infty\}\) the
posterior means converge almost surely, and the no-atom conclusion in
\eqref{C-collapse-limit} removes, with probability zero, the paths with
\(b_\infty=\beta\); this proves the setwise identity of Theorem~\ref{thm:smooth-global}. With
\(B=\Theta^S_\psi\), the joint Borel event
\(\{\vartheta\in\Theta^S_\psi\}\setminus\mathcal W\) is
\(P^{\mu,\psi}\)-null, and Lemma~\ref{lem:disintegration}
gives \(P^\psi_\vartheta(\mathcal W)=1\) for \(\mu\)-almost
every \(\vartheta\in\Theta^S_\psi\). A nondegenerate Gaussian prior has a
strictly positive Lebesgue density, so its null sets are the Lebesgue-null
sets.
\end{proof}

\subsection{Proof of Corollary~\ref{cor:covariance}}\label{app:covariance}
\begin{proof}[Proof of Corollary~\ref{cor:covariance}]
The map \((u,\bar x,w)\mapsto\Sigma(u,\bar x,w)\) of \eqref{eq:sigma} is
polynomial, does not depend on \((a,b)\), and satisfies
\(\Sigma(0,\bar x,0)=0\). Fix \(\vartheta_0\in\Theta^T_\psi\), let
\(\mathcal N\) be the open neighborhood of \((\vartheta_0,0)\) on which
Definition~\ref{def:regular} provides the extension \(\widetilde\Psi\), and
denote \(\widetilde\Psi\) again by \(\Psi\). Take an
open neighborhood of \((\vartheta_0,0)\) in \(\mathcal N\), of the form
\(\mathcal A_0\times\mathcal V\) with \(\mathcal A_0\subseteq\R^2\) and
\(\mathcal V\subseteq\mathbb S\) open.

\emph{Openness.} The function \(h=\Psi(\cdot,0)\) is \(C^1\) on
\(\mathcal A_0\), so \(h_b-h\,h_a\) is nonzero on a neighborhood of
\(\vartheta_0\) contained in \(\mathcal A_0\); since
\(\mathcal A_0\times\mathcal V\) is a neighborhood of \((\vartheta,0)\) for
every \(\vartheta\in\mathcal A_0\), on which \(\Psi\) represents \(\psi\),
this neighborhood lies in \(\Theta^T_\psi\).

\emph{\ref{cond:I1}.} If \((a,b)\in\mathcal A_0\), \(\bar x\in X\) and feasible states
\(s_n\to(a,b,0,\bar x,0)\), then \(\Sigma(u_n,\bar x_n,w_n)\to0\), so for
large \(n\)
\[
 (a_n,b_n,\Sigma(u_n,\bar x_n,w_n))\in\mathcal A_0\times\mathcal V,
 \qquad
 \psi(s_n)=\Psi(a_n,b_n,\Sigma(u_n,\bar x_n,w_n)),
\]
and continuity of \(\Psi\) gives \(\psi(s_n)\to\Psi(a,b,0)=h(a,b)\), which lies in
the closed set \(X\). So \(h\) is the complete-information limit of \(\psi\) on
\(\mathcal A_0\), and \(x^\star(\vartheta_0)=h(\vartheta_0)\).

\emph{\ref{cond:I2}.} On the open set
\(\mathcal O=\{s\in\R^5:(a,b)\in\mathcal A_0,\ \Sigma(u,\bar x,w)\in\mathcal V\}\),
which contains \(s^\star(\vartheta_0)\), the map
\(\bar\psi(s)=\Psi(a,b,\Sigma(u,\bar x,w))\) equals \(\psi\) at feasible
states. By the chain rule, it is \(C^1\) with locally Lipschitz first
derivatives.

\emph{\ref{cond:I3}.} Since \(\Sigma\) does not depend on \((a,b)\),
\((\bar\psi_a,\bar\psi_b)(s^\star(\vartheta_0))=(h_a,h_b)(\vartheta_0)\), so
\(\gamma(\vartheta_0)=h_b(\vartheta_0)-h(\vartheta_0)h_a(\vartheta_0)\ne0\).

Hence \(\Theta^T_\psi\subseteq\Theta^I_\psi\subseteq\Theta^S_\psi\), with
\(x^\star=h\) and \(\gamma=h_b-h\,h_a\) on \(\Theta^T_\psi\), and \(\Theta^T_\psi\) is
open, hence Borel. Theorem~\ref{thm:smooth-global}, restricted to
\(\Theta^T_\psi\), gives the stated conclusions.
\end{proof}

\renewcommand{\bibsection}{\section*{\refname}\pdfbookmark[1]{References}{part1-article-references}}
\bibliographystyle{econ}
\bibliography{part1}

\clearpage
\pagenumbering{arabic}
\setcounter{section}{0}
\setcounter{subsection}{0}
\setcounter{theorem}{0}
\setcounter{equation}{0}
\setcounter{figure}{0}
\setcounter{table}{0}
\setcounter{footnote}{0}
\renewcommand{\thesection}{S\arabic{section}}
\renewcommand{\thefigure}{S\arabic{figure}}
\renewcommand{\thetable}{S\arabic{table}}
\pdfbookmark[0]{Supplement}{part1-supplement}
\phantomsection\label{supplement}
\thispagestyle{plain}
\begin{center}
\vspace*{1.4cm}
{\LARGE Supplement to ``\ArticleTitle''\par}
\vspace{1.2cm}
{\large \AuthorOne\textsuperscript{1}, \AuthorTwo\textsuperscript{2},
and \AuthorThree\textsuperscript{3}\par}
\bigskip
{\normalsize
\textsuperscript{1}\AffiliationOne\par
\textsuperscript{2}\AffiliationTwo\par
\textsuperscript{3}\AffiliationThree\par}
\end{center}

\begin{abstract}
This supplement provides three technical results in Sections~\ref{supp:general-policies}--\ref{supp:exclusion}, namely the posterior results, the trapping lemma, and the
abstract exclusion theorem with its proof.  They are stated in
enough generality that the companion paper \citep{kky2026partii} can use them as well.
  Sections~\ref{supp:myopic}
to~\ref{supp:targeting} prove Corollary~\ref{cor:myopic}, verify the
examples and the two rules of Section~\ref{sec:rules}, and analyze the
targeting payoff.
\end{abstract}

\section{Posterior results for general policies}\label{supp:general-policies}

This section states Lemmas~\ref{lem:posterior-interface}
and~\ref{lem:disintegration} for policies that may randomize or depend on
the whole history.  Sections~\ref{supp:trapping} and~\ref{supp:exclusion}
use this setting.

A \emph{policy kernel} is a Borel randomized posterior-state Markov kernel:
a Borel stochastic kernel \(\Pi(s,dx)\) from the feasible states to
\(X=[\underline x,\overline x]\).  On an augmented probability space let
\(\zeta_1,\zeta_2,\ldots\) be i.i.d.\ uniform random variables on \([0,1]\)
such that \(\vartheta\), the i.i.d.\ \(N(0,1)\) sequence
\((\varepsilon_t)_{t\ge1}\) and the seed sequence \((\zeta_t)_{t\ge1}\) are
mutually independent, and realize \(\Pi\) by a jointly Borel map \(\psi\)
from the product of the feasible states and \([0,1]\) to \(X\), with
\(\psi(s,\zeta)\) distributed as \(\Pi(s,\cdot)\) for every feasible \(s\)
when \(\zeta\) is uniform.  Such a map exists by \citet[Lemma~4.22]{k2021},
applied with the Borel space \(X\) \citep[Theorem~1.8]{k2021}.  The action
is \(x_{t+1}=\psi(s_t,\zeta_{t+1})\).  Put
\[
 \mathcal H_t\coloneqq\sigma(\zeta_1,y_1,\ldots,\zeta_t,y_t),
 \qquad
 \mathcal F_t\coloneqq\mathcal H_t\vee\sigma(\zeta_{t+1}),
 \qquad t\ge0,
\]
with \(\mathcal H_0\) trivial, and
\(\mathcal F_\infty\coloneqq\sigma\left(\bigcup_t\mathcal H_t\right)\).  The
history \(\mathcal H_t\) records the seeds and outcomes through date \(t\);
the decision \(\sigma\)-field \(\mathcal F_t\) adds the seed of the next
action, so \(x_{t+1}\) is \(\mathcal F_t\)-measurable.  The path space is
\(\Omega\coloneqq([0,1]\times X\times\R)^{\mathbb N}\), with coordinates
\((\zeta_t,x_t,y_t)_{t\ge1}\); \(\mathcal Q\) is the law of
\(\bigl((\varepsilon_t),(\zeta_t)\bigr)\); \(\Phi(\vartheta',e,v)\) is
generated by \(x_t=\psi(s_{t-1},v_t)\) and \(y_t=\alpha'+\beta'x_t+e_t\);
\(P^\Pi_{\vartheta'}\coloneqq\mathcal Q\circ\Phi(\vartheta',\cdot)^{-1}\);
and \(P^{\mu,\Pi}\) is the joint law of
\((\vartheta,\Phi(\vartheta,\varepsilon,\zeta))\).  The law of
\((\vartheta,(x_t,y_t)_{t\ge1})\) under \(P^{\mu,\Pi}\), and that of
\((x_t,y_t)_{t\ge1}\) under \(P^\Pi_{\vartheta'}\), depend on \(\Pi\) alone,
not on the realization.  The fitted-mean error
\(\xi_t=\alpha+\beta\bar x_t-a_t-b_t\bar x_t\), the weight
\(\kappa_t=A_t/(A_t+1)\) and the limits \((a_\infty,b_\infty)\) are as in
Appendix~\ref{app:posterior}.

\begin{lemma}[Posterior interface for policy kernels]
\label{S1-lem:posterior-general}
Let \(\Pi\) be a policy kernel.  Then, for every \(t\), \(s_t\) is
\(\mathcal H_t\)-measurable and \(x_{t+1}\) is \(\mathcal F_t\)-measurable
up to \(P^{\mu,\Pi}\)-null sets, and
parts~\ref{lem:posterior-interface:a}--\ref{lem:posterior-interface:c} of Lemma~\ref{lem:posterior-interface} hold under
\(P^{\mu,\Pi}\), with the decision \(\sigma\)-field \(\mathcal F_t\) of this
section; in part~\ref{lem:posterior-interface:a} the conditional law of \(\vartheta\) given
\(\mathcal F_t\) equals its conditional law given \(\mathcal H_t\), and in
part~\ref{lem:posterior-interface:b} \(\eta_{t+1}\) is \(\mathcal H_{t+1}\)-measurable up to
\(P^{\mu,\Pi}\)-null sets.  On the augmented probability space, where
\(\mathcal H_t\) and \(\mathcal F_t\) are generated by the seeds and
outcomes themselves, these measurability statements hold exactly.  The same
holds when \([\underline x,\overline x]\) is replaced, here and in the path
space, by an arbitrary set \(X\subseteq\R\), bounded or not, and the actions
are \(x_{t+1}=\psi(s_t,\zeta_{t+1})\) for a jointly Borel map \(\psi\) into
\(X\).
\end{lemma}

\begin{lemma}[Fixed-truth transfer for policy kernels]
\label{S1-lem:disintegration-general}
Let \(\Pi\) be a policy kernel.  Then \(\Phi\) is Borel, and parts~\ref{lem:disintegration:i}--\ref{lem:disintegration:iii}
of Lemma~\ref{lem:disintegration} hold with \(P^{\mu,\Pi}\) and
\(P^\Pi_\vartheta\) in place of \(P^{\mu,\psi}\) and \(P^\psi_\vartheta\).
The same holds for an arbitrary set \(X\subseteq\R\) of actions, as in
Lemma~\ref{S1-lem:posterior-general}.
\end{lemma}

\begin{proof}
We prove both lemmas.  By induction, \(s_t\) is a Borel function of
\((\zeta_1,y_1,\ldots,\zeta_t,y_t)\) on the augmented space and on the image
of \(\Phi\), hence \(P^{\mu,\Pi}\)-almost surely on \(\Omega\), and
\[
 \mathcal H_t\subseteq\sigma(\vartheta,\varepsilon_{\le t},\zeta_{\le t}),
 \qquad
 \mathcal F_t\subseteq\sigma(\vartheta,\varepsilon_{\le t},\zeta_{\le t+1}).
\]
By mutual independence, \(\varepsilon_{t+1}\) is independent of
\(\sigma(\vartheta)\vee\mathcal F_t\) and \(\zeta_{t+1}\) of
\(\sigma(\vartheta)\vee\mathcal H_t\).  The second fact gives
\(E[f(\vartheta)\mid\mathcal F_t]=E[f(\vartheta)\mid\mathcal H_t]\) for every
bounded Borel \(f\) (check it on the \(\pi\)-system of sets
\(H\cap\{\zeta_{t+1}\in G\}\), \(H\in\mathcal H_t\)), so the fresh seed adds no
parameter information.  With these two facts, the proofs of
Lemmas~\ref{lem:posterior-interface} and~\ref{lem:disintegration} apply
verbatim, with \(\mathcal F_t\) the decision \(\sigma\)-field, the update
from \(\mathcal F_t\) to \(\mathcal H_{t+1}=\mathcal F_t\vee\sigma(y_{t+1})\)
in place of that from \(\mathcal F_t\) to \(\mathcal F_{t+1}\), and
\(\mathcal Q\) the joint law of noise and seeds.
\end{proof}

\emph{Decision rules.}  A decision rule \(\psi\) is the Dirac kernel
\(\Pi_\psi(s,\cdot)=\delta_{\psi(s)}\), realized by \(\psi(s,\zeta)=\psi(s)\).
The seeds then enter no action, outcome or posterior state, and
\(P^{\mu,\Pi_\psi}\) and \(P^{\Pi_\psi}_\vartheta\) are the laws
\(P^{\mu,\psi}\) and \(P^\psi_\vartheta\) of Appendix~\ref{app:posterior} on
every event determined by \(\vartheta\), the actions and the outcomes.
Every result of Sections~\ref{supp:trapping} and~\ref{supp:exclusion}
therefore applies to a decision rule.

\begin{remark}[History-dependent policies]\label{A-rem:predictable}
Lemmas~\ref{S1-lem:posterior-general}
and~\ref{S1-lem:disintegration-general} hold, with the same proofs, for
every \emph{Borel predictable policy}: a sequence of Borel maps
\(\psi_t\), \(t\ge0\), with
\[
 x_{t+1}=\psi_t(s_0,\zeta_1,y_1,\ldots,\zeta_t,y_t,\zeta_{t+1}),
\]
taking values in a set \(X'\subseteq\R\), which replaces \(X\)
in the path space \(\Omega\), and with \(\Phi\) generated by
\(x_t=\psi_{t-1}(s_0,v_1,y_1,\ldots,v_{t-1},y_{t-1},v_t)\).  Such a policy
may depend on the whole history, and it is given directly by the maps
\(\psi_t\), so no realization step is needed.  The proofs use only that
\(x_{t+1}\) is \(\mathcal F_t\)-measurable, the two independence facts
above, martingale convergence, L\'evy's upward theorem, the strong law of
large numbers, and that \(\Phi\) is Borel.
\end{remark}

\section{The trapping lemma}\label{supp:trapping}

This section states the trapping lemma for policy kernels, as in
Section~\ref{supp:general-policies}, that track a smooth map.
Appendix~\ref{app:local} applies it to a decision rule that coincides with a
smooth map on the relevant collar, so that the tracking error is zero.

Throughout, the truth \(\vartheta=(\alpha,\beta)\) is fixed.  We write a
state in the coordinates \((a,b,u,\bar x,V)\), where \(V=1/w\) is the slope
precision; \((a,b,u,\bar x,w)\mapsto(a,b,u,\bar x,1/w)\) is a bijection of
the feasible states onto \(\{u>0,\ V>0\}\).  A map \(\varphi\) of these
coordinates is evaluated as \(\varphi(s_t)\coloneqq\varphi(a_t,b_t,u_t,\bar x_t,V_t)\),
and \(s_t\in\mathcal C\) means \((a_t,b_t,u_t,\bar x_t,V_t)\in\mathcal C\);
\(A^{-1}\) and \(u\) are used interchangeably.  The filtrations
\((\mathcal H_t)\), \((\mathcal F_t)\), the residual
\(\xi_t=\alpha+\beta\bar x_t-a_t-b_t\bar x_t\) and \(\kappa_t=A_t/(A_t+1)\) are
as in Section~\ref{supp:general-policies}, and
\(\sigma_t^2=1+A_t^{-1}+V_t^{-1}D_t^2\) is the prediction-error variance of
Lemma~\ref{S1-lem:posterior-general}\ref{lem:posterior-interface:b}; at a state \(s\) with action \(x\)
and \(D=x-\bar x\), \(\sigma^2\coloneqq1+A^{-1}+V^{-1}D^2\).  Write
\(\boldsymbol m=(a,b)\), \(\boldsymbol e_a\coloneqq(1,0)\),
\(\boldsymbol n(\bar x)\coloneqq(-\bar x,1)\), and let \(\nabla_{\boldsymbol m}\) be
the derivative in \(\boldsymbol m\) with \((u,\bar x,V)\) held fixed.

Under a fixed-truth law \(P^\Pi_\vartheta\) of a policy kernel \(\Pi\), the
actions are \(x_{t+1}=\psi(s_t,\zeta_{t+1})\), and the structural shocks
\(\varepsilon_t=y_t-\alpha-\beta x_t\) are i.i.d.\ \(N(0,1)\) and independent
of the seeds, so \(\varepsilon_{t+1}\) is independent of \(\mathcal F_t\).
The recursions \eqref{eq:design} and \eqref{eq:means} are algebraic
identities in the realized actions and outcomes
(Lemma~\ref{S1-lem:posterior-general}\ref{lem:posterior-interface:a}), so they hold pathwise under every
fixed-truth law.  By Lemma~\ref{S1-lem:posterior-general}\ref{lem:posterior-interface:b},
\((A_t+1)V_{t+1}=A_tV_t\sigma_t^2\), and \eqref{eq:means} gives
\(b_{t+1}-b_t=V_t^{-1}D_tI_{t+1}/\sigma_t^2\).  With
\(\bar x_{t+1}=\bar x_t+D_t/(A_t+1)\), a direct computation from
\eqref{eq:means} gives
\(a_{t+1}-a_t=(A_t^{-1}-\bar x_tV_t^{-1}D_t)I_{t+1}/\sigma_t^2\), so
\begin{equation}
 \boldsymbol m_{t+1}-\boldsymbol m_t
 =\frac{A_t^{-1}\boldsymbol e_a+V_t^{-1}D_t\boldsymbol n(\bar x_t)}
   {\sigma_t^2}\,I_{t+1}.
 \label{B-eq:mean-increment}
\end{equation}

\subsection{Tracking a smooth map}

Let \(\Pi\) be a policy kernel, realized by fresh independent seeds as in
Section~\ref{supp:general-policies}; a decision rule is the Dirac case.  Let
\(\varphi\) be a Borel map of the coordinates \((a,b,u,\bar x,V)\), put
\(D^\varphi(s)=\varphi(s)-\bar x\), and let \(\mathcal C\) be a Borel set of
states and \(C_*\) a finite constant.

\begin{definition}[Tracking]\label{B-def:tracking}
Given a nondegenerate Gaussian initial prior \(\mu\), a policy kernel
\(\Pi\) \emph{tracks \(\varphi\) on \(\mathcal C\) with constant \(C_*\)}
for \(\mu\) if, \(P^{\mu,\Pi}\)-almost surely, at every date \(t\) with
\(s_t\in\mathcal C\),
\begin{equation}
 \lvert x_{t+1}-\varphi(s_t)\rvert
 \le C_*\bigl(A_t^{-1}+\lvert D^\varphi(s_t)\rvert\bigr)^2.
 \label{B-eq:trap-tracking}
\end{equation}
\end{definition}

Tracking is a property of a kernel together with one initial prior.  For a
fixed initial prior and a kernel that tracks \(\varphi\), all probabilities and
expectations below use \(P^\Pi_\vartheta\), under which
\eqref{B-eq:trap-tracking} still holds at every such date almost surely. It
is a countable collection of finite-date observable events that holds
prior-predictively almost surely, and Lemma~\ref{B-lem:finite-history}
transfers it to every fixed truth.  If the bound in
\eqref{B-eq:trap-tracking} holds for every action in the support of
\(\Pi(s,\cdot)\) at every feasible \(s\in\mathcal C\), in particular for the
Dirac kernel \(\Pi_\psi\) of a decision rule \(\psi\) with
\(\lvert\psi(s)-\varphi(s)\rvert\le C_*(A^{-1}+\lvert D^\varphi(s)\rvert)^2\)
there, then \(\Pi\) tracks \(\varphi\) on \(\mathcal C\) with constant
\(C_*\) for every initial prior.

\begin{auxlemma}[Equivalence of finite observable-history laws]
\label{B-lem:finite-history}
Keep one nondegenerate Gaussian initial prior \(\mu\) and one policy kernel
\(\Pi\) fixed, realized by the fresh uniform seeds of
Section~\ref{supp:general-policies}, which are independent of the parameter
and the structural-noise sequence.  For every truth
\(\vartheta=(\alpha,\beta)\) and every finite date \(n\), the law of the
observable outcomes and seeds through that date, including the next
decision seed, \((y_1,\dots,y_n,\zeta_1,\ldots,\zeta_{n+1})\), under
\(P^\Pi_\vartheta\) is mutually absolutely continuous with its law under
\(P^{\mu,\Pi}\).  Consequently, every countable collection of
truth-independent observable finite-date support properties that holds
\(P^{\mu,\Pi}\)-almost surely holds \(P^\Pi_\vartheta\)-almost surely for
every \(\vartheta\).
\end{auxlemma}

\begin{proof}
Under every truth, the actions are the same Borel functions
\(x_j=\psi(s_{j-1},\zeta_j)=x_j(y_{<j},\zeta_{\le j})\) of earlier outcomes,
seeds and the initial prior.  Relative to Lebesgue measure on \(\R^n\) times
the uniform measure on the seeds, the law under \(\vartheta\) has density
\[
 q_\vartheta(y_1,\ldots,y_n,\zeta_1,\ldots,\zeta_{n+1})
 =\prod_{j=1}^n
   \frac{1}{\sqrt{2\pi}}
   \exp\left(-\frac12
      \bigl(y_j-\alpha-\beta x_j(y_{<j},\zeta_{\le j})\bigr)^2\right),
\]
which lies in \((0,(2\pi)^{-n/2}]\).  So the mixture density
\(q_\mu=\int q_\vartheta\,\mu(d\vartheta)\) is also strictly positive and
finite, and both laws are equivalent to the reference
measure.  A countable union over dates and properties proves the last
assertion.
\end{proof}

\subsection{The trapping lemma}

The \(C^{1,1}\) norm of a map on a set is the maximum of the sup norms of the
map and of its first derivatives on that set and the Lipschitz constant of
its first derivatives there.

\begin{lemma}[Trapping and open-prior entry]
\label{lem:trapping}
Fix the truth \(\vartheta=(\alpha,\beta)\) and a target
\((a^e,b^e,x^e)\) with \(a^e+b^ex^e=\alpha+\beta x^e\).
Suppose there are \(r_*>0\), \(u_*\in(0,1)\), \(V_*<\infty\), a finite
\(C_*\), a positive \(\lambda^0_*\) and a map \(\varphi\) of the
coordinates \((a,b,u,\bar x,V)\)
such that, for every \(V^e\ge V_*\), on the two-sided collar
\[
 \mathcal C(V^e)=\bigl\{(a,b,u,\bar x,V):
 \lvert a-a^e\rvert+\lvert b-b^e\rvert+\lvert\bar x-x^e\rvert<r_*,\
 \lvert u\rvert<u_*,\ V^e/2\le V\le2V^e\bigr\},
\]
with \(D^\varphi=\varphi-\bar x\) and \(\lambda^0=\varphi_b-\bar x\varphi_a\),
\begin{enumerate}
\item[\condtarget{lem:trapping:a}{(a)}] \(\varphi\) is \(C^{1,1}\) with \(C^{1,1}\) norm at most \(C_*\);
\item[\condtarget{lem:trapping:b}{(b)}] \(\varphi(a^e,b^e,0,x^e,V^e)=x^e\);
\item[\condtarget{lem:trapping:c}{(c)}] \(\lvert\lambda^0\rvert\ge\lambda^0_*\).
\end{enumerate}
Then for every \(r\in(0,r_*)\), with \(r<\lvert b^e-\beta\rvert/2\) when
\(b^e\ne\beta\), there is \(\bar V_r\in[V_*,\infty)\) with the following
property.  For every \(V^e\ge\bar V_r\) and every \(\varepsilon\in(0,1)\)
there are \(\overline A<\infty\) and \(r_0>0\) such that, for every
nondegenerate Gaussian initial prior with
\[
 A_0>\overline A,\qquad
 d_0\coloneqq\lvert a_0-a^e\rvert+\lvert b_0-b^e\rvert+\lvert\bar x_0-x^e\rvert
 +\lvert V_0-V^e\rvert<r_0,
\]
and every policy kernel \(\Pi\) that tracks \(\varphi\) on
\(\mathcal C_*=\bigcup_{V^e\ge V_*}\mathcal C(V^e)\) with constant \(C_*\)
for that initial prior, the following hold simultaneously with
\(P^\Pi_\vartheta\)-probability at least \(1-\varepsilon\):
\begin{enumerate}
\item[\condtarget{lem:trapping:i}{(i)}] \(\lvert a_t-a^e\rvert+\lvert b_t-b^e\rvert+\lvert\bar x_t-x^e\rvert<r\)
and \(V^e/2\le V_t\le2V^e\) for every \(t\ge0\);
\item[\condtarget{lem:trapping:ii}{(ii)}] \(V_\infty<\infty\), \(b_t\to b_\infty\),
\(\bar x_t\to\bar x_\infty\), \(x_{t+1}\to\bar x_\infty\), \(\xi_t\to0\)
and \(a_t\to\alpha+\beta\bar x_\infty-b_\infty\bar x_\infty\), with
\(\lvert b_\infty-\beta\rvert\ge\lvert b^e-\beta\rvert-r>0\) when
\(b^e\ne\beta\).
\end{enumerate}
These priors form a nonempty open set of feasible Gaussian priors.  The
thresholds \(\bar V_r\), \(\overline A\) and \(r_0\) do not depend on the
initial prior or the policy kernel.
\end{lemma}

In particular, hypothesis \ref{lem:trapping:a} holds on a collar \(\mathcal C(V^e)\) with
\(V^e>0\) if \(\varphi\) is \(C^2\) there with derivatives of order at most
two bounded by \(C_*\), in the operator norms for one fixed norm on \(\R^5\)
that define the \(C^{1,1}\) norm.  This is the case if the \(C^{2,1}\) norm
\(\lVert\varphi\rVert_{C^{2,1}}\) on the collar, the maximum of the sup norms
of the map and of its derivatives of order at most two there and the
Lipschitz constant of its second derivatives there, is at most \(C_*\).
Indeed, the collar is then convex with nonempty interior, and the mean value
inequality makes the first derivatives \(C_*\)-Lipschitz.  Replacing \(V_*\)
by \(\max\{V_*,1\}\) does not weaken the conclusion and makes every collar in
the hypotheses nondegenerate.

\begin{remark}[Trapping for history-dependent policies]\label{B-rem:seeded}
Definition~\ref{B-def:tracking} and Lemmas~\ref{B-lem:finite-history}
and~\ref{lem:trapping} hold verbatim, with the same proofs, when \(\Pi\) is
replaced by a Borel predictable policy of Remark~\ref{A-rem:predictable}
with values in \(X\) and i.i.d.\ uniform seeds, and \(P^{\mu,\Pi}\) and
\(P^\Pi_\vartheta\) by its prior-predictive and fixed-truth laws.  In
particular, the thresholds \(\bar V_r\), \(\overline A\) and \(r_0\) of
Lemma~\ref{lem:trapping} do not depend on the initial prior or the policy.
The proofs use the policy only through three facts, which hold for such a
policy: under every truth, \(x_{t+1}\) is one Borel function of \(s_0\),
\(\zeta_1,\ldots,\zeta_{t+1}\) and \(y_1,\ldots,y_t\), so it is
\(\mathcal F_t\)-measurable; the recursions of this section are identities
in the realized actions and outcomes, by
Lemma~\ref{S1-lem:posterior-general}\ref{lem:posterior-interface:a}--\ref{lem:posterior-interface:b} and
Remark~\ref{A-rem:predictable}; and, under every fixed truth,
\(\varepsilon_{t+1}\) is \(N(0,1)\) and independent of \(\mathcal F_t\),
because \(\vartheta\), the noise and the seeds are mutually independent
and, by the first fact, \(\mathcal F_t\) is generated by
\(\varepsilon_1,\ldots,\varepsilon_t\) and \(\zeta_1,\ldots,\zeta_{t+1}\).
The first and third facts give the density in the proof of
Lemma~\ref{B-lem:finite-history}, which then transfers tracking to every
fixed truth as above.  All three give the conditional estimates in the proof
of Lemma~\ref{lem:trapping}, whose constants depend on neither the initial
prior nor the policy.
\end{remark}

The rest of this section proves Lemma~\ref{lem:trapping}.  Fix \(V^e\ge V_*\).  Constants
\(C\) may depend on \(V^e\), the target, the truth and
\(r_*,u_*,V_*,C_*,\lambda^0_*\), but not on the initial prior or the kernel;
the estimates are made for one fixed initial prior in the displayed set and
one kernel that tracks \(\varphi\) for it.  The collar is
\(\mathcal C(V^e)\subseteq\mathcal C_*\).  Define the \textit{shifted gap}
\begin{equation}
 c_D^{\mathrm{loc}}(s)=\frac{\varphi_a}{V^{-1}\lambda^0},\qquad
 D^{\varphi,\circ}(s)=D^\varphi(s)+A^{-1}c_D^{\mathrm{loc}}(s).
 \label{B-eq:Dcirc}
\end{equation}
By \ref{lem:trapping:a} and \ref{lem:trapping:c}, \(c_D^{\mathrm{loc}}\) is bounded and Lipschitz on the
collar, and hence so is \(D^{\varphi,\circ}\).  The collar is convex, so the
mean value inequality, applied to \(z'\mapsto\varphi(z')-\mathrm D\varphi(z)(z'-z)\),
gives the first-order Taylor bound
\begin{equation}
 \lvert\varphi(z')-\varphi(z)-\mathrm D\varphi(z)(z'-z)\rvert
 \le C_*\lvert z'-z\rvert^2
 \qquad(z,z'\in\mathcal C(V^e)).
 \label{B-eq:taylor}
\end{equation}
The estimates below use \ref{lem:trapping:a} only through these facts, at pairs of states
that both lie in the collar, and the bounds on \(\varphi\) and
\(\mathrm D\varphi\).  Off the collar, put \(D^{\varphi,\circ}\coloneqq0\).

\subsubsection{The fixed-truth intercept residual}

Under the fixed truth, \(I_{t+1}=y_{t+1}-a_t-b_tx_{t+1}
=\xi_t+(\beta-b_t)D_t+\varepsilon_{t+1}\).  The pathwise recursion
\(A_{t+1}\xi_{t+1}=A_t\xi_t-\varepsilon_{t+1}\) of
Lemma~\ref{S1-lem:posterior-general}\ref{lem:posterior-interface:a}, with \(A_t=A_0+t\), gives
\(\xi_t=\bigl(A_0\xi_0-\sum_{j=1}^t\varepsilon_j\bigr)/(A_0+t)\).  Hence
\begin{align}
 E\sum_{t\ge0}A_t^{-1}\lvert\xi_t\rvert
 &\le C\left(\lvert\xi_0\rvert+\sqrt{A_0^{-1}}\right),
 \label{B-eq:xi-sum}\\
 E\left[\sup_{t\ge0}\lvert\xi_t\rvert^2\right]
 &\le C(\xi_0^2+A_0^{-1}).
 \label{B-eq:xi-sup}
\end{align}
For \eqref{B-eq:xi-sum}, sum
\(A_t^{-1}E\lvert\xi_t\rvert\le\lvert\xi_0\rvert A_t^{-2}/A_0^{-1}+A_t^{-3/2}\).
For \eqref{B-eq:xi-sup}, split time into the blocks \(t\le A_0\) and
\(2^jA_0<t\le2^{j+1}A_0\) and apply the \(L^2\) maximal inequality to the
Gaussian partial sums on each block; the block bounds sum geometrically.

\subsubsection{Innovation cancellation in the shifted gap}

By \eqref{B-eq:trap-tracking}, which holds almost surely under the
fixed-truth law, the realized action at a feasible state \(s\) of the collar
is \(x=\varphi(s)+e\) with \(\lvert e\rvert\le C(A^{-1}+\lvert D^\varphi\rvert)^2\).
On the collar, by \eqref{B-eq:Dcirc},
\begin{equation}
 \lvert D\rvert\le C(A^{-1}+\lvert D^{\varphi,\circ}\rvert),\qquad
 \lvert e\rvert\le C(A^{-1}+\lvert D^{\varphi,\circ}\rvert)^2.
 \label{B-eq:gap-bounds}
\end{equation}
By \eqref{B-eq:mean-increment}, the mean moves along
\(A^{-1}\boldsymbol e_a+V^{-1}D\boldsymbol n(\bar x)\).  Since
\(\nabla_{\boldsymbol m}D^\varphi\,\boldsymbol e_a=\varphi_a\),
\(\nabla_{\boldsymbol m}D^\varphi\,\boldsymbol n(\bar x)=\lambda^0\),
\(D=D^\varphi+e\) and
\(V^{-1}\lambda^0D^{\varphi,\circ}=V^{-1}\lambda^0D^\varphi+A^{-1}\varphi_a\),
the shift cancels the \(A^{-1}\) term:
\[
 \nabla_{\boldsymbol m}D^\varphi\,
 [A^{-1}\boldsymbol e_a+V^{-1}D\boldsymbol n(\bar x)]
 =V^{-1}\lambda^0D^{\varphi,\circ}+V^{-1}\lambda^0e
 =V^{-1}\lambda^0D^{\varphi,\circ}+O\bigl(A^{-2}+(D^{\varphi,\circ})^2\bigr),
\]
where the derivatives hold the original coordinates \(u,\bar x,V\) fixed
before evaluation at the actual state.  Only derivatives of \(\varphi\) enter
here; \(c_D^{\mathrm{loc}}\) is not differentiated.  The deterministic
increments of \(A^{-1}\), \(\bar x\) and \(V\) are
\(O(A^{-2}+(D^{\varphi,\circ})^2)\).  When both \(s_t\) and \(s_{t+1}\) lie in
\(\mathcal C(V^e)\), the Taylor bound \eqref{B-eq:taylor} controls the
second-order part of the increment of \(\varphi\), including the mean jump
for large \(\varepsilon\), and the Lipschitz bound on \(c_D^{\mathrm{loc}}\)
controls the increment of \(A^{-1}c_D^{\mathrm{loc}}\), which is
\((A_{t+1}^{-1}-A_t^{-1})c_D^{\mathrm{loc}}(s_{t+1})
+A_t^{-1}\bigl(c_D^{\mathrm{loc}}(s_{t+1})-c_D^{\mathrm{loc}}(s_t)\bigr)\).
Therefore, whenever \(s_t\) and \(s_{t+1}\) lie in \(\mathcal C(V^e)\),
\begin{equation}
 D_{t+1}^{\varphi,\circ}
 =(1+\gamma_t^{\mathrm{loc}}\varepsilon_{t+1})D_t^{\varphi,\circ}
 +R_{t+1}^{\mathrm{loc}},
 \qquad
 \gamma_t^{\mathrm{loc}}=\frac{V_t^{-1}\lambda^0(s_t)}{\sigma_t^2},
 \label{B-eq:gap-recursion}
\end{equation}
\begin{equation}
 \lvert R_{t+1}^{\mathrm{loc}}\rvert
 \le C\lvert\xi_t\rvert\lvert D_t^{\varphi,\circ}\rvert
 +C\bigl(A_t^{-2}+(D_t^{\varphi,\circ})^2\bigr)(1+\varepsilon_{t+1}^2).
 \label{B-eq:remainder}
\end{equation}
So the gap is a multiplicative process with additive forcing of order
\(A_t^{-2}\).  If \(s_t\in\mathcal C(V^e)\) and
\(s_{t+1}\notin\mathcal C(V^e)\), then \(D_{t+1}^{\varphi,\circ}=0\).
Hence, whenever \(s_t\in\mathcal C(V^e)\),
\begin{equation}
 \lvert D_{t+1}^{\varphi,\circ}\rvert
 \le\lvert1+\gamma_t^{\mathrm{loc}}\varepsilon_{t+1}\rvert
 \lvert D_t^{\varphi,\circ}\rvert
 +C\lvert\xi_t\rvert\lvert D_t^{\varphi,\circ}\rvert
 +C\bigl(A_t^{-2}+(D_t^{\varphi,\circ})^2\bigr)(1+\varepsilon_{t+1}^2).
 \label{B-eq:gap-step}
\end{equation}

\subsubsection{Fractional-moment contraction}

\begin{auxlemma}[Robust Gaussian fractional moments]
\label{B-lem:gaussian-moment}
Fix \(\nu\in(0,1)\), \(0<\gamma_{\min}\le\gamma_{\max}<\infty\), a compact
positive range for \(\sigma_G\), and an integer \(m_{\mathrm{env}}\ge0\).
There are \(\epsilon_{G,0},c_G,C_G>0\) such that
\[
 E\bigl[\lvert 1+\gamma_G\epsilon_GI+R_{\mathrm{pert}}(I)\rvert^{\nu}\bigr]
 \le 1-c_G\epsilon_G^2+C_GB_{\mathrm{pert}}
\]
whenever \(I=\sigma_GZ\), \(Z\sim N(0,1)\),
\(\gamma_{\min}\le\lvert\gamma_G\rvert\le\gamma_{\max}\),
\(0<\epsilon_G\le\epsilon_{G,0}\), and
\(\lvert R_{\mathrm{pert}}(I)\rvert\le B_{\mathrm{pert}}
(1+\lvert I\rvert^{m_{\mathrm{env}}})\).
Here \(B_{\mathrm{pert}}\ge0\), and \(R_{\mathrm{pert}}\) may be an
arbitrary measurable function.  The constants depend only on \(\nu\),
\(\gamma_{\min}\), \(\gamma_{\max}\), the range of \(\sigma_G\) and
\(m_{\mathrm{env}}\); hence the same conclusion holds conditionally when
\(\gamma_G\), \(\epsilon_G\), \(\sigma_G\), \(B_{\mathrm{pert}}\) and
\(R_{\mathrm{pert}}\) are measurable with respect to the conditioning
\(\sigma\)-field, \(Z\) is standard normal given that \(\sigma\)-field, and
the displayed ranges hold.  In that conditional statement the left side is
the generalized conditional expectation of the nonnegative variable
\(\lvert 1+\gamma_G\epsilon_GI+R_{\mathrm{pert}}(I)\rvert^{\nu}\), with
values in \([0,\infty]\); it requires no integrability, and in particular
\(B_{\mathrm{pert}}\) need not be integrable.
\end{auxlemma}

\begin{proof}
Since \((x+h)^{\nu}\le x^{\nu}+\nu x^{\nu-1}h\) for \(x>0\), \(h\ge0\),
outside the null event \(1+\gamma_G\epsilon_GI=0\),
\begin{equation}
 \lvert 1+\gamma_G\epsilon_GI+R_{\mathrm{pert}}(I)\rvert^{\nu}
 \le\lvert 1+\gamma_G\epsilon_GI\rvert^{\nu}
 +\nu B_{\mathrm{pert}}(1+\lvert I\rvert^{m_{\mathrm{env}}})
  \lvert 1+\gamma_G\epsilon_GI\rvert^{\nu-1}.
 \label{B-eq:gm-split}
\end{equation}
Put \(t=\gamma_G\epsilon_G\).  The second factor in \eqref{B-eq:gm-split} has
uniformly bounded expectation. This is trivial on \(\lvert 1+tI\rvert\ge1/2\), and
on the complement \(\lvert I\rvert\in[\frac1{2\lvert t\rvert},\frac3{2\lvert t\rvert}]\),
the Gaussian density is at most \(Ce^{-c/t^2}\) uniformly over \(\sigma_G\),
and the substitution \(z=1+tI\) bounds the contribution by
\(Ce^{-c/t^2}(\lvert t\rvert^{-1}+\lvert t\rvert^{-m_{\mathrm{env}}-1})
\int_{-1/2}^{1/2}\lvert z\rvert^{\nu-1}dz\), bounded as \(t\to0\).  For the
first term, Taylor's theorem on \(\lvert tI\rvert\le1/2\), symmetry for the
truncated linear term and Gaussian tails on the complement give, uniformly in
the prescribed ranges,
\[
 E\bigl[\lvert 1+\gamma_G\epsilon_GI\rvert^{\nu}\bigr]
 =1-\frac{\nu(1-\nu)}2\gamma_G^2\sigma_G^2\epsilon_G^2+O(\epsilon_G^3)
 +O(e^{-c/\epsilon_G^2})
 \le1-c_G\epsilon_G^2
\]
for \(\epsilon_{G,0}\) small enough.  This gives the conditional bound when
the left variable is integrable.  Otherwise apply that case to
\(R_{\mathrm{pert}}\mathbf 1_{G_n}\) and \(B_{\mathrm{pert}}\mathbf 1_{G_n}\),
\(G_n=\{B_{\mathrm{pert}}\le n\}\), whose left variable is dominated by
\(1+\gamma_{\max}\epsilon_{G,0}\lvert I\rvert
+n(1+\lvert I\rvert^{m_{\mathrm{env}}})\) and equals the original one on
\(G_n\); the conditional expectation of a nonnegative variable is local on
events of the conditioning \(\sigma\)-field, and \(G_n\) increases to the
whole space.
\end{proof}

The cost \(C_GB_{\mathrm{pert}}\), first-power rather than
\(C_GB_{\mathrm{pert}}^{\nu}\), is essential.  The case
\(\gamma_{\min}=\gamma_{\max}=\sigma_G=1\), \(m_{\mathrm{env}}=0\),
\(R_{\mathrm{pert}}\equiv0\), \(B_{\mathrm{pert}}=0\), \(\gamma_G=\sgn c\),
\(\epsilon_G=\lvert c\rvert\) of Lemma~\ref{B-lem:gaussian-moment} gives, for
every \(\nu\in(0,1)\), constants \(c_*(\nu)\coloneqq\epsilon_{G,0}\) and
\(c_{\mathrm{loc}}(\nu)\coloneqq c_G\) such that, with \(Z\sim N(0,1)\),
\begin{equation}
 E[\lvert 1+cZ\rvert^{\nu}]\le1-c_{\mathrm{loc}}(\nu)c^2
 \qquad(0<\lvert c\rvert\le c_*(\nu)).
 \label{B-eq:gm}
\end{equation}

Fix \(\nu_{\mathrm{loc}}\in(1/2,1)\), say \(\nu_{\mathrm{loc}}=3/4\), and
write \(c_*=c_*(\nu_{\mathrm{loc}})\),
\(c_{\mathrm{loc}}=c_{\mathrm{loc}}(\nu_{\mathrm{loc}})\).  For \(V^e\)
sufficiently large, \(0<c_{\min}\le\lvert\gamma_t^{\mathrm{loc}}\rvert\le c_*\)
throughout the collar. The lower bound uses \ref{lem:trapping:c} and \(V\le2V^e\), and the upper
bound uses \ref{lem:trapping:a} and \(V\ge V^e/2\).  Thus \eqref{B-eq:gm} gives a fixed
contraction deficit \(\kappa_{\mathrm{loc}}>0\).  Apply
\(\lvert x+y\rvert^{\nu_{\mathrm{loc}}}\le\lvert x\rvert^{\nu_{\mathrm{loc}}}
+\lvert y\rvert^{\nu_{\mathrm{loc}}}\) to
\eqref{B-eq:gap-step}, and choose fixed bounds
\(\lvert\xi\rvert\le C_\xi^{\mathrm{trap}}\) and
\(\lvert D^{\varphi,\circ}\rvert\le C_D^{\mathrm{trap}}\) so small that
\(C(C_\xi^{\mathrm{trap}})^{\nu_{\mathrm{loc}}}
\lvert D^{\varphi,\circ}\rvert^{\nu_{\mathrm{loc}}}\) and
\(C\lvert D^{\varphi,\circ}\rvert^{2\nu_{\mathrm{loc}}}\) absorb at most half
the contraction.  Then, before exit,
\begin{equation}
 E\bigl[\lvert D_{t+1}^{\varphi,\circ}\rvert^{\nu_{\mathrm{loc}}}
 \bigm|\mathcal F_t\bigr]
 \le(1-\kappa_{\mathrm{loc}}/2)\lvert D_t^{\varphi,\circ}\rvert^{\nu_{\mathrm{loc}}}
 +CA_t^{-2\nu_{\mathrm{loc}}}.
 \label{B-eq:contraction}
\end{equation}
Here the structural innovation \(\varepsilon_{t+1}\), not the standardized
prediction error \(I_{t+1}/\sigma_t\), is conditionally standard normal.

\subsubsection{A quantitative nonexit bound}

Fix the radius \(r\in(0,r_*)\) of the lemma.  Choose \(\bar V_r\ge V_*\) so
large that the contraction estimates above apply on
\(V\in[V^e/2,2V^e]\) whenever \(V^e\ge\bar V_r\), and fix such a finite
\(V^e\).  Choose small fixed bounds on \(b-b^e\) and \(\bar x-x^e\), and keep
sufficiently small bounds on \(\xi\) and \(D^{\varphi,\circ}\).  Since the
target satisfies \(a^e+b^ex^e=\alpha+\beta x^e\),
\[
 a-a^e=(\beta-b^e)(\bar x-x^e)-(b-b^e)\bar x-\xi,
\]
so these bounds can be chosen so that their closed collar lies in
\(\mathcal C(V^e)\) and satisfies
\(\lvert a-a^e\rvert+\lvert b-b^e\rvert+\lvert\bar x-x^e\rvert<r\).  Take
\(A_0\) large enough to stay in the feasible \(A^{-1}\)-range, which
\(A_t^{-1}\) never leaves since it decreases.  Let \(\tau_{\mathrm{trap}}\)
be the first date at which a bound on \(b-b^e\), \(\bar x-x^e\), \(\xi\) or
\(D^{\varphi,\circ}\) is reached, or \(V\ge2V^e\).  Before
\(\tau_{\mathrm{trap}}\) the state lies in \(\mathcal C(V^e)\), so a date at
which it leaves the collar is a first exit through \(b\), \(\bar x\), \(\xi\)
or \(V\).  Take initial values with
fixed positive margins from these boundaries and \(3V^e/4<V_0<5V^e/4\);
since \(V_t\) is nondecreasing, it never crosses \(V^e/2\).

Iterating \eqref{B-eq:contraction} up to \(N\wedge\tau_{\mathrm{trap}}\)
gives, with \(2\nu_{\mathrm{loc}}>1\),
\begin{gather}
 E\bigl[\lvert D_{N\wedge\tau_{\mathrm{trap}}}^{\varphi,\circ}
 \rvert^{\nu_{\mathrm{loc}}}\bigr]
 +\frac{\kappa_{\mathrm{loc}}}2\,E\sum_{t<N\wedge\tau_{\mathrm{trap}}}
 \lvert D_t^{\varphi,\circ}\rvert^{\nu_{\mathrm{loc}}}
 \le C_{\mathrm{init}}^{\mathrm{loc}}
 \coloneqq\lvert D_0^{\varphi,\circ}\rvert^{\nu_{\mathrm{loc}}}
 +C\sum_{t\ge0}A_t^{-2\nu_{\mathrm{loc}}},
 \label{B-eq:sum-bound}\\
 \sum_{t\ge0}A_t^{-2\nu_{\mathrm{loc}}}
 \le A_0^{-2\nu_{\mathrm{loc}}}
 +\frac{A_0^{-(2\nu_{\mathrm{loc}}-1)}}{2\nu_{\mathrm{loc}}-1}\longrightarrow0.
 \label{B-eq:forcing}
\end{gather}
By \eqref{B-eq:gap-bounds}, \((D^{\varphi,\circ})^2\le
(C_D^{\mathrm{trap}})^{2-\nu_{\mathrm{loc}}}\lvert D^{\varphi,\circ}\rvert^{\nu_{\mathrm{loc}}}\)
and \(\bar x_{t+1}-\bar x_t=D_t/(A_t+1)\),
\begin{gather}
 E\sum_{t<\tau_{\mathrm{trap}}}D_t^2
 \le C\left((C_D^{\mathrm{trap}})^{2-\nu_{\mathrm{loc}}}C_{\mathrm{init}}^{\mathrm{loc}}
 +\sum_tA_t^{-2}\right),
 \label{B-eq:D2-sum}\\
 E\sum_{t<\tau_{\mathrm{trap}}}\lvert\bar x_{t+1}-\bar x_t\rvert
 \le C\left(A_0^{-1}(C_D^{\mathrm{trap}})^{1-\nu_{\mathrm{loc}}}
 C_{\mathrm{init}}^{\mathrm{loc}}+\sum_tA_t^{-2}\right).
 \label{B-eq:xbar-sum}
\end{gather}
The right side of \eqref{B-eq:D2-sum} tends to zero as
\(A_0^{-1},D_0^{\varphi,\circ}\to0\); since
\(V_{t+1}-V_t=\kappa_tD_t^2\le D_t^2\), it controls excessive growth of
\(V\).  Equation~\eqref{B-eq:xbar-sum} controls \(\bar x\)-exit.

By \eqref{B-eq:mean-increment},
\(b_{t+1}-b_t=\sigma_t^{-2}V_t^{-1}D_t[\xi_t+(\beta-b_t)D_t+\varepsilon_{t+1}]\).
Its stopped martingale part has expected quadratic variation bounded by the
right side of \eqref{B-eq:D2-sum}.  By \eqref{B-eq:xi-sum} and
\eqref{B-eq:gap-bounds}, its total drift has expectation at most
\(C\bigl((C_D^{\mathrm{trap}})^{1-\nu_{\mathrm{loc}}}C_{\mathrm{init}}^{\mathrm{loc}}
+\lvert\xi_0\rvert+\sqrt{A_0^{-1}}+E\sum_{t<\tau_{\mathrm{trap}}}D_t^2\bigr)\).
The maximal inequality for the martingale and Markov's inequality for the
drift control slope exit.  Equation~\eqref{B-eq:xi-sup} controls
\(\xi\)-exit, and \eqref{B-eq:sum-bound} controls \(D^{\varphi,\circ}\)-exit,
including an overshoot at the exit date.  These, with the precision,
average-action and slope exits, exhaust all first exits.  A union bound with
\(\nu_{\mathrm{loc}}=3/4\) gives, for \(A_0\ge1\) and
\(\lvert\xi_0\rvert\le1\),
\begin{equation}
 P^\Pi_\vartheta(\tau_{\mathrm{trap}}<\infty)
 \le C\bigl(
 \lvert D_0^{\varphi,\circ}\rvert^{3/4}+\lvert\xi_0\rvert+A_0^{-1/2}
 \bigr),
 \label{B-eq:exit-prob}
\end{equation}
with \(C\) uniform over initial states with the fixed margins. By
\eqref{B-eq:forcing} the forcing sum is at most \(CA_0^{-1/2}\),
\(\sum_tA_t^{-2}\le CA_0^{-1}\), \(\xi_0^2\le\lvert\xi_0\rvert\), and the
fixed margins absorb the denominators from Markov's inequality and the
maximal inequality.

On nonexit, \eqref{B-eq:sum-bound}--\eqref{B-eq:D2-sum} give
\(\sum_t\lvert D_t^{\varphi,\circ}\rvert^{\nu_{\mathrm{loc}}}<\infty\) and
\(\sum_tD_t^2<\infty\) almost surely; the stopped slope martingale
converges, its drift is absolutely summable, and \(\bar x_t\) converges by
\eqref{B-eq:xbar-sum}.  Hence \(V_\infty<\infty\), \(b_t\to b_\infty\),
\(\bar x_t\to\bar x_\infty\) and \(x_t\to\bar x_\infty\).  If
\(b^e\ne\beta\), then \(\lvert b_t-b^e\rvert<r<\lvert\beta-b^e\rvert/2\), so
\(\lvert b_\infty-\beta\rvert\ge\lvert b^e-\beta\rvert-r>0\).  Also
\(\xi_t\to0\), so \(a_t\to\alpha+\beta\bar x_\infty-b_\infty\bar x_\infty\).

\subsubsection{Open feasible-prior entry}

At \((a_0,b_0,\bar x_0,V_0)=(a^e,b^e,x^e,V^e)\) the target condition gives
\(\xi_0=0\), and \ref{lem:trapping:b} gives \(D^\varphi=D^{\varphi,\circ}=0\) at the reference
point \((a^e,b^e,A^{-1}=0,x^e,V^e)\), which lies in the collar, where
\(D^{\varphi,\circ}\) is Lipschitz.  So, with \(d_0\) as in the lemma, there is
\(C\), uniform on a small fixed neighborhood, with
\(\lvert\xi_0\rvert\le Cd_0\) and
\(\lvert D_0^{\varphi,\circ}\rvert\le C(d_0+A_0^{-1})\).  Choose
\(\overline A\) large and \(r_0\) small so that every state with
\(A_0>\overline A\) and \(d_0<r_0\) has the fixed initial margins above.
Then \eqref{B-eq:exit-prob} gives, uniformly over these states,
\[
 P^\Pi_\vartheta(\tau_{\mathrm{trap}}<\infty)
 \le C\bigl(r_0^{3/4}+r_0+\overline A^{-1/2}\bigr),
\]
and, shrinking \(r_0\) and enlarging \(\overline A\) if necessary, this is at
most the prescribed \(\varepsilon\).  So the nonexit event, on which the
convergence above holds almost surely, has probability at least
\(1-\varepsilon\).

For every \(A_0^{-1}>0\) the prior covariance is positive definite, since
its determinant is
\((A_0^{-1}+\bar x_0^2V_0^{-1})V_0^{-1}-\bar x_0^2V_0^{-2}=A_0^{-1}V_0^{-1}\).
The map from \((A^{-1},\bar x,V^{-1})\) to positive-definite covariance
matrices is a continuous bijection with continuous inverse, and every initial inequality
above can be chosen strict.  Hence \(A_0>\overline A\) and \(d_0<r_0\) define
a nonempty open set of feasible Gaussian priors, each with nonexit
probability at least \(1-\varepsilon\).  This proves Lemma~\ref{lem:trapping}.

\begin{remark}[Trapping at one slope precision]\label{B-rem:single}
The proof above fixes \(V^e\) first and uses hypotheses \ref{lem:trapping:a}--\ref{lem:trapping:c} and the
tracking condition only on the collar \(\mathcal C(V^e)\).  Its only
largeness condition on \(V^e\), the upper bound on
\(\lvert\gamma_t^{\mathrm{loc}}\rvert\), depends on neither the target nor
the radius \(r\).  The proof therefore gives the following form, which
Appendix~\ref{app:local} uses.  Fix the truth \(\vartheta=(\alpha,\beta)\),
an action \(x^e\), \(r_*>0\), \(u_*\in(0,1)\), a finite \(C_*\) and a
positive \(\lambda^0_*\).  There is \(\widehat V<\infty\) such that, for every
\(V^e\ge\widehat V\), every target \((a^e,b^e)\) with
\(a^e+b^ex^e=\alpha+\beta x^e\), and every Borel map \(\varphi\) that
satisfies \ref{lem:trapping:a}--\ref{lem:trapping:c} on the single collar \(\mathcal C(V^e)\), the following
holds.  For every \(r\in(0,r_*)\), with \(r<\lvert b^e-\beta\rvert/2\) when
\(b^e\ne\beta\), and every \(\varepsilon\in(0,1)\), there are
\(\overline A<\infty\) and \(r_0>0\) such that the priors in the displayed
set of Lemma~\ref{lem:trapping} form a nonempty open set, and, for each of
them and every policy kernel \(\Pi\) that tracks \(\varphi\) on
\(\mathcal C(V^e)\) with constant \(C_*\) for that prior, conclusions \ref{lem:trapping:i}
and \ref{lem:trapping:ii} of Lemma~\ref{lem:trapping} hold simultaneously with
\(P^\Pi_\vartheta\)-probability at least \(1-\varepsilon\).  By
Remark~\ref{B-rem:seeded}, the same holds for the policies of that remark,
with their prior-predictive and fixed-truth laws.
\end{remark}

\section{The abstract exclusion theorem}\label{supp:exclusion}

This section states and proves the exclusion theorem that
Appendix~\ref{app:global} uses. The theorem is stated for a policy kernel
\(\Pi\) in the setting of Section~\ref{supp:general-policies}. Its hypothesis
is a \textit{coherent exclusion package}: a finite list of deterministic
conditions under which the actions track a smooth reference map of the
posterior state, up to a quadratic error. Appendix~\ref{app:global} applies
the theorem to a decision rule \(\psi\), through the Dirac kernel
\(\Pi_\psi\), with a reference map that equals \(\psi\) itself.

Throughout, \(\Pi\) is a fixed policy kernel, realized by the seeds of
Section~\ref{supp:general-policies}. Every \(P\) and \(E\) refers to the
prior-predictive law \(P^{\mu,\Pi}\) unless a truth is subscripted, and
\(P^\Pi_\vartheta\) is the fixed-truth law of
Lemma~\ref{S1-lem:disintegration-general}. As in Appendix~\ref{app:global},
\(D_t=x_{t+1}-\bar x_t\) is the action gap,
\(I_{t+1}=y_{t+1}-a_t-b_tx_{t+1}\) is the prediction error,
\(\kappa_t=A_t/(A_t+1)\), and \(r_t=V_t/A_t\) is the precision ratio;
\(E_\infty=\{V_\infty=\infty\}\) is the complete-learning event. By
\eqref{eq:design}, \(V_{t+1}=V_t+\kappa_tD_t^2\). On feasible states we use
the independent local coordinates \(u=A^{-1}\), \(w=V^{-1}\) and
\(\ell_V=\log w=-\log V\), on chart domains with \(w>0\). By
\eqref{C-average-action}, \(\bar x_t\) lies in
\(X_{\mathrm{avg}}\coloneqq\operatorname{co}(\{\bar x_0\}\cup X)\) on every
path, so \(\lvert D_t\rvert\le C_D\coloneqq\sup\{\lvert x-\bar x\rvert:x\in X,\
\bar x\in X_{\mathrm{avg}}\}<\infty\) and \(V_t\le V_0+C_D^2t\) (this bound is
not used in Lemma~\ref{lem:anchored-collapse}).

\begin{lemma}[Anchored consequences of complete-learning collapse]
\label{lem:anchored-collapse}
Let \(B\subseteq\mathbb R^2\) and
let \({x^\star}:B\to X\). Suppose that,
\(P^{\mu,\Pi}\)-almost surely on
\(E_\infty\cap\{\vartheta=(\alpha,\beta)\in B\}\),
\(x_{t+1}\to {x^\star}(\vartheta)\). Then, under
\({P}^{\mu,\Pi}\), on this event,
\begin{equation}
 (a_t,b_t,A_t^{-1},V_t^{-1})\to(\alpha,\beta,0,0),\quad
 \bar x_t\to {x^\star}(\vartheta),\quad D_t\to0,\quad r_t\to0
 \label{C-collapse}
\end{equation}
almost surely.  The same holds for a policy kernel with values in an
arbitrary Borel set \(X\subseteq\R\), bounded or not.
\end{lemma}

\begin{proof}
On the displayed event, \(A_t^{-1},V_t^{-1}\to0\) and
Lemma~\ref{S1-lem:posterior-general}\ref{lem:posterior-interface:c} gives, almost surely,
\((a_t,b_t,A_t^{-1},V_t^{-1})\to(\alpha,\beta,0,0)\).
Almost surely on the event, the hypothesis, the identity
\eqref{C-average-action} and Ces\`aro convergence imply \(\bar x_t\to {x^\star}(\vartheta)\), hence
\(D_t=x_{t+1}-\bar x_t\to0\). Finally,
\[
 V_t=V_0+\sum_{j=0}^{t-1}\kappa_jD_j^2
 \le V_0+\sum_{j=0}^{t-1}D_j^2.
\]
Since \(D_j^2\to0\), its Ces\`aro mean tends to zero, and therefore
\(V_t/A_t\to0\). This proves \eqref{C-collapse}.
\end{proof}

The normalized actual action gap is \(D_t/r_t=A_tD_t/V_t\).
It enters the slope update as
\(b_{t+1}-b_t=\kappa_t\chi_t A_t^{-1}(D_t/r_t)I_{t+1}\),
where \(\chi_t=V_t/V_{t+1}\). The next lemma shows that complete learning at a
vanishing precision ratio needs this normalized gap to be unbounded.

\begin{lemma}[Complete learning requires an unbounded normalized revision rate]
\label{lem:revision-necessity}
On every part of \(E_\infty\) on which \(V_t/A_t\to0\),
\begin{equation}
 \sup_t\left\lvert \frac{D_t}{r_t}\right\rvert =\infty.
 \label{C-p-unbounded}
\end{equation}
\end{lemma}

\begin{proof}
Suppose instead that \(\lvert D_t/r_t\rvert \le K\) for every \(t\ge N\). Then
\(D_t^2\le K^2V_t^2/A_t^2\).
By the exact information recursion,
\[
 \frac1{V_t}-\frac1{V_{t+1}}
 =\frac{V_{t+1}-V_t}{V_tV_{t+1}}
 \le \frac{D_t^2}{V_t^2}
 \le \frac{K^2}{A_t^2}.
\]
On \(E_\infty\), \(V_m^{-1}\to0\). Summing from \(t=n\) to \(m-1\) and letting
\(m\to\infty\) yields
\[
 \frac1{V_n}\le K^2\sum_{t=n}^{\infty}A_t^{-2}\le \frac{CK^2}{A_n}.
\]
Hence \(V_n/A_n\ge (CK^2)^{-1}\) for every sufficiently large \(n\), which
contradicts the hypothesis \(V_n/A_n\to0\). Thus \eqref{C-p-unbounded} holds.
\end{proof}

Write \(z_b=(a,b,u,\bar x)\), \(w=V^{-1}\), and \(\ell_V=\log w\).
A smooth reference map is a state-based reference action map; the smoothness
requirements below concern this map, and the policy kernel \(\Pi\) enters
only through clauses \ref{package:approximation} and
\ref{package:anchoring}.

\begin{definition}[Coherent exclusion package]
\label{def:package}
Fix \(\eta\in(6/7,1]\), the compact action interval \(X\), the nondegenerate
Gaussian initial prior \(\mu\), and the fixed Borel randomized
posterior-state Markov kernel \(\Pi\) realized by the fresh independent
seeds of Section~\ref{supp:general-policies}.
Let \(\Theta_0\subset\mathbb R^2\) be Borel.
A \emph{coherent exclusion package} on \(\Theta_0\) consists of a countable
Borel cover \(\Theta_0=\bigcup_m\Theta_m\) and the following deterministic data
for each \(m\).
\begin{enumerate}[label=\textup{(P\arabic*)},ref=\textup{(P\arabic*)}]
\item\label{package:geometry}
Choose bounded rectangles \(\mathcal B_m^{\rm in}\Subset \mathcal B_m^{\rm mid}\)
in \(z_b\), products of open intervals in \((a,b,\bar x)\) with the intervals
\([0,u_m^{\rm in})\) and \([0,u_m^{\rm mid})\) in \(u\), and \(\ell_m^{\rm in}<\ell_m^{\rm out}\). Here and in
every application of this definition, closures, interiors and compact containment of
these rectangles are taken relative to the half-space \(\{u\ge0\}\) of the bounded
coordinates. Set
\[
 \mathcal S_m^{\mathrm{tail}}=\mathcal B_m^{\rm mid}\times(-\infty,\ell_m^{\rm out}),\qquad
 \mathcal K_m^0=\overline{\mathcal B}_m^{\rm in}\times(-\infty,\ell_m^{\rm in}].
\]

\item\label{package:regularity}
One state-based map \(\varphi_m\) on \(\mathcal S_m^{\mathrm{tail}}\) has a finite
\(C^{1,\eta}\) norm in the coordinates \((z_b,\ell_V)=(a,b,u,\bar x,\ell_V)\), and
on \(\mathcal K_m^0\) its first derivatives \(\varphi_{m,a}\) and \(\varphi_{m,b}\)
have a common Lipschitz constant in \(\ell_V\). Here the \(C^{1,\eta}\) norm
of a map on a set is the maximum of the sup norms of the map and of its first
derivatives on that set and the \(\eta\)-H\"older constant of its first
derivatives there. Put
\[
 D^{\varphi_m}=\varphi_m-\bar x,\qquad
 \kappa=(1+u)^{-1},\qquad \bar x_m^{+,\varphi}=\bar x+\kappa u D^{\varphi_m}.
\]
On \(\mathcal K_m^0\), for fixed finite \(C_m\) and positive \(\lambda_m\),
\[
 \lvert \varphi_{m,b}-\bar x_m^{+,\varphi}\varphi_{m,a}\rvert \ge\lambda_m,\qquad
 \lvert \varphi_{m,\bar x}\rvert +\lvert \partial_{\ell_V}\varphi_m\rvert
 \le C_m(u+e^{\ell_V}).
\]

\item\label{package:corridor}
Choose positive thresholds
\(\varepsilon_{u,m},\varepsilon_{v,m},\varepsilon_{\theta,m},\varepsilon_{d,m}\) and put
\[
 \mathcal K_m=\mathcal K_m^0\cap\left\{
 0<u\le\varepsilon_{u,m},\
 e^{\ell_V}\le\varepsilon_{v,m},\
 ue^{-\ell_V}\le\varepsilon_{\theta,m},\
 \lvert D^{\varphi_m}\rvert \le\varepsilon_{d,m}
 \right\}.
\]
The thresholds satisfy the finitely many inequalities that
Section~\ref{S2-section-5} requires of them (for \(\eta=1\), read at one fixed
\(\eta'\in(6/7,1)\); see the paragraph after
Theorem~\ref{thm:exclusion}), which depend only on \(\eta\), \(X\),
\(\mu\), the data of \ref{package:geometry}--\ref{package:regularity} and the
constant \(C_{2,m}\) of \ref{package:approximation}; they are fixed once, before
any path is observed, and are never decreased later.

\item\label{package:approximation}
The policy kernel tracks \(\varphi_m\), that is, for one finite \(C_{2,m}\),
\({P}^{\mu,\Pi}\)-almost surely, at every date \(t\) with
\(s_t\in\mathcal K_m\),
\begin{equation}
 \lvert x_{t+1}-\varphi_m(s_t)\rvert
       \le C_{2,m}\bigl(A_t^{-1}+\lvert D^{\varphi_m}(s_t)\rvert \bigr)^2.
 \label{eq:tracking}
\end{equation}

\item\label{package:anchoring}
\({P}^{\mu,\Pi}\)-almost surely on \(\{V_t\to\infty,\ \vartheta\in\Theta_m\}\),
\(V_t/A_t\to0\) and the state eventually remains in \(\mathcal K_m\).
\end{enumerate}
\end{definition}

The following theorem is the main result of this section. Its proof occupies
Sections~\ref{S2-section-2}--\ref{S2-section-9}.

\begin{theorem}[Almost-sure exclusion of complete learning]
\label{thm:exclusion}
If the fixed prior \(\mu\) and the policy kernel \(\Pi\) admit a coherent exclusion package
on \(\Theta_0\) \textup{(}Definition~\textup{\ref{def:package})}, then
\[
 {P}^{\mu,\Pi}
 (V_t\to\infty,\ (\alpha,\beta)\in\Theta_0)=0.
\]
Consequently complete learning has fixed-truth probability zero for
\(\mu\)-almost every truth in \(\Theta_0\), equivalently for Lebesgue-almost
every truth in \(\Theta_0\).
\end{theorem}

For \(\eta=1\) the theorem is the case of one fixed \(\eta'\in(6/7,1)\) applied
to the same package. If \(L_1\) is the Lipschitz constant of \(\mathrm D\varphi_m\)
on \(\mathcal S_m^{\mathrm{tail}}\), then
\[
 \lvert \mathrm D\varphi_m(z)-\mathrm D\varphi_m(z')\rvert
 \le\max\{L_1,2\sup\lvert \mathrm D\varphi_m\rvert \}\,\lvert z-z'\rvert ^{\eta'}
\]
(separate \(\lvert z-z'\rvert \le1\) from \(\lvert z-z'\rvert >1\)), so a finite
\(C^{1,1}\) norm bounds the \(C^{1,\eta'}\) norm; \(\eta\) enters the remaining
clauses only through the thresholds of \ref{package:corridor}, which are then
fixed by the inequalities of Section~\ref{S2-section-5} at \(\eta'\).

\begin{remark}[Exclusion for history-dependent policies]\label{S3-rem:seeded}
Lemma~\ref{lem:anchored-collapse}, Definition~\ref{def:package} and
Theorem~\ref{thm:exclusion} hold verbatim, with the same proofs, when \(\Pi\)
is replaced by a Borel predictable policy of Remark~\ref{A-rem:predictable}
with values in \(X\) and i.i.d.\ uniform seeds, and \(P^{\mu,\Pi}\) and
\(P^\Pi_\vartheta\) by its prior-predictive and fixed-truth laws; for
Lemma~\ref{lem:anchored-collapse}, the values may again lie in an arbitrary
Borel set \(X\subseteq\R\).  The proof of
Lemma~\ref{lem:anchored-collapse} uses the policy only through
Lemma~\ref{S1-lem:posterior-general}\ref{lem:posterior-interface:c}, which holds for such a policy by
Remark~\ref{A-rem:predictable}, and through \eqref{eq:design} and
\eqref{C-average-action}, which are identities in the realized actions.
\end{remark}

\subsection{Outline of the proof and the continuation}

The proof argues by contradiction. Clause~\ref{package:anchoring} gives
\(r_t\to0\) on the covered complete-learning event, so
Lemma~\ref{lem:revision-necessity} requires \(D_t/r_t\) to be unbounded. A
stopped weighted estimate for a centered reference-map gap
(Sections~\ref{S2-section-2}--\ref{S2-section-6}) instead forces
\(V_t\ge c_\omega A_t^{2/3}\) eventually, hence \(\sum_tV_t^{-2}<\infty\)
(Section~\ref{S2-section-7}), and a finite-clock argument then makes
\(D_t/r_t\) bounded (Section~\ref{S2-section-8}). Countable localization and
Lemma~\ref{S1-lem:disintegration-general} finish the proof
(Section~\ref{S2-section-9}).

Fix a coherent exclusion package and one index \(m\), suppressed where no
confusion arises. In physical-time formulas, \(E_t\) conditions on the
decision \(\sigma\)-field \(\mathcal F_t=\mathcal H_t\vee\sigma(\zeta_{t+1})\),
after the current action is known. The smooth reference map is evaluated at
the actual state. Write \(D^\varphi=\varphi(s)-\bar x\), \(r=V/A\),
\(\ell_V=-\log V\), \(\mathcal D=D^\varphi/r-c_{\mathrm{sm}}\) and
\(\varpi=A^{-1/3}V^{1/4}\); the center \(c_{\mathrm{sm}}\) is constructed in
Section~\ref{S2-section-3}.

\paragraph{Continuation.}
Fix a coordinatewise smooth saturation \(S_m^{\mathrm{coord}}\) of
\((a,b,\bar x)\) into the corresponding sides of \(\mathcal B_m^{\rm mid}\),
with bounded derivatives, equal to the identity near the corresponding sides
of \(\overline{\mathcal B}_m^{\rm in}\), and leaving \(u\) fixed; since
\(u_m^{\rm in}<u_m^{\rm mid}\), it maps \(\{z_b:0\le u\le u_m^{\rm in}\}\)
into \(\mathcal B_m^{\rm mid}\), and it exists for every choice of rectangles
in \ref{package:geometry}. Use the single continuation
\(\varphi_m^{\rm ext}(z_b,\ell_V)=\varphi_m(S_m^{\mathrm{coord}}(z_b),\ell_V)\).
Every feasible successor from \(\mathcal K_m^0\) has
\(0<u^+=\kappa u<u\le u_m^{\rm in}\) and
\(\ell_V^+\le\ell_V\le\ell_m^{\rm in}<\ell_m^{\rm out}\), so saturation gives
one uniform \(C^{1,\eta}\) bound on the complete successor-segment hull. The
continuation agrees with \(\varphi_m\) at current and surviving successor
states, and its Taylor bound holds for every Gaussian realization, with a
fixed polynomial envelope having all finite conditional moments. It defines
the full-successor gap \(D^{\varphi,+}=\varphi^{\mathrm{ext}}(s^+)-\bar x^+\).

\paragraph{Entry, exit and constants.}
For a deterministic entry date \(N\), let
\(\mathcal E_{m,N}\coloneqq \{s_N\in \mathcal K\}\) and
\(\tau^{\mathrm{space}}\coloneqq \inf\{t\ge N:s_t\notin \mathcal K\}\), whose
indices \((m,N)\) are restored in Section~\ref{S2-section-9}. These are
observable, i.e., \(\mathcal E_{m,N}\in\mathcal H_N\), and \(\tau^{\mathrm{space}}\)
is an \(\mathcal H_t\)-stopping time, hence an \(\mathcal F_t\)-stopping time.
At an exit-causing successor, define \(D^\varphi/r\) using
\(\varphi^{\rm ext}\) and the center using its fixed global extension;
thereafter freeze \(D^\varphi/r\), \(\mathcal D\) and \(\varpi\) at those exit
values. Constants may depend on the fixed package, the initial posterior and
\(N\), but not on a lower cutoff for \(\ell_V=\log(V^{-1})\).

\subsection{Preliminary estimates}\label{S2-section-2}

Derivatives use the independent coordinates \(z_\ell=(z_b,\ell_V)\),
holding the other coordinates fixed; \(\mathrm D_{z_b}c_h\) also fixes \(h\).
We differentiate before evaluating at \(u=A^{-1}\), \(\ell_V=-\log V\).
Coefficients use \(A^{-1},V^{-1}\) on paths and \(u,e^{\ell_V}\) on extended
domains, including \(u=0\). At a pre-exit date abbreviate
\(e=x-\varphi\), \(D=D^\varphi+e\), \(\kappa=A/(A+1)\),
\(\chi=(1+\kappa V^{-1}D^2)^{-1}\) and
\(\chi^\varphi=(1+\kappa V^{-1}(D^\varphi)^2)^{-1}\); the normalized actual
gap \(D/r\) and reference gap \(D^\varphi/r\) remain distinct.

The thresholds of \ref{package:corridor} ensure, first of all, the coarse
restrictions
\begin{equation}
A^{-1},\ V^{-1},\ r,\ \lvert D^\varphi\rvert \le1,\qquad \kappa,\chi\ge\tfrac12;
\label{S2-eq-2-1}
\end{equation}
Section~\ref{S2-section-5} lists the further requirements. The bound
\(\lvert D\rvert\le C_D\) remains, and \eqref{eq:tracking} gives
\begin{equation}
\lvert e\rvert \le C(A^{-2}+(D^\varphi)^2).
\label{S2-eq-2-2}
\end{equation}
The threshold \(\varepsilon_{d,m}\) ensures \(C\lvert D^\varphi\rvert\le1/2\)
on the corridor; with
\(\lvert D^\varphi\rvert \le \lvert D\rvert +CA^{-2}+C(D^\varphi)^2\) this
gives
\begin{equation}
\left.
\begin{aligned}
\lvert D\rvert &\le C(\lvert D^\varphi\rvert +A^{-2}),&
\lvert D^\varphi\rvert &\le 2\lvert D\rvert +CA^{-2},\\
\left\lvert \frac{D}{r}-\frac{D^\varphi}{r}\right\rvert &\le C\lvert D^\varphi\rvert \left\lvert \frac{D^\varphi}{r}\right\rvert +CV^{-1}A^{-1},&
\left\lvert \frac{D}{r}\right\rvert &\le C\left\lvert \frac{D^\varphi}{r}\right\rvert +CV^{-1}A^{-1},
\end{aligned}
\quad \right\}
\label{S2-eq-2-3}
\end{equation}
and in particular
\begin{equation}
D^2\le C((D^\varphi)^2+A^{-4}),\qquad (D^\varphi)^2\le C(D^2+A^{-4}).
\label{S2-eq-2-4}
\end{equation}
By Lemma~\ref{S1-lem:posterior-general}\ref{lem:posterior-interface:b},
\(I\mid\mathcal F_t\sim N(0,\sigma^2)\) with \(\sigma^2=1+A^{-1}+V^{-1}D^2\),
and on the coarse corridor
\begin{equation}
1\le \sigma^2\le 2+C_D^2.
\label{S2-eq-2-5}
\end{equation}
Thus every fixed polynomial in \(I\) has uniformly bounded conditional
moments, for every Gaussian realization, including those causing exit. The
fractional-moment estimates below use Lemma~\ref{B-lem:gaussian-moment}.

\paragraph{Information interpolation and its deterministic budget.}
Set \(p_{\mathrm{info}}=4-2/\eta>1\). The identity
\(V^{-(1+\eta)}\lvert D\rvert ^\eta
=(V^{-2})^{1-\eta/2}(V^{-p_{\mathrm{info}}}D^2)^{\eta/2}\)
and weighted Young's inequality give, for every \(\epsilon_Y>0\),
\begin{equation}
V^{-(1+\eta)}\lvert D\rvert ^\eta
\le\epsilon_Y V^{-2}+C_{\epsilon_Y} V^{-p_{\mathrm{info}}}D^2.
\label{S2-eq-2-8}
\end{equation}
This term has a deterministic pre-exit budget. Suppose the entry corridor
requires \(V_N\ge v_\ast>0\), let \(L_\ast=1+C_D^2/v_\ast\), and let
\(\kappa_\ast=1/2\), a lower bound for \(\kappa_t\) by \eqref{S2-eq-2-1}.
Before exit, \(D_t^2=\kappa_t^{-1}(V_{t+1}-V_t)\) and
\(V_{t+1}/V_t\le L_\ast\); comparing the sum with an integral gives, for every
finite stopping horizon,
\begin{equation}
\sum_{N\le t<\tau^{\mathrm{space}}}V_t^{-p_{\mathrm{info}}}D_t^2
\le
\kappa_\ast^{-1}L_\ast^{p_{\mathrm{info}}}
\int_{V_N}^{V_{\tau^{\mathrm{space}}}}v^{-p_{\mathrm{info}}}\,dv
\le
\frac{\kappa_\ast^{-1}L_\ast^{p_{\mathrm{info}}}}{p_{\mathrm{info}}-1}
v_\ast^{\,1-p_{\mathrm{info}}}
\eqqcolon C_{\mathrm{info}}.
\label{S2-eq-2-9}
\end{equation}
The bound holds with any additional stop and, by monotone convergence, for
an infinite horizon; it includes the last pre-exit increment, also when its
outcome causes exit.

\subsection{Center construction}\label{S2-section-3}

Fix \(k_h=7/8\). Since \(\eta>6/7\),
\(\frac{3}{4\eta}<k_h<\frac1{2-\eta}\), so
\(\eta_0\coloneqq k_h\eta\) and \(\eta_2\coloneqq 1-k_h(2-\eta)\) satisfy
\(\frac34<\eta_0<1\) and \(0<\eta_2<\eta_0\). At current states define
\(\bar x^{+,\varphi}=\bar x+\kappa uD^\varphi\),
\(\lambda^{+,\varphi}=\varphi_b-\bar x^{+,\varphi}\varphi_a\), and the exact
finite-precision center
\[
c_{\mathrm{ex}}=-\frac{\varphi_a}{\chi^\varphi\lambda^{+,\varphi}}.
\]
On \(\mathcal K_m^0=\mathcal B_c\times(-\infty,\ell_m^{\rm in}]\), with the
collar \(\mathcal B_c=\overline{\mathcal B}_m^{\rm in}\) containing every
current \(z_b\), put \(c_{\rm col}=c_{\mathrm{ex}}\). It is well defined and
bounded, since \(\lvert\lambda^{+,\varphi}\rvert\ge\lambda_m\) by the
transversality clause of \ref{package:regularity} and
\(\chi^\varphi=(1+\kappa e^{\ell_V}(D^\varphi)^2)^{-1}\) is bounded away from
zero there (the rectangle is bounded, \(\kappa\le1\) and
\(e^{\ell_V}\le e^{\ell_m^{\rm in}}\)). By \ref{package:regularity},
\(c_{\rm col}\) has a uniform \(\eta\)-H\"older constant \(L\) in \(z_b\) and a
uniform Lipschitz constant in \(\ell_V\), both independent of a lower
\(\ell_V\)-cutoff. Extend it by
\(c_{\mathrm{ext}}(z_b,\ell_V)=\inf_{z\in \mathcal B_c}
(c_{\rm col}(z,\ell_V)+L\lVert z_b-z\rVert ^\eta)\), clipped to a fixed
interval containing the range of \(c_{\rm col}\); since
\(\lVert z_b-z\rVert^\eta\) is a metric power, \(c_{\mathrm{ext}}\) equals
\(c_{\rm col}\) on \(\mathcal B_c\) and keeps both constants. Mollify only in
\(z_b\), \(c_h(z_b,\ell_V)=\int \varrho_h^{\mathrm{sm}}(z)c_{\mathrm{ext}}(z_b-z,\ell_V)\,dz\),
with a nonnegative smooth compactly supported kernel of integral one.
Subtracting \(c_{\mathrm{ext}}(z_b,\ell_V)\) inside each zero-integral
derivative kernel gives, for \(0<h\le2\),
\begin{equation}
\begin{gathered}
\lvert c_h\rvert \le C,\qquad
\lvert c_h-c_{\mathrm{ex}}\rvert \le Ch^\eta\quad (\text{at current states}),\\
\lVert \mathrm D_{z_b}c_h\rVert \le Ch^{\eta-1},\qquad
\lVert \mathrm D_{z_b}^2c_h\rVert \le Ch^{\eta-2},\qquad
\lvert \partial_hc_h\rvert \le Ch^{\eta-1},
\end{gathered}
\label{S2-eq-3-4}
\end{equation}
and an \(\ell_V\)-Lipschitz constant independent of \(h\). Set
\begin{equation}
h=r^{k_h},\qquad
c_{\mathrm{sm}}=c_h(z_b,\ell_V),\qquad
\mathcal D=\frac{D^\varphi}{r}-c_{\mathrm{sm}}.
\label{S2-eq-3-5}
\end{equation}
This center is state-based and seed-independent; its \(\ell_V\)-motion is
handled by the value-level Lipschitz bound, so the estimates below use only
derivatives of the center in \(z_b\) and in \(h\). The identity
\(D^\varphi=r(\mathcal D+c_{\mathrm{sm}})\) gives
\begin{equation}
\lvert D^\varphi\rvert \le r(\lvert \mathcal D\rvert +C),\qquad
(D^\varphi)^2\le Cr(\lvert D^\varphi\rvert +r)\lvert \mathcal D\rvert +Cr^2.
\label{S2-eq-3-6}
\end{equation}

\subsection{Exact recursion and centered normal form}\label{S2-section-4}

The state increments are
\begin{gather*}
(A^+)^{-1}-A^{-1}=-\kappa A^{-2},\qquad
\bar x^+-\bar x=\kappa A^{-1}D,\qquad
\ell_V^+-\ell_V=\log\chi,\\
b^+-b=\kappa V^{-1}\chi DI,\qquad
a^+-a=\kappa A^{-1}I-\bar x^+\kappa V^{-1}\chi DI.
\end{gather*}
Taylor-expand the fixed \(\varphi^{\rm ext}\) along the entire actual
successor segment (at the current state its derivatives equal those of
\(\varphi\)), put \(\lambda^+=\varphi_b-\bar x^+\varphi_a\), and multiply the
successor gap by \((V^+)^{-1}/(A^+)^{-1}=\chi V^{-1}/(\kappa A^{-1})\). This
gives the exact identity
\begin{equation}
\frac{D^{\varphi,+}}{r^+}
=\chi \frac{D^\varphi}{r}
+\chi V^{-1}\left(\varphi_a+\chi\lambda^+\frac{D}{r}\right)I
-\chi V^{-1}e+B_{\mathrm{cov}}+R_\varphi,
\label{S2-eq-4-2}
\end{equation}
where \(R_\varphi\) is \(\chi V^{-1}/(\kappa A^{-1})\) times the Taylor
remainder and
\[
B_{\mathrm{cov}}=
\chi A^{-1}\varphi_{\bar x} \frac{D}{r}
-\chi V^{-1}A^{-1}\varphi_u
+\frac{\chi}{\kappa}\frac1r\varphi_{\ell_V}\log\chi.
\]
The leading coefficient is \(\chi\), not \(\chi/\kappa\), because
\(\frac{\chi}{\kappa}\frac{D^\varphi}{r}-\chi V^{-1}D
=\chi \frac{D^\varphi}{r}-\chi V^{-1}e\). Since
\(\lVert \Delta(z_b,\ell_V)\rVert\le C(A^{-1}+V^{-1}\lvert D\rvert )(1+\lvert I\rvert )\),
the \(C^{1,\eta}\) bound on the successor-segment hull and \eqref{S2-eq-2-3}
give
\begin{equation}
\lvert R_\varphi\rvert
\le C\left(V^{-1}A^{-\eta}+
V^{-(1+\eta)}\lvert D\rvert ^\eta\left\lvert \frac{D^\varphi}{r}\right\rvert \right)\mathcal P_0(I),
\qquad \mathcal P_0(I)=1+I^2,
\label{S2-eq-4-4}
\end{equation}
also at an exit-causing successor, where it bounds the extended
reference-map gap, not the next actual-policy gap.

\paragraph{Center increment.}
The precision ratio satisfies
\begin{equation}
r^+=\frac{\kappa}{\chi}r,\qquad
r^+-r=-\kappa A^{-1}r+\kappa^2A^{-1}D^2,
\label{S2-eq-4-5}
\end{equation}
so \(\lvert r^+-r\rvert \le CA^{-1}(r+D^2)\) and \(h^+/h\) is bounded above
and below by fixed positive constants. Decompose exactly
\begin{equation}
c_{\mathrm{sm}}^+-c_{\mathrm{sm}}
=B_c^{\mathrm{noise}} I+B_c^{\mathrm{drift}}+R_c+B_c^\ell+B_c^h,
\label{S2-eq-4-6}
\end{equation}
where, with derivatives at \((z_b,\ell_V,h)\), the predictable noise
coefficient and the drift are
\(B_c^{\mathrm{noise}}=\kappa A^{-1}c_{h,a}+\kappa V^{-1}\chi D(c_{h,b}-\bar x^+c_{h,a})\)
and \(B_c^{\mathrm{drift}}=-\kappa A^{-2}c_{h,u}+\kappa A^{-1}Dc_{h,\bar x}\);
\(R_c\) is the second-order Taylor remainder of \(c_h(\cdot,\ell_V)\) from
\(z_b\) to \(z_b^+\); \(B_c^\ell=c_h(z_b^+,\ell_V^+)-c_h(z_b^+,\ell_V)\); and
\(B_c^h=c_{h^+}(z_b^+,\ell_V^+)-c_h(z_b^+,\ell_V^+)\). By \eqref{S2-eq-2-3},
\eqref{S2-eq-3-4}, \eqref{S2-eq-4-5}, \(A^{-1}=V^{-1}r\),
\(r^{1-k_h(1-\eta)}\le r^{\eta_0}\), the \(\ell_V\)-Lipschitz bound,
\(r^{k_h(\eta-2)}A^{-2}=V^{-2}r^{1+\eta_2}\) and
\(r\lvert\mathcal D\rvert\le\lvert D^\varphi\rvert+Cr\),
\begin{equation}
\begin{gathered}
\lvert B_c^{\mathrm{noise}}\rvert
\le CV^{-1}r^{\eta_0}(1+\lvert \mathcal D\rvert ),\qquad
\lvert B_c^{\mathrm{drift}}\rvert \le CA^{-1}r^{\eta_0}\lvert \mathcal D\rvert +CA^{-1},\\
\lvert B_c^\ell\rvert \le CV^{-1}D^2,\qquad
\lvert B_c^h\rvert\le C(A^{-1}r^{\eta_0}+V^{-1}r^{\eta_0}D^2),\\
\lvert R_c\rvert
\le CV^{-2}r^{\eta_2}(\lvert D^\varphi\rvert +r)\lvert \mathcal D\rvert \mathcal P_0(I)+CA^{-1}\mathcal P_0(I).
\end{gathered}
\label{S2-eq-4-8}
\end{equation}

\paragraph{Predictable Gaussian splitting.}
Subtract \eqref{S2-eq-4-6} from \eqref{S2-eq-4-2}. After extracting
\(\chi^2V^{-1}\lambda^+\mathcal D\), the coefficient of \(I\) is exactly
\[
B_I
=\chi V^{-1}(\varphi_a+\chi\lambda^+c_{\mathrm{sm}})
+\chi^2V^{-1}\lambda^+\frac e r-B_c^{\mathrm{noise}}.
\]
The center is built for the exact cancellation
\[
\varphi_a+\chi^\varphi\lambda^{+,\varphi} c_{\mathrm{ex}}=0.
\]
The error of replacing \(\chi\lambda^+\) by \(\chi^\varphi\lambda^{+,\varphi}\)
is controlled by
\begin{equation}
\lvert \lambda^+-\lambda^{+,\varphi}\rvert
\le CA^{-1}\lvert e\rvert ,
\qquad
\lvert \chi-\chi^\varphi\rvert
\le CV^{-1}\lvert e\rvert (\lvert D\rvert +\lvert D^\varphi\rvert ),
\label{S2-eq-4-16}
\end{equation}
and contributes at most
\(CV^{-1}(A^{-2}+(D^\varphi)^2)\le CV^{-1}r(\lvert D^\varphi\rvert +r)\lvert \mathcal D\rvert +CA^{-1}\)
to \(B_I\), by \eqref{S2-eq-2-2} and \eqref{S2-eq-3-6}; the mollifier error
contributes \(CV^{-1}r^{\eta_0}\) by \eqref{S2-eq-3-4}. With
\eqref{S2-eq-4-8} and the bound
\(\lvert e/r\rvert\le C(\lvert D^\varphi\rvert +r)\lvert \mathcal D\rvert +C(r+V^{-1}A^{-1})\),
which follows from \eqref{S2-eq-2-3} and \eqref{S2-eq-3-6}, this gives
\[
\lvert B_I\rvert
\le CV^{-1}\bigl(\lvert D^\varphi\rvert +r+r^{\eta_0}\bigr)\lvert \mathcal D\rvert
+C(A^{-1}+V^{-1}r^{\eta_0}).
\]
Split this predictable coefficient as
\(B_I=R_I^{\mathrm{rel}}\mathcal D+R_I^{\mathrm{add}}\), where
\(R_I^{\mathrm{add}}\) is the clip of \(B_I\) to the interval given by the
additive part of the bound and \(R_I^{\mathrm{rel}}=(B_I-R_I^{\mathrm{add}})/\mathcal D\)
(zero at \(\mathcal D=0\)). Both parts are predictable, and
\begin{equation}
\lvert R_I^{\mathrm{rel}}\rvert
\le CV^{-1}(\lvert D^\varphi\rvert +r+r^{\eta_0}),\qquad
\lvert R_I^{\mathrm{add}}\rvert
\le C(A^{-1}+V^{-1}r^{\eta_0}).
\label{S2-eq-4-18}
\end{equation}
The predictable Gaussian coefficient
\(\gamma_{\mathrm{NF}}=\chi\lambda^++R_I^{\mathrm{rel}}/(\chi V^{-1})
=\chi\lambda^+ +O(\lvert D^\varphi\rvert +r+r^{\eta_0})\)
satisfies, by transversality, \eqref{S2-eq-4-16}, and the thresholds of
\ref{package:corridor},
\begin{equation}
0<\gamma_{\min}\le\lvert \gamma_{\mathrm{NF}}\rvert \le \gamma_{\max}<\infty;
\label{S2-eq-4-20}
\end{equation}
for instance, if the thresholds give
\(\lvert \chi\lambda^+-\lambda^{+,\varphi}\rvert \le\lambda_m/4\) and
\(\lvert \gamma_{\mathrm{NF}}-\chi\lambda^+\rvert \le\lambda_m/4\) on the
corridor, then \(\gamma_{\min}=\lambda_m/2\) is valid.

\paragraph{The remaining terms.}
Put \(B_{\mathrm{cmp}}=-\chi V^{-1}e\). The remaining exact expression is
\[
R_{\mathrm{rem}}
=(\chi-1)c_{\mathrm{sm}}+B_{\mathrm{cmp}}+B_{\mathrm{cov}}+R_\varphi
-B_c^{\mathrm{drift}}-R_c-B_c^\ell-B_c^h.
\]
Besides the bounds \eqref{S2-eq-4-8} on \(B_c^{\mathrm{drift}}\) and \(R_c\),
a direct computation from \eqref{S2-eq-2-2}--\eqref{S2-eq-2-4},
\eqref{S2-eq-3-6}, \(V^{-1}r=A^{-1}\) and \(r^{\eta_0}\le1\) gives
\begin{gather*}
\bigl\lvert(\chi-1)c_{\mathrm{sm}}+B_{\mathrm{cmp}}
      -B_c^\ell-B_c^h\bigr\rvert
\le CA^{-1}(\lvert D^\varphi\rvert+r)
      \lvert \mathcal D\rvert+CA^{-1},\\
\lvert B_{\mathrm{cov}}\rvert
 \le CA^{-1}(A^{-1}+V^{-1})\lvert \mathcal D\rvert
+CV^{-2}\bigl[\lvert D^\varphi\rvert+A^{-2}
      +r(\lvert D^\varphi\rvert+r)\bigr]\lvert \mathcal D\rvert
      +CA^{-1}\mathcal P_0(I).
\end{gather*}
For \(B_{\mathrm{cov}}\) the derivative bounds of \ref{package:regularity}
give \(\lvert\varphi_{\bar x}\rvert+\lvert \partial_{\ell_V}\varphi\rvert\le C(A^{-1}+V^{-1})\),
so with \(\lvert\log\chi\rvert\le CV^{-1}D^2\) and \(A^{-1}/r=V^{-1}\) its
three terms are at most \(CA^{-1}(A^{-1}+V^{-1})(\lvert \mathcal D\rvert+1)\),
\(CV^{-1}A^{-1}\) and \(CV^{-2}D^2(1+1/r)\); in the last, \eqref{S2-eq-2-3}
gives
\begin{equation}
V^{-2}\frac{D^2}{r}=V^{-2}\lvert D\rvert \left\lvert \frac{D}{r}\right\rvert
\le CV^{-2}(\lvert D^\varphi\rvert +A^{-2})\left(\left\lvert \frac{D^\varphi}{r}\right\rvert +V^{-1}A^{-1}\right),
\label{S2-eq-4-23}
\end{equation}
and \(D^\varphi/r=\mathcal D+c_{\mathrm{sm}}\) with \eqref{S2-eq-3-6} finishes
the bound. With
\(\epsilon_{\mathrm{loc}}=A^{-1}+V^{-1}+\lvert D^\varphi\rvert +r^{\eta_2}\),
and since \(A^{-2}\le A^{-1}\) and \(r,r^{\eta_0}\le r^{\eta_2}\) on the coarse
corridor, the coefficient of \(\lvert \mathcal D\rvert\) in each of these four
bounds is at most \(C\epsilon_{\mathrm{loc}}(A^{-1}+V^{-2})\).

It remains to treat \(R_\varphi\). By \eqref{S2-eq-4-4} and
\eqref{S2-eq-2-8}, its part proportional to \(\mathcal D\) is at most
\((\epsilon_Y V^{-2}+C_{\epsilon_Y} V^{-p_{\mathrm{info}}}D^2)\lvert \mathcal D\rvert \mathcal P_0(I)\).
Its bounded-center part is at most
\(CV^{-1}A^{-\eta}(1+\lvert \mathcal D\rvert ^\eta)\mathcal P_0(I)\), by
\eqref{S2-eq-2-3} and \(D^\varphi=r(\mathcal D+c_{\mathrm{sm}})\), and
Young's inequality, with
\((V^{-1}A^{-\eta})^{1/(1-\eta)}(V^{-2})^{-\eta/(1-\eta)}=V^{-1}r^{\eta/(1-\eta)}\),
\(\eta/(1-\eta)>1\) and \(r\le1\), gives
\[
CV^{-1}A^{-\eta} \lvert \mathcal D\rvert ^\eta \mathcal P_0(I)
\le
\epsilon_Y V^{-2}\lvert \mathcal D\rvert
+C_{\epsilon_Y} V^{-1}r^{\eta/(1-\eta)}\mathcal P_0(I)^{1/(1-\eta)},
\]
whose last coefficient is at most \(C_{\epsilon_Y}A^{-1}\).

Choose once and for all an integer
\(m_{\mathrm{env}}\ge \lceil 1/(1-\eta)\rceil+2\) and set
\(\mathcal P(I)=(1+I^2)^{m_{\mathrm{env}}}\). It dominates all envelopes
above, and its degree is independent of \(\epsilon_Y\). Hence
\(\lvert R_{\mathrm{rem}}\rvert\le B_{\mathrm{rem}}^{\mathrm{rel}}\mathcal P(I)\lvert \mathcal D\rvert +B_{\mathrm{rem}}^{\mathrm{add}}\mathcal P(I)\),
with
\(B_{\mathrm{rem}}^{\mathrm{rel}}=C\epsilon_{\mathrm{loc}}(A^{-1}+V^{-2})+\epsilon_Y V^{-2}+C_{\epsilon_Y} V^{-p_{\mathrm{info}}}D^2\)
and \(B_{\mathrm{rem}}^{\mathrm{add}}=C_{\epsilon_Y}(A^{-1}+V^{-1}A^{-\eta})\).
Split \(R_{\mathrm{rem}}\) as \(B_I\) was, now allowing the split to depend on
\(I\): \(R_{\mathrm{rem}}=R_{\mathrm{NF}}^{\mathrm{rel}}(I)\mathcal D+R_{\mathrm{rem}}^{\mathrm{add}}(I)\),
with
\(\lvert R_{\mathrm{NF}}^{\mathrm{rel}}(I)\rvert \le B_{\mathrm{rem}}^{\mathrm{rel}}\mathcal P(I)\)
and
\(\lvert R_{\mathrm{rem}}^{\mathrm{add}}(I)\rvert \le B_{\mathrm{rem}}^{\mathrm{add}}\mathcal P(I)\).
With \(R_{\mathrm{NF}}^{\mathrm{add}}(I)=R_I^{\mathrm{add}}I+R_{\mathrm{rem}}^{\mathrm{add}}(I)\)
we obtain the exact normal form
\begin{equation}
\mathcal D^+=\chi(1+\gamma_{\mathrm{NF}}V^{-1}I)\mathcal D+R_{\mathrm{NF}}^{\mathrm{rel}}(I)\mathcal D+R_{\mathrm{NF}}^{\mathrm{add}}(I),
\label{S2-eq-NF}
\end{equation}
with
\begin{equation}
\begin{aligned}
\lvert R_{\mathrm{NF}}^{\mathrm{rel}}(I)\rvert
&\le
\bigl[
C\epsilon_{\mathrm{loc}}(A^{-1}+V^{-2})+\epsilon_Y V^{-2}
+C_{\epsilon_Y} V^{-p_{\mathrm{info}}}D^2
\bigr]\mathcal P(I),
\\
\lvert R_{\mathrm{NF}}^{\mathrm{add}}(I)\rvert
&\le C_{\epsilon_Y}
(A^{-1}+V^{-1}r^{\eta_0}+V^{-1}A^{-\eta})\mathcal P(I).
\end{aligned}
\label{S2-eq-4-27}
\end{equation}

\subsection{Weighted estimate and finite acyclic threshold choice}\label{S2-section-5}

Fix \(\nu_\varpi=24/25\). The exact scale update
\(\varpi^+/\varpi=\kappa^{1/3}\chi^{-1/4}\) and \eqref{S2-eq-NF} give
\begin{equation}
\varpi^+\mathcal D^+
=\kappa^{1/3}\chi^{3/4}
\left(1+\gamma_{\mathrm{NF}}V^{-1}I+\frac{R_{\mathrm{NF}}^{\mathrm{rel}}(I)}{\chi}\right)\varpi\mathcal D
+\varpi^+R_{\mathrm{NF}}^{\mathrm{add}}(I).
\label{S2-eq-5-1}
\end{equation}

\paragraph{Additive summability.}
On \(V\ge1\), \(r\le1\), \(A\ge1\), each of
\(\varpi A^{-1}=A^{-13/12}r^{1/4}\),
\(\varpi V^{-1}r^{\eta_0}=A^{-13/12}r^{\eta_0-3/4}\) and
\(\varpi V^{-1}A^{-\eta}=A^{-\eta-1/3}V^{-3/4}\) is at most \(A^{-13/12}\),
the last because \(\eta>3/4\). Since also \(\varpi^+\le C\varpi\) and
\(13\nu_\varpi/12=26/25>1\),
\(E_t[\lvert \varpi^+R_{\mathrm{NF}}^{\mathrm{add}}(I)\rvert ^{\nu_\varpi}]\le C_{\epsilon_Y} A^{-26/25}\).

\paragraph{Explicit contraction margins.}
Write \(m_\varpi=\kappa^{\nu_\varpi/3}\chi^{3\nu_\varpi/4}\). On
\(A^{-1}\le1\) and \(\chi\ge1/2\), \(m_\varpi\ge C_\varpi^{\min}>0\). Moreover
\(1-\kappa^{\nu_\varpi/3}
=\int_0^{A^{-1}} (\nu_\varpi/3)(1+t)^{-\nu_\varpi/3-1}\,dt
\ge (\nu_\varpi/3)2^{-\nu_\varpi/3-1}A^{-1}\), so, since \(\chi\le1\), for a
fixed \(c_{\rm cal}>0\),
\begin{equation}
m_\varpi\le\kappa^{\nu_\varpi/3}\le1-c_{\rm cal}A^{-1}.
\label{S2-eq-5-4}
\end{equation}
Apply Lemma~\ref{B-lem:gaussian-moment} to the factor in \eqref{S2-eq-5-1},
with \(\epsilon_G=V^{-1}\), \(\gamma_G=\gamma_{\mathrm{NF}}\) and
\(R_{\mathrm{pert}}=R_{\mathrm{NF}}^{\mathrm{rel}}(I)/\chi\), using
\eqref{S2-eq-4-20}, \eqref{S2-eq-4-27} and the fixed polynomial
\(\mathcal P\). Subadditivity of \(x\mapsto \lvert x\rvert ^{\nu_\varpi}\) in
\eqref{S2-eq-5-1} gives
\[
\begin{aligned}
E_t[\lvert \varpi^+\mathcal D^+\rvert ^{\nu_\varpi}]
\le {}&
\Bigl(
1-c_{\rm cal}A^{-1}
-C_\varpi^{\min}c_GV^{-2}
+C_0\epsilon_{\mathrm{loc}}(A^{-1}+V^{-2})
\\
&\qquad+C_1\epsilon_Y V^{-2}+C_{\epsilon_Y} V^{-p_{\mathrm{info}}}D^2
\Bigr)
\lvert \varpi\mathcal D\rvert ^{\nu_\varpi}
+C_{\epsilon_Y} A^{-26/25},
\end{aligned}
\]
with constants \(C_0,C_1\) independent of \(\epsilon_Y\).

The choices are finite and acyclic, in this order.
(i)~The calendar margin \(c_{\rm cal}\) of \eqref{S2-eq-5-4} and
\(C_\varpi^{\min}\) are fixed by \(\nu_\varpi\); \(\epsilon_{G,0},c_G,C_G\)
come from Lemma~\ref{B-lem:gaussian-moment} with the ranges
\eqref{S2-eq-2-5} and \eqref{S2-eq-4-20}; \(C_0,C_1\) are the constants of
the last display. All of them depend only on \(\eta\), \(k_h\),
\(\nu_\varpi\), the original package, the continuation, the center
extension, the mollifier and the polynomial degree.
(ii)~Choose \(\epsilon_Y\) with \(C_1\epsilon_Y\le C_\varpi^{\min}c_G/4\).
(iii)~Choose \(\delta_E>0\) with
\(C_0\delta_E\le\min\{c_{\rm cal}/2,\, C_\varpi^{\min}c_G/4\}\).
(iv)~Fix the thresholds of \ref{package:corridor} so that, on
\(\mathcal K_m\),
\[
A^{-1},\ V^{-1},\ \lvert D^\varphi\rvert \le\delta_E/4,\qquad
r\le(\delta_E/4)^{1/\eta_2},\qquad V^{-1}\le \epsilon_{G,0},
\]
together with the coarse requirements that give
\eqref{S2-eq-2-1}--\eqref{S2-eq-2-5}, the two-way bounds between actual
and reference gaps, and the fixed Gaussian-coefficient ranges
\eqref{S2-eq-4-20}.
The constants \(C_{\epsilon_Y}\) enlarge only additive and
information-budget terms and enter no contraction choice, so the thresholds
are fixed once, as \ref{package:corridor} requires, and are never decreased
later. Since then \(\epsilon_{\mathrm{loc}}\le\delta_E\), we have proved, at
every pre-exit date and including its full successor,
\begin{equation}
E_t[\lvert \varpi^+\mathcal D^+\rvert ^{\nu_\varpi}]
\le
\bigl[
1-c(A^{-1}+V^{-2})
+C_{\epsilon_Y} B_t^{\mathrm{info}}
\bigr]\lvert \varpi\mathcal D\rvert ^{\nu_\varpi}
+C_{\epsilon_Y} A_t^{-26/25},
\label{S2-eq-5-6}
\end{equation}
where \(B_t^{\mathrm{info}}=V_t^{-p_{\mathrm{info}}}D_t^2\).

\subsection{Original-law stopping, entry bounds and exit values}\label{S2-section-6}

Set
\[
L_t^{\varpi}=\mathbf1_{\mathcal E_{m,N}}
\left\lvert \varpi_{t\wedge\tau^{\mathrm{space}}}\mathcal D_{t\wedge\tau^{\mathrm{space}}}^{\rm ext}\right\rvert ^{\nu_\varpi},\qquad
M_t^{\mathrm{info}}=
\prod_{j=N}^{t-1}
\left(1+C_{\epsilon_Y}\mathbf1_{\mathcal E_{m,N}}\mathbf1_{\{j<\tau^{\mathrm{space}}\}}B_j^{\mathrm{info}}\right).
\]
By \eqref{S2-eq-2-9}, \(1\le M_t^{\mathrm{info}}\le e^{C_{\epsilon_Y} C_{\mathrm{info}}}\)
for every \(t\). Bounded-time integrability holds before any limit,
because \(\varphi^{\rm ext}\) and the center are bounded,
\(\bar x_t\in X_{\mathrm{avg}}\), \(A_t=A_0+t\) and
\(V_0\le V_t\le V_0+C_D^2t\), so the extended stopped scalar is bounded by a
deterministic constant on every bounded horizon. At entry, \(A_N\) is
deterministic, \(V_N\) lies between the entry threshold and that bound, and
\(D_N^{\varphi}\), \(c_{\mathrm{sm},N}\) are bounded; hence
\(L_N^{\varpi}\le C_N^{(0)}\mathbf1_{\mathcal E_{m,N}}\).

Let \(B_t^{\varpi}=C_{\epsilon_Y} A_t^{-26/25}\). On
\(\mathcal E_{m,N}\cap\{t<\tau^{\mathrm{space}}\}\), \eqref{S2-eq-5-6} with its
negative drift discarded reads
\(E_t[L_{t+1}^{\varpi}]\le(M_{t+1}^{\mathrm{info}}/M_t^{\mathrm{info}})L_t^{\varpi}+B_t^{\varpi}\);
dividing by the predictable \(M_{t+1}^{\mathrm{info}}\ge1\) shows that
\[
M_t^{\varpi}\coloneqq \frac{L_t^{\varpi}}{M_t^{\mathrm{info}}}
+\mathbf1_{\mathcal E_{m,N}}\sum_{j=t}^\infty B_j^{\varpi}
\]
satisfies \(E_t[M_{t+1}^{\varpi}]\le M_t^{\varpi}\) there. At and after exit,
and off \(\mathcal E_{m,N}\), this remains true because \(L^\varpi\) and
\(M^{\mathrm{info}}\) are frozen and the deterministic tail decreases. Since
\(M^\varpi\) is adapted, nonnegative and, as \(\sum_jB_j^{\varpi}<\infty\),
bounded by a deterministic constant on every bounded horizon, it is a
nonnegative supermartingale. A nonnegative supermartingale converges almost
surely to a finite limit \citep[Theorem~4.2.12, p.~193]{d2019}, so
\(\sup_tM_t^{\varpi}<\infty\) almost surely, and the bound on
\(M^{\mathrm{info}}\) gives \(\sup_tL_t^{\varpi}<\infty\) almost surely.

\subsection{Deriving the complete-learning rate}\label{S2-section-7}

On \(\mathcal E_{m,N}\cap\{\tau^{\mathrm{space}}=\infty\}\), boundedness of
\(\varpi\mathcal D\) and of the center, and \eqref{S2-eq-2-3}, give
\(\lvert D_t/r_t\rvert \le C_\omega A_t^{1/3}V_t^{-1/4}\); the bounded-center
term is absorbed because \(V_t\le A_t\) and \(A_t\ge1\) on the corridor, so
\(\varpi_t\le A_t^{-1/12}\le1\). The exact information recursion gives
\(V_t^{-1}-V_{t+1}^{-1}=\kappa_t\chi_tA_t^{-2}(D_t/r_t)^2\le A_t^{-2}(D_t/r_t)^2\),
and therefore
\begin{equation}
\sqrt{V_t^{-1}}-\sqrt{V_{t+1}^{-1}}
=\frac{V_t^{-1}-V_{t+1}^{-1}}
{\sqrt{V_t^{-1}}+\sqrt{V_{t+1}^{-1}}}
\le\frac{(D_t/r_t)^2}{A_t^2\sqrt{V_t^{-1}}}
\le C_\omega A_t^{-4/3}.
\label{S2-eq-7-2}
\end{equation}
Only now restrict attention to \(V_t\to\infty\), so \(V_t^{-1}\to0\). Summing
\eqref{S2-eq-7-2} to infinity gives
\(\sqrt{V_n^{-1}}\le C_\omega\sum_{t=n}^\infty A_t^{-4/3}\le C_\omega' A_n^{-1/3}\).
Hence, eventually,
\begin{equation}
V_n\ge c_\omega A_n^{2/3},
\qquad\text{and so}\qquad \sum_n V_n^{-2}<\infty.
\label{S2-eq-7-3}
\end{equation}

\subsection{Finite-clock boundedness}\label{S2-section-8}

\begin{auxlemma}[Finite-clock boundedness]\label{S2-lem:finite-clock} Let a scalar adapted process satisfy, before a stopping time \(\tau\),
\[
\mathsf X^+=\chi \mathsf X+j+\sigma_X I+R^X,
\]
where \(\chi,j,\sigma_X,B^{\mathrm{noise}},B_X^{\mathrm{rel}},B_X^{\mathrm{add}}\) are predictable,
\(0<\chi\le1\), \(\lvert j\rvert \le C_X(1-\chi)\), \(\lvert \sigma_X\rvert \le CB^{\mathrm{noise}}(1+\lvert \mathsf X\rvert )\),
and
\(E_t [I]=0\), \(E_t [I^2]\le C\),
\(\lvert R^X\rvert \le(B_X^{\mathrm{rel}}\lvert \mathsf X\rvert +B_X^{\mathrm{add}})G^{\mathrm{env}}\), \(B_X^{\mathrm{rel}},B_X^{\mathrm{add}}\ge0\), and \(E_t[(G^{\mathrm{env}})^2]\le C\).
The remainder may depend arbitrarily on \(I\) and other successor randomness.
At and after \(\tau\), the coefficients need only be predictable.

Freeze \(\mathsf X\) at the stopping time, including its full exit value. If the initial value is finite, then
\begin{equation}
\sum_{t<\tau}[(B_t^{\mathrm{noise}})^2+B_{X,t}^{\mathrm{rel}}+B_{X,t}^{\mathrm{add}}+(B_{X,t}^{\mathrm{rel}}+B_{X,t}^{\mathrm{add}})^2]<\infty
\label{S2-eq-8-2}
\end{equation}
implies that \(\mathsf X_t\) is almost surely bounded. This statement holds on the summability event.
\end{auxlemma}

\begin{proof}
Put \(\Psi_{C_X}(\mathsf X)=\operatorname{dist}(\mathsf X,[-C_X,C_X])^2\). Its
derivative is globally \(2\)-Lipschitz, so
\begin{equation}
\Psi_{C_X}(x+h)\le \Psi_{C_X}(x)+\Psi_{C_X}'(x)h+h^2,
\label{S2-eq-8-3}
\end{equation}
and \(\Psi_{C_X}(\chi \mathsf X+j)\le\chi^2\Psi_{C_X}(\mathsf X)\le \Psi_{C_X}(\mathsf X)\),
because \(\chi \mathsf X+j\in\chi \mathsf X+(1-\chi)[-C_X,C_X]\). Apply
\eqref{S2-eq-8-3} at the predictable point \(x=\chi \mathsf X+j\); the term
linear in \(\sigma_X I\) has conditional mean zero. With
\(\lvert \mathsf X\rvert ^2\le 2C_X^2+2\Psi_{C_X}(\mathsf X)\),
\(\lvert \Psi_{C_X}'(x)\rvert \le2\sqrt{\Psi_{C_X}(\mathsf X)}\),
Cauchy--Schwarz and \((r+s)^2\le2r^2+2s^2\),
\[
E_t[\Psi_{C_X}(\mathsf X^+)]
\le \Psi_{C_X}(\mathsf X)+CB_t^{\mathrm{loc}}(1+\Psi_{C_X}(\mathsf X)),
\]
where \(B_t^{\mathrm{loc}}\) is the summand of \eqref{S2-eq-8-2}. Hence \(1+E_t[\Psi_{C_X}(\mathsf X_{t+1})]\le(1+CB_t^{\mathrm{loc}})(1+\Psi_{C_X}(\mathsf X_t))\)
before the stopping time, and trivially at and after it, where \(\mathsf X\)
is frozen and \(B_t^{\mathrm{loc}}\) is set to zero. Therefore
\[
\mathsf Z_t=\frac{1+\Psi_{C_X}(\mathsf X_t)}
{(1+\Psi_{C_X}(\mathsf X_N))\prod_{N\le j<t}(1+CB_j^{\mathrm{loc}})},\qquad t\ge N,
\]
is adapted and nonnegative with \(E_t[\mathsf Z_{t+1}]\le \mathsf Z_t\), the
denominator at \(t+1\) being \(\mathcal F_t\)-measurable. Since
\(\mathsf Z_N=1\), induction gives \(E[\mathsf Z_t]\le1\), so \(\mathsf Z\)
is a nonnegative supermartingale and converges almost surely to a finite
limit \citep[Theorem~4.2.12, p.~193]{d2019}. On \eqref{S2-eq-8-2} with a
finite initial value, the denominator of \(\mathsf Z_t\) is at most
\((1+\mathsf X_N^2)\exp(C\sum_jB_j^{\mathrm{loc}})<\infty\), so
\(\sup_t\Psi_{C_X}(\mathsf X_t)<\infty\), and
\(\mathsf X_t^2\le2C_X^2+2\Psi_{C_X}(\mathsf X_t)\) gives the almost-sure
boundedness of \(\mathsf X\).
\end{proof}

\paragraph{Application to the uncentered reference-map gap.}
Apply Lemma~\ref{S2-lem:finite-clock} to \(\mathsf X_t=D_t^\varphi/r_t\) and
\eqref{S2-eq-4-2}, with \(B_t^{\mathrm{noise}}=V_t^{-1}\). The Gaussian
coefficient of \eqref{S2-eq-4-2} is at most
\(CV^{-1}(1+\lvert D^\varphi/r\rvert)\). For the drift, \eqref{S2-eq-2-2}--\eqref{S2-eq-2-4}
and \(1-\chi=\kappa\chi V^{-1}D^2\) give
\begin{equation}
\lvert B_{\mathrm{cmp}}\rvert =\lvert \chi V^{-1}e\rvert
\le C\chi V^{-1}D^2+CV^{-1}A^{-2}
\le C_X(1-\chi)+CV^{-1}A^{-2};
\label{S2-eq-8-6}
\end{equation}
let \(j\) be the clip of the predictable \(B_{\mathrm{cmp}}\) to
\([-C_X(1-\chi),C_X(1-\chi)]\), so the rest is at most \(CV^{-1}A^{-2}\). The
covariance drift has the coarse bound
\begin{equation}
\lvert B_{\mathrm{cov}}\rvert \le CV^{-2}\left\lvert \frac{D^\varphi}{r}\right\rvert +CA^{-1}V^{-1}.
\label{S2-eq-8-8}
\end{equation}
For its first term use \(A^{-1}\le V^{-1}\) and \eqref{S2-eq-2-3}; its last
term is at most \(CV^{-2}D^2(1+1/r)\le CV^{-2}D^2/r\) by
Section~\ref{S2-section-4} and \(r\le1\), and \eqref{S2-eq-4-23} with
\(\lvert D^\varphi\rvert ,A^{-1},V^{-1}\le1\) bounds this by
\(CV^{-2}\lvert D^\varphi/r\rvert +CV^{-1}A^{-1}\). Combining
\eqref{S2-eq-4-4}, \eqref{S2-eq-8-6} and \eqref{S2-eq-8-8},
Lemma~\ref{S2-lem:finite-clock} applies with a fixed polynomial Gaussian
envelope and
\begin{equation}
B_{X,t}^{\mathrm{rel}}=C(V_t^{-2}+V_t^{-(1+\eta)}\lvert D_t\rvert ^\eta),
\qquad
B_{X,t}^{\mathrm{add}}=CV_t^{-1}A_t^{-\eta};
\label{S2-eq-8-9}
\end{equation}
the terms \(A^{-1}V^{-1}\) and \(V^{-1}A^{-2}\) are covered by
\(V^{-1}A^{-\eta}\), since \(A^{-1}\le1\) and \(\eta<1\).

The lemma is applied with the stopping time \(\tau^{\mathrm{space}}\), the
process frozen at its full exit value, and the finite initial value
\(D_N^\varphi/r_N\). Its condition \eqref{S2-eq-8-2}, a sum over all dates
on \(\{\tau^{\mathrm{space}}=\infty\}\), holds on
\(\mathcal E_{m,N}\cap\{\tau^{\mathrm{space}}=\infty\}\cap\{\sum_tV_t^{-2}<\infty\}\).
Indeed, \(\sum_t(B_t^{\mathrm{noise}})^2=\sum_tV_t^{-2}<\infty\);
\eqref{S2-eq-2-8}--\eqref{S2-eq-2-9} give
\(\sum_t B_{X,t}^{\mathrm{rel}}\le C(1+\epsilon_Y)\sum_tV_t^{-2}+CC_{\epsilon_Y}C_{\mathrm{info}}<\infty\);
by Cauchy--Schwarz,
\(\sum_t B_{X,t}^{\mathrm{add}}\le C(\sum_tV_t^{-2})^{1/2}(\sum_{t\ge N}A_t^{-2\eta})^{1/2}<\infty\),
the last series being deterministic and finite because
\(A_t^{-1}=(A_0+t)^{-1}\) and \(2\eta>1\); and on the corridor
\(B_{X,t}^{\mathrm{rel}},B_{X,t}^{\mathrm{add}}\) are uniformly bounded, so
\((B_{X,t}^{\mathrm{rel}}+B_{X,t}^{\mathrm{add}})^2\le C(B_{X,t}^{\mathrm{rel}}+B_{X,t}^{\mathrm{add}})\).
Lemma~\ref{S2-lem:finite-clock} therefore gives
\(\sup_{t\ge N}\lvert D_t^\varphi/r_t\rvert<\infty\) almost surely on that
event, and \eqref{S2-eq-2-3} gives the boundedness of \(D_t/r_t\) there.

\subsection{Contradiction, localization and disintegration}\label{S2-section-9}

\begin{proof}[Proof of Theorem~\ref{thm:exclusion}]
On
\(\mathcal E_{m,N}\cap\{\tau_{m,N}^{\mathrm{space}}=\infty\}\cap\{V_t\to\infty\}\),
Sections~\ref{S2-section-6}--\ref{S2-section-7} give the finite clock
\(\sum_tV_t^{-2}<\infty\), and Section~\ref{S2-section-8} then makes
\(D_t/r_t\) bounded. Every finite-date \(D_t/r_t\) is finite, so tail
boundedness contradicts Lemma~\ref{lem:revision-necessity} on
\(r_t\to0\). Therefore
\begin{equation}
P^{\mu,\Pi}\left(\mathcal E_{m,N},\, \tau_{m,N}^{\mathrm{space}}=\infty,\, V_t\to\infty,\, r_t\to0\right)=0.
\label{S2-eq-9-2}
\end{equation}

Now restore the package indices \(m\). For each covered truth set
\(\Theta_m\), \ref{package:anchoring} places the complete-learning event
\(\{V_t\to\infty,\ \vartheta\in\Theta_m\}\), up to a \(P^{\mu,\Pi}\)-null
set, in
\(\bigcup_{N\ge0}
\{s_N\in \mathcal K_m,\, \tau_{m,N}^{\mathrm{space}}=\infty,\, V_t\to\infty,\, r_t\to0\}\).
On that event \(r_t=V_t/A_t\to0\) by its precision clause, and its residence
clause gives an entry date \(N\) with \(s_N\in\mathcal K_m\) and
\(\tau_{m,N}^{\mathrm{space}}=\infty\). Every event in this countable union
is null by \eqref{S2-eq-9-2}. Taking the countable union over \(m\) proves
\begin{equation}
P^{\mu,\Pi}(V_t\to\infty,\, \vartheta\in\Theta_0)=0.
\label{S2-eq-9-3}
\end{equation}
By the disintegration identity of
Lemma~\ref{S1-lem:disintegration-general}, \eqref{S2-eq-9-3} gives
\(0=\int_{\Theta_0}P^{\Pi}_{\vartheta}(V_t\to\infty)\,\mu(d\vartheta)\).
Thus complete learning has fixed-truth probability zero for \(\mu\)-almost
every covered truth. A nondegenerate Gaussian prior has an everywhere
positive Lebesgue density, so its null sets are the Lebesgue-null sets, and
the exceptional set of truths is also Lebesgue-null.

The exceptional set depends on the fixed initial prior and the fixed kernel.
\end{proof}

\section{Myopic decision rules}\label{supp:myopic}

This section proves Corollary~\ref{cor:myopic}.  \(U\) satisfies
\ref{ass:payoff:U1}--\ref{ass:payoff:U3}.  Let \(Z\sim N(0,1)\),
\(G(x,m)\coloneqq E[U(x,m+Z)]\) and \(H(x;a,b)\coloneqq G(x,a+bx)\), so
that \(H(\cdot;\vartheta)=g^{\mathrm{full}}(\cdot;\vartheta)\), and write
\(r(\vartheta)\) for the maximizer of \(H(\cdot;\vartheta)\) on \(X\) at every
parameter where it is unique; on \(\Theta^R\), \(r=x^\star\), and \(Q\) and
\(M\) are evaluated at \(r\) as in the main text.  Let \(\mathbb S\) be the set of symmetric \(2\times2\) matrices, as
in Definition~\ref{def:regular}.  For \(x\in X\) and
\(p=(a,b,\Sigma)\in\R^2\times\mathbb S\), put
\(q(x;\Sigma)\coloneqq(1,x)\Sigma(1,x)^\top\),
\[
 \mathcal J(x;p)\coloneqq E\Bigl[U\Bigl(x,\ a+bx+\sqrt{1+q(x;\Sigma)}\,Z\Bigr)\Bigr],
\]
on \(\mathcal D\coloneqq\{(x,p)\in X\times\R^2\times\mathbb S:1+q(x;\Sigma)>0\}\),
which is relatively open in \(X\times\R^2\times\mathbb S\).  With
\(m\coloneqq a+bx\) and \(\varsigma\coloneqq\sqrt{1+q(x;\Sigma)}\),
\begin{equation}\label{My-eq:density}
 \mathcal J(x;p)=\int_\R U(x,y)\,\frac{1}{\sqrt{2\pi}\,\varsigma}
 \exp\left(-\frac{(y-m)^2}{2\varsigma^2}\right)dy .
\end{equation}
At a
feasible state \(s\), with \(\Sigma=\Sigma(u,\bar x,w)\) of
\eqref{eq:sigma}, \(q(x;\Sigma)=u+w(x-\bar x)^2\) is the posterior variance
of the mean outcome at \(x\), so the posterior predictive law of the outcome
is \(N(a+bx,1+q)\) and \(\mathcal J(x;a,b,\Sigma(u,\bar x,w))\) is the
posterior expected payoff of \(x\).  At \(\Sigma=0\),
\(\mathcal J(x;a,b,0)=H(x;a,b)\).  The decision rule \(\psi_0\) is a Borel
selection of \(s\mapsto\arg\max_{x\in X}\mathcal J(x;a,b,\Sigma(u,\bar x,w))\);
every statement below holds for every such selection.

\begin{lemma}[The myopic objective]\label{My-lem:objective}
\begin{enumerate}[label=(\alph*)]
\item \(G\) and \(H\) are \(C^3\) on \(X\times\R\) and\label{My-lem:objective:a}
\(X\times\R^2\), respectively, with one-sided derivatives in \(x\) at the endpoints of \(X\).
\item \(\mathcal J\) is \(C^3\) on \(\mathcal D\), with one-sided derivatives\label{My-lem:objective:b}
in \(x\) at the endpoints of \(X\), and its derivatives of order at most three
are obtained by differentiating under the integral in
\eqref{My-eq:density}.
\item A Borel selection \(\psi_0\) exists.\label{My-lem:objective:c}
\end{enumerate}
\end{lemma}

\begin{proof}
In \eqref{My-eq:density}, \(p\) enters only through the Gaussian density,
which is smooth in \((m,\varsigma)\), and \(m\) and \(\varsigma\) are smooth
in \((x,p)\) on \(\mathcal D\); the action \(x\) enters also through \(U\),
which is \(C^3\) in \(x\) on \(X\).  So every derivative of order \(k\le3\) of the
integrand in \((x,p)\) is a finite sum of terms
\(\partial_x^jU(x,y)\,P(y)\exp\bigl(-(y-m)^2/(2\varsigma^2)\bigr)\) with
\(j\le k\) and \(P\) a polynomial of degree at most \(2k\) whose coefficients
are continuous in \((x,p)\).  The functions \(\partial_x^jU(x,\cdot)\) are Borel, as
pointwise limits of difference quotients, one-sided at an endpoint of \(X\).  On a compact subset of
\(\mathcal D\), \(\lvert m\rvert\le M_0\) and
\(\varsigma\in[\varsigma_0,\varsigma_1]\) with \(\varsigma_0>0\), so, by
\ref{ass:payoff:U3}, each term is dominated by a multiple of
\(e^{c_U\lvert y\rvert}(1+\lvert y\rvert)^{6}\exp\{-(\lvert y\rvert-M_0)_+^2/(2\varsigma_1^2)\}\),
which is integrable.  Differentiation under the integral and continuity of the
derivatives follow from dominated convergence, applied at an endpoint of \(X\)
to one-sided difference quotients in \(x\); this gives
\ref{My-lem:objective:b}, and \ref{My-lem:objective:a} as the case
\(\Sigma=0\).  No derivative of \(U\) in the outcome is used.  For example,
with \(Y=m+\varsigma Z\),
\begin{gather*}
 \mathcal J_x=E[U_x(x,Y)]+\frac{b}{\varsigma}E[U(x,Y)Z]
 +\frac{q_x(x;\Sigma)}{2\varsigma^2}E\bigl[U(x,Y)(Z^2-1)\bigr],\\
 G_m(x,m)=E[U(x,m+Z)Z].
\end{gather*}
For \ref{My-lem:objective:c}, \((s,x)\mapsto\mathcal J(x;a,b,\Sigma(u,\bar x,w))\) is continuous
on the open set of feasible states times the compact set \(X\), and the
measurable maximum theorem \citep[Theorem~18.19]{ab2006} gives a Borel
selection.
\end{proof}

\begin{lemma}[Regular parameter sets]\label{My-lem:regular-sets}
\(\Theta^I\) is open and \(\Theta^B\) is Borel; hence \(\Theta^R\) is Borel.
\end{lemma}

\begin{proof}
Let \(\vartheta_0\in\Theta^I\), \(r_0=r(\vartheta_0)\).  By continuity,
\(H_{xx}<0\) and \(G_m(x,\alpha+\beta x)\ne0\) on
\(\overline I\times\overline B\) for an open interval \(I\ni r_0\) with
\(\overline I\subseteq\operatorname{int}X\) and a closed ball \(\overline B\)
about \(\vartheta_0\).  The gap
\(\delta\coloneqq H(r_0;\vartheta_0)-\max_{X\setminus I}H(\cdot;\vartheta_0)\)
is positive, and uniform continuity of \(H\) on \(X\times\overline B\) gives
a ball \(B'\subseteq\overline B\) about \(\vartheta_0\) on which
\(\lvert H(\cdot;\vartheta)-H(\cdot;\vartheta_0)\rvert<\delta/3\) on \(X\).  For
\(\vartheta\in B'\) every maximizer on \(X\) therefore lies in \(I\), where
\(H(\cdot;\vartheta)\) is strictly concave, so it is unique, interior, and has
\(Q<0\) and \(M\ne0\).  Hence \(B'\subseteq\Theta^I\).  The upper-endpoint
part of \(\Theta^B\) is
\(\bigcap_n\{m^+_n>0\}\cap\{H_x(\overline x;\cdot)>0\}\), with the continuous
functions
\(m^+_n(\vartheta)\coloneqq\min_{x\in[\underline x,\overline x-1/n]}
\{H(\overline x;\vartheta)-H(x;\vartheta)\}\), where \(n\) ranges over the
integers with \(n>1/(\overline x-\underline x)\); this is a countable
intersection of open sets.  The lower-endpoint part is analogous, with the
same range of \(n\).
\end{proof}

\begin{lemma}[The local myopic maximizer]\label{My-lem:branch}
Let \(\vartheta^\circ\in\R^2\) be such that \(H(\cdot;\vartheta^\circ)\) has a
unique maximizer \(r^\circ=r(\vartheta^\circ)\) on \(X\), with
\(r^\circ\in\operatorname{int}X\) and
\(Q(\vartheta^\circ)\coloneqq H_{xx}(r^\circ;\vartheta^\circ)<0\); every
\(\vartheta^\circ\in\Theta^I\) qualifies.  There are
an open neighborhood \(\mathcal U\) of \((\vartheta^\circ,0)\) in
\(\R^2\times\mathbb S\), an open interval \(I\ni r^\circ\) with
\(\overline I\subseteq\operatorname{int}X\), and a \(C^2\) map
\(\phi:\mathcal U\to I\) such that:
\begin{enumerate}[label=(\roman*)]
\item \(X\times\mathcal U\subseteq\mathcal D\), \(\mathcal J_{xx}<0\) on\label{My-lem:branch:i}
\(I\times\mathcal U\), and for every \(p\in\mathcal U\), \(\phi(p)\) is the
unique maximizer of \(\mathcal J(\cdot;p)\) on \(X\) and the unique zero of
\(\mathcal J_x(\cdot;p)\) in \(I\);
\item \(\phi(a,b,0)=r(a,b)\), the unique maximizer of \(H(\cdot;a,b)\) on\label{My-lem:branch:ii}
\(X\), whenever \((a,b,0)\in\mathcal U\), and there
\begin{equation}\label{My-eq:gamma}
 r_b-r\,r_a=-\frac{M}{Q},
\end{equation}
with \(Q\coloneqq H_{xx}(r;a,b)<0\) and \(M\coloneqq G_m(r,a+br)\) at
\(r=r(a,b)\).
\end{enumerate}
\end{lemma}

\begin{proof}
Choose a compact neighborhood \(K\) of \((\vartheta^\circ,0)\) with
\(1+q\ge1/2\) on \(X\times K\).  Since
\(\mathcal J_{xx}(r^\circ;\vartheta^\circ,0)=Q(\vartheta^\circ)<0\), continuity
gives an open interval \(I\ni r^\circ\) with \(\overline I\subseteq\operatorname{int}X\) and a neighborhood \(\mathcal U_0\subseteq K\) of
\((\vartheta^\circ,0)\) with \(\mathcal J_{xx}\le Q(\vartheta^\circ)/2\) on
\(\overline I\times\mathcal U_0\).  The implicit function theorem for
\(\mathcal J_x(x;p)=0\) on the open set
\(\mathcal D\cap(\operatorname{int}X\times\R^2\times\mathbb S)\), at
\((r^\circ;\vartheta^\circ,0)\), where
\(\mathcal J_x=H_x=0\), gives a \(C^2\) zero \(\phi\) with values in \(I\) on a
neighborhood of \((\vartheta^\circ,0)\); by strict concavity it is the unique
zero in \(I\) and the unique maximizer on \(\overline I\).  The Berge maximum
theorem \citep[Theorem~17.31]{ab2006} makes \(p\mapsto\arg\max_X\mathcal
J(\cdot;p)\) upper hemicontinuous, and it equals \(\{r^\circ\}\) at
\((\vartheta^\circ,0)\); so on a smaller neighborhood \(\mathcal U\) every
maximizer on \(X\) lies in \(I\), and therefore equals \(\phi(p)\).  This is
\ref{My-lem:branch:i}.  At \(\Sigma=0\), \(\mathcal J=H\), so \(\phi(a,b,0)\) is the unique
maximizer of \(H(\cdot;a,b)\) on \(X\), which is \(r(a,b)\).
Differentiating \(H_x(r(a,b);a,b)=0\), with
\(H_x=G_x+bG_m\), gives \(Q\,r_a+G_{xm}+bG_{mm}=0\) and
\(Q\,r_b+G_m+r(G_{xm}+bG_{mm})=0\), all evaluated at
\((r,a+br)\); subtracting \(r\) times the first from the second gives
\eqref{My-eq:gamma}.
\end{proof}

\begin{proof}[Proof of Corollary~\ref{cor:myopic}]
\emph{Part \ref{cor:myopic:ii}, interior truths.}  Let \(\vartheta\in\Theta^I\), with the
data \(\mathcal U\), \(I\), \(\phi\) of Lemma~\ref{My-lem:branch} at
\(\vartheta^\circ=\vartheta\).  At every feasible state \(s\) with
\(p=(a,b,\Sigma(u,\bar x,w))\in\mathcal U\), the action \(\psi_0(s)\)
maximizes \(\mathcal J(\cdot;p)\) on \(X\), whose unique maximizer is
\(\phi(p)\) by Lemma~\ref{My-lem:branch}\ref{My-lem:branch:i}; so
\(\psi_0(s)=\phi(a,b,\Sigma(u,\bar x,w))\).  The map \(\phi\) is \(C^2\) on
the open set \(\mathcal U\), hence \(C^1\) with locally Lipschitz first
derivatives, and \(h=\phi(\cdot,0)=r\) near \(\vartheta\) satisfies
\(h_b-h\,h_a=-M/Q\ne0\) at \(\vartheta\) by Lemma~\ref{My-lem:branch}\ref{My-lem:branch:ii}.  So
\(\vartheta\in\Theta^T_{\psi_0}\) by Definition~\ref{def:regular}, with
complete-information action \(r=x^\star\) and \(\gamma=-M/Q\).

\emph{Part \ref{cor:myopic:ii}, boundary truths.}  Let \(\vartheta\in\Theta^B\) with
\(r(\vartheta)=\overline x\); the lower endpoint is symmetric.  Since
\(H_x(\overline x;\vartheta)>0\) and \(\overline x\) is the unique maximizer,
there is \(c>0\) with \(H(\overline x;\vartheta)-H(x;\vartheta)\ge
c(\overline x-x)\) on \(X\), near \(\overline x\) by the positive derivative
and on the rest of \(X\) by compactness.  By uniform continuity of
\(\mathcal J_x\) on \(X\times K\), for a compact neighborhood \(K\) of
\((\vartheta,0)\) with \(1+q\ge1/2\) on \(X\times K\), and because
\(\mathcal J_x(\cdot;\vartheta,0)=H_x(\cdot;\vartheta)\), there is \(r_C>0\)
such that \(\lvert\mathcal J_x(x;\sigma(s))-H_x(x;\vartheta)\rvert<c/2\) on \(X\) at
every feasible state \(s\) with \(\lvert a-\alpha\rvert,\lvert b-\beta\rvert,u,w<r_C\) and
\(\operatorname{dist}(\bar x,X)<r_C\); here
\(\sigma(s)\coloneqq(a,b,\Sigma(u,\bar x,w))\), whose covariance entries
\(u+\bar x^2w\), \(-\bar xw\), \(w\) are then small.  At such a state,
\(\mathcal J(\overline x;\sigma(s))-\mathcal J(x;\sigma(s))
=\int_x^{\overline x}\mathcal J_x(\cdot;\sigma(s))\ge\frac c2(\overline x-x)\)
for \(x<\overline x\), so \(\psi_0(s)=\overline x\).
Thus \(\vartheta\in\Theta^C_{\psi_0}\).

\emph{Part \ref{cor:myopic:ii}, conclusion.}  Hence \(\Theta^R\subseteq\Theta_{\psi_0}\),
and \(\Theta^R\) is Borel by Lemma~\ref{My-lem:regular-sets}.
Theorem~\ref{thm:global} applied to \(\psi_0\), restricted to
\(\Theta^R\) by monotonicity, gives the three displayed conclusions.  The
exceptional null set may depend on the prior and on the selection.

\emph{Part \ref{cor:myopic:i}.}  Let \((a^e,b^e,x^e)\) be an interior regular Berk--Nash
equilibrium, so that \(\vartheta^\circ=(a^e,b^e)\in\Theta^I\) and
\(x^e=r(\vartheta^\circ)\), with the data of Lemma~\ref{My-lem:branch} and
\(\sigma\) as above.  The \(C^2\) map
\(\bar\psi_0\coloneqq\phi\circ\sigma\) on the open neighborhood
\(\sigma^{-1}(\mathcal U)\) of \(z^e\) equals \(\psi_0\) at its feasible
states.  Choose \(w_1>0\) with
\(\sigma(a^e,b^e,0,x^e,w)\in\mathcal U\) for \(\lvert w\rvert<w_1\).  There
\(\sigma(a^e,b^e,0,x^e,w)=(\vartheta^\circ,w\,nn^\top)\) with \(n=(-x^e,1)\),
so \(q(x;w\,nn^\top)=w(x-x^e)^2\).  Since \(\mathcal J(x;p)\) depends on
\(\Sigma\) only through \(q(x;\Sigma)\),
\(\mathcal J(x;\vartheta^\circ,w\,nn^\top)
=\mathcal J\bigl(x;\vartheta^\circ,\operatorname{diag}(w(x-x^e)^2,0)\bigr)\)
for \(x\in X\).  By Lemma~\ref{My-lem:objective}\ref{My-lem:objective:b},
\(\mathcal J\) is \(C^1\) on \(\mathcal D\), and \(w(x-x^e)^2\) and its
derivative vanish at \(x^e\); so the chain rule gives
\(\mathcal J_x(x^e;\vartheta^\circ,w\,nn^\top)=\mathcal J_x(x^e;\vartheta^\circ,0)
=H_x(x^e;\vartheta^\circ)=0\), and
Lemma~\ref{My-lem:branch}\ref{My-lem:branch:i} gives \(\bar\psi_0(a^e,b^e,0,x^e,w)=x^e\) for
\(\lvert w\rvert<w_1\).  These points lie in the open domain
\(\sigma^{-1}(\mathcal U)\) of the continuous \(\bar\psi_0\), so
\(\psi_0(s_n)\to x^e\) along every sequence of feasible states converging to
one of them.  With no-surprise, the case \(w=0\)
makes \((a^e,b^e,x^e)\) a complete-information Berk--Nash equilibrium under
\(\psi_0\) in the sense of Definition~\ref{def:bn-psi}, and each \(w\in(0,w_1)\)
makes \((a^e,b^e,x^e,w)\) a Berk--Nash equilibrium under \(\psi_0\).  Since \(\sigma\) is the
identity in \((a,b)\) and its covariance part does not depend on \((a,b)\),
\(\gamma^e=\phi_b-x^e\phi_a=r_b-r\,r_a=-M^e/Q^e\ne0\) at
\(\vartheta^\circ\), by Lemma~\ref{My-lem:branch}\ref{My-lem:branch:ii}.  Since \(\phi\) is
\(C^2\) and \(\sigma\) is polynomial, \(\bar\psi_0\) is \(C^2\), so its first
derivatives are locally Lipschitz, so Theorem~\ref{thm:smooth-local} applies,
and, since \((a^e,b^e,x^e,w)\) is a Berk--Nash equilibrium under \(\psi_0\) for
\(0<w<w_1\), its last assertion centers the priors at \((a^e,b^e,x^e,V^e)\) for
every sufficiently large \(V^e\).

\emph{Part \ref{cor:myopic:iii}.}  Let \((a^e,b^e,x^e)\) be a complete-information Berk--Nash equilibrium under
\(\psi_0\).  The feasible states
\(s_n=(a^e,b^e,1/n,x^e,1/n)\) converge to \(z^e\), so
\(\psi_0(s_n)\to x^e\) by Definition~\ref{def:bn-psi}; since \(X\) is
closed, \(x^e\in X\).  Also \(\sigma(s_n)\to(a^e,b^e,0)\).  Passing to the
limit in \(\mathcal J(\psi_0(s_n);\sigma(s_n))\ge\mathcal J(x;\sigma(s_n))\),
\(x\in X\), by continuity of \(\mathcal J\) on \(\mathcal D\), and using
\(\mathcal J(\cdot;a^e,b^e,0)=H(\cdot;a^e,b^e)\), shows that \(x^e\)
maximizes \(H(\cdot;a^e,b^e)\) on \(X\).  The no-surprise condition is not
used.
\end{proof}

\section{Example verifications}\label{supp:examples}

This section verifies the hypotheses of Theorems~\ref{thm:smooth-local}
and~\ref{thm:global} for the three examples and the two rules of
Section~\ref{sec:rules}.  Neither theorem requires the rule to be optimal.
Write \(\Pi_X(z)\coloneqq\min\{\overline x,\max\{\underline x,z\}\}\),
\(\sigma(s)\coloneqq(a,b,\Sigma(u,\bar x,w))\) with \(\Sigma\) of
\eqref{eq:sigma}, and \(n_x\coloneqq(-x,1)^\top\), so that
\(\sigma(a,b,0,x,w)=(a,b,w\,n_xn_x^\top)\).  Every rule below has the
following form.

\begin{lemma}[Projected smooth rules]\label{Ex-lem:projected}
Let \(\mathcal G\subseteq\R^2\times\mathbb S\) be open, let
\(g:\mathcal G\to\R\) be \(C^1\) with locally Lipschitz first derivatives,
and suppose that
\(\psi(s)=\Pi_X(g(\sigma(s)))\) at every feasible state \(s\) with
\(\sigma(s)\in\mathcal G\).  Write \(\tilde h(a,b)\coloneqq g(a,b,0)\), the complete-information action before projection.
\begin{enumerate}[label=(\alph*)]
\item Let \((\vartheta,0)\in\mathcal G\).  If \(\tilde h(\vartheta)\in\operatorname{int}X\)\label{Ex-lem:projected:a}
and \(\tilde h_b\ne\tilde h\,\tilde h_a\) at \(\vartheta\), then \(\vartheta\in\Theta^T_\psi\),
with complete-information action \(h=\tilde h\) near \(\vartheta\) and \(\gamma=\tilde h_b-\tilde h\,\tilde h_a\) at
\(\vartheta\).  If
\(\tilde h(\vartheta)\notin X\), then \(\vartheta\in\Theta^C_\psi\).
\item Let \(g\) be \(C^\infty\), and let \((a^e,b^e,x^e)\) satisfy the\label{Ex-lem:projected:b}
no-surprise condition, with
\(x^e\in\operatorname{int}X\), \((a^e,b^e,0)\in\mathcal G\), and
\(g(a^e,b^e,w\,n_{x^e}n_{x^e}^\top)=x^e\) for every sufficiently small
\(w\ge0\).  Then \(\psi\) admits the \(C^\infty\) extension \(\bar\psi=g\circ\sigma\)
to an open neighborhood of \(z^e=(a^e,b^e,0,x^e,0)\);
\((a^e,b^e,x^e)\) is a complete-information Berk--Nash equilibrium under
\(\psi\) with response coefficient \(\gamma^e=\tilde h_b-x^e \tilde h_a\) at
\((a^e,b^e)\); and \((a^e,b^e,x^e,w)\) is a Berk--Nash equilibrium under
\(\psi\) for every sufficiently small \(w\ge0\).
\end{enumerate}
\end{lemma}

\begin{proof}
\ref{Ex-lem:projected:a} By continuity, \(g\in\operatorname{int}X\) on an open neighborhood
\(\mathcal N\subseteq\mathcal G\) of \((\vartheta,0)\), so
\(\psi(s)=g(\sigma(s))\) at every feasible state \(s\) with
\(\sigma(s)\in\mathcal N\), and \(g\) is \(C^1\) with locally Lipschitz first
derivatives there,
so Definition~\ref{def:regular} applies with \(\Psi=g\) on \(\mathcal N\).  If \(\tilde h(\vartheta)>\overline x\),
continuity of \(g\) at \((\vartheta,0)\) gives \(r_C>0\) with
\(g(\sigma(s))>\overline x\), hence \(\psi(s)=\overline x\), at every feasible
\(s\) with \(\lvert a-\alpha\rvert,\lvert b-\beta\rvert,u,w<r_C\) and
\(\operatorname{dist}(\bar x,X)<r_C\), since the entries of \(\Sigma(s)\) are
then small.  So \(\psi\) has the constant complete-information limit
\(\overline x\) on \(\{\lvert a-\alpha\rvert,\lvert b-\beta\rvert<r_C\}\), and
\(\vartheta\) is locally constant with \(e=\overline x\) and \(r=r_C\); the case \(\tilde h(\vartheta)<\underline x\) is symmetric.  \ref{Ex-lem:projected:b} The map
\(\bar\psi\coloneqq g\circ\sigma\) on the open set
\(\sigma^{-1}(\{g\in\operatorname{int}X\})\ni z^e\) is a \(C^\infty\)
extension, and \(\bar\psi(a^e,b^e,0,x^e,w)=x^e\) for small \(w\ge0\), where
\((a^e,b^e,0,x^e,w)\) lies in that open set, so, by continuity of
\(\bar\psi\) and the no-surprise condition, these are the asserted
equilibria; since
\(\sigma\) is the identity in \((a,b)\) and its covariance part does not
depend on \((a,b)\), \(\bar\psi_a=g_a\) and
\(\bar\psi_b=g_b\) at \(z^e\), which equal \(\tilde h_a\) and \(\tilde h_b\).
\end{proof}

When a rule depends on the state only through \((a,b)\), the condition on
\(w\) in \ref{Ex-lem:projected:b} reduces to \(\tilde h(a^e,b^e)=x^e\).  If \(\R^2\setminus\Theta_\psi\) is contained in
a Lebesgue-null set \(L\), then \(\mu(L)=0\) for every nondegenerate
Gaussian \(\mu\) (Lemma~\ref{lem:disintegration}\ref{lem:disintegration:iii}), and
Theorem~\ref{thm:global} gives
\begin{equation}\label{Ex-eq:generic}
 P^{\mu,\psi}(V_t\to\infty)=0,
 \qquad
 P^{\mu,\psi}(\mathcal W)=1,
\end{equation}
together with \(P^{\mu,\psi}(\mathcal W,\vartheta\in B)=\mu(B)\) for every
Borel \(B\) and \(P^\psi_\vartheta(\mathcal W)=1\) for Lebesgue-almost every
\(\vartheta\in\R^2\): the \emph{generic-failure conclusions} for \(\psi\).  At
each complete-information Berk--Nash equilibrium of \ref{Ex-lem:projected:b}
with \(\gamma^e\ne0\), Theorem~\ref{thm:smooth-local} applies, with priors
centered at \((a^e,b^e,x^e,V^e)\) for every sufficiently large \(V^e\), since
\((a^e,b^e,x^e,w)\) is a Berk--Nash equilibrium under \(\psi\) for small
\(w>0\) by \ref{Ex-lem:projected:b}.

\subsection{Advertising}\label{Ex-sec:advertising}

With \(U=y-x^2/2\) (Example~\ref{ex:advertising}), the rule is
\(\psi(s)=\Pi_X(b)\), with \(g=b\), \(\tilde h_a=0\) and \(\tilde h_b=1\), so \(\gamma=1\)
everywhere.  For every truth and every slope \(c\in\operatorname{int}X\),
\((a^c,c,c)\) with \(a^c=\alpha+(\beta-c)c\) is a complete-information
Berk--Nash equilibrium under \(\psi\) with \(\gamma^c=1\), and Theorem~\ref{thm:smooth-local} applies, so every
incorrect slope in \(\operatorname{int}X\) is locally persistent.  Truths with
\(\beta\in\operatorname{int}X\) are interior regular, truths with
\(\beta\notin X\) locally constant, and the generic-failure conclusions hold
with \(L=\{\beta\in\{\underline x,\overline x\}\}\).

\emph{A rule of thumb.}  For \(\psi_{\ell,\tau}(s)=\Pi_X(\ell+\tau b)\),
\(\tau\ne0\), \(g=\ell+\tau b\) and \(\gamma=\tau\).  For \(\tau<0\) the
rule is not the myopic optimum of any payoff \(y-C(x)\) with \(C\) convex,
since that optimum is nondecreasing in \(b\).  Every slope \(c\) with
\(x^c=\ell+\tau c\in\operatorname{int}X\) gives a complete-information Berk--Nash equilibrium
\((\alpha+(\beta-c)x^c,c,x^c)\) under \(\psi_{\ell,\tau}\) with \(\gamma^c=\tau\); truths with
\(\ell+\tau\beta\in\operatorname{int}X\) are interior regular, those with
\(\ell+\tau\beta\notin X\) locally constant, and the generic-failure
conclusions hold with \(L=\{\ell+\tau\beta\in\{\underline x,\overline x\}\}\).
With \(\tau=0\) every truth would be locally constant, so generic failure
would still hold, but no equilibrium would have \(\gamma^e\ne0\).

\subsection{Pricing}\label{Ex-sec:pricing}

With \(U=(x-k)y\) (Example~\ref{ex:pricing}), the myopic rule is, on
\(\{b<0\}\), \(\Pi_X(x^u(a,b))\) with \(x^u=k/2-a/(2b)\); for \(b>0\) it
chooses \(\overline x\) if \(T(a,b)\coloneqq a+b(\underline x+\overline x-k)>0\)
and \(\underline x\) otherwise; and for \(b=0\) it chooses \(\overline x\) if
\(a>0\) and \(\underline x\) otherwise.  On \(\mathcal G=\{b<0\}\),
\(g=x^u\), \(\tilde h_a=-1/(2b)\) and \(\tilde h_b=a/(2b^2)\), so
\[
 \tilde h_b-\tilde h\,\tilde h_a=\frac{a+b\tilde h}{2b^2}=-\frac{\tilde h-k}{2b}.
\]
\emph{Equilibria.}  For a slope \(c<0\) with \(c+\beta\ne0\),
\(x^c=(kc-\alpha)/(c+\beta)\) and \(a^c=\alpha+(\beta-c)x^c\) solve
\(x^c=x^u(a^c,c)\) and no-surprise; if \(x^c\in\operatorname{int}X\), it is a
complete-information Berk--Nash equilibrium under \(\psi\) with
\[
 \gamma^c=\frac{a^c+cx^c}{2c^2}=\frac{\alpha+\beta k}{2c(c+\beta)},
\]
nonzero exactly under the viability condition \(\alpha+\beta k\ne0\), that is, true
expected demand at marginal cost is nonzero.
\emph{Regular truths.}  If \(\beta<0\) and \(x^u(\vartheta)\in\operatorname{int}X\)
with \(x^u\ne k\), the truth is interior regular; if \(\beta<0\) and
\(x^u(\vartheta)\notin X\), locally constant.  If \(\beta>0\) and
\(T(\vartheta)\ne0\), the rule is constant near \(\vartheta\) by continuity of
\(T\).  If \(\beta=0\) and \(\alpha>0\), then for \(r_C\le\alpha/2\) small
enough the rule chooses \(\overline x\) at every \((a,b)\) with
\(\lvert a-\alpha\rvert,\lvert b\rvert<r_C\). This holds at \(b=0\) since \(a>0\), at \(b>0\) since
\(T>\alpha/2-r_C\lvert\underline x+\overline x-k\rvert>0\), and at \(b<0\) since
\(x^u>k/2+\alpha/(4r_C)>\overline x\); \(\alpha<0\) is symmetric.  So the
nonregular truths lie on the four lines \(\alpha=\beta(k-2\underline x)\),
\(\alpha=\beta(k-2\overline x)\), \(\alpha=-\beta k\) and
\(\alpha=-\beta(\underline x+\overline x-k)\), together with the origin, and the
generic-failure conclusions hold, in particular on the downward-sloping truths
\(\{\beta<0\}\).

\subsection{Inflation}\label{Ex-sec:inflation}

With \(U=-[(y-y^*)^2+\omega x^2]\), \(\omega>0\)
(Example~\ref{ex:inflation}), posterior expected loss is quadratic in \(x\),
and the passive rule \citep[equation~(17)]{wieland2000monetary} is
\(\psi=\Pi_X\circ g\) with
\[
 g(a,b,\Sigma)\coloneqq\frac{-\Sigma_{12}+b(y^*-a)}{\Sigma_{22}+b^2+\omega},
 \qquad
 \tilde h(a,b)=\varphi^{CE}(a,b)\coloneqq\frac{b(y^*-a)}{b^2+\omega},
\]
on \(\mathcal G=\{\Sigma_{22}>-\omega\}\).  In the state coordinates,
\(g\circ\sigma=\bar x+\lambda(\varphi^{CE}-\bar x)\) with
\(\lambda=(b^2+\omega)/(w+b^2+\omega)\in(0,1)\), so the passive action moves only
part of the way from the average action to the certainty-equivalent action.
With \(\Delta\coloneqq y^*-\alpha\) and \(K_c\coloneqq\omega+c\beta\ne0\),
\(x^c=c\Delta/K_c\) and \(a^c=\alpha+(\beta-c)x^c\) satisfy
\(c(y^*-a^c)=(c^2+\omega)x^c\), hence
\(g(a^c,c,w\,n_{x^c}n_{x^c}^\top)=[wx^c+(c^2+\omega)x^c]/(w+c^2+\omega)=x^c\)
for every \(w>-\omega\).  If \(x^c\in\operatorname{int}X\), it is a
complete-information Berk--Nash equilibrium under \(\psi\) with
\[
 \gamma^c=\tilde h_b-x^c\tilde h_a=\frac{\omega\Delta}{K_c(c^2+\omega)},
\]
nonzero exactly when \(\Delta\ne0\); with the calibration
\((\alpha,\beta,y^*,\omega)=(9,-0.7,0,0.14)\) of
\citet[Table~1]{wieland2000monetary}, this holds at every \(c<0\) with
\(x^c\in\operatorname{int}X\).  At a truth, with
\(r^u\coloneqq\varphi^{CE}(\vartheta)\),
\(\tilde h_b-\tilde h\,\tilde h_a=\omega\Delta/(\beta^2+\omega)^2\), so truths with
\(r^u\in\operatorname{int}X\) and \(\Delta\ne0\) are interior regular and
truths with \(r^u\notin X\) are locally constant.  The nonregular truths lie in
\(\{\Delta=0\}\cup\{r^u=\underline x\}\cup\{r^u=\overline x\}\); for each
\(\beta\ne0\) each condition selects at most one \(\alpha\), so this set is
Lebesgue-null by Fubini's theorem, and the generic-failure conclusions hold.

\emph{The certainty-equivalent rule.}  The rule
\(\psi^{CE}(s)=\Pi_X(\varphi^{CE}(a,b))\) ignores parameter uncertainty and
is more aggressive than the passive rule.  It has \(g=\tilde h=\varphi^{CE}\), the
same function \(\tilde h\), and therefore the same equilibria, response
coefficients, regular truths and generic-failure conclusions, so the
incomplete-learning conclusions do not depend on the caution of the passive
rule.

\begin{remark}[Response coefficients of the myopic rules]\label{Ex-rem:myopic}
For the myopic rules above, the curvature \(Q\) and the outcome coefficient
\(M\) of Corollary~\ref{cor:myopic}, at the parameter \((a^c,c)\) and the
action \(x^c\), are \(Q=-1\), \(M=1\) (advertising); \(Q=2c\), \(M=x^c-k\)
(pricing); and \(Q=-2(c^2+\omega)\), \(M=-2(a^c+cx^c-y^*)\) (inflation).  In
each case \(\gamma^c=-M/Q\), as Corollary~\ref{cor:myopic} predicts; the
verifications above use neither this identity nor optimality.
\end{remark}

\section{Targeting payoffs}\label{supp:targeting}

This section proves the statements of Remark~\ref{rem:targeting}.  Let
\(X=[-1,1]\) and \(U(x,y)=-y^2\), so \(G(x,m)=-1-m^2\) and
\(g^{\mathrm{full}}(x;(a,b))=-1-(a+bx)^2\).  If \(b\ne0\) and
\(-a/b\in\operatorname{int}X\), the unique maximizer is \(x^\star=-a/b\), with
\begin{equation}\label{S5-eq:regularity}
 \partial_{xx}g^{\mathrm{full}}(x^\star;(a,b))=-2b^2<0,
 \qquad
 G_m(x^\star,a+bx^\star)=0 .
\end{equation}
Posterior expected payoff is
\(-1-(a_t+b_tx)^2-u_t-w_t(x-\bar x_t)^2\), a strictly concave quadratic, so
the myopic rule is
\begin{equation}\label{S5-eq:policy}
 \psi_0(a,b,u,\bar x,w)\coloneqq\Pi_X\left(\frac{-ab+w\bar x}{b^2+w}\right),
\end{equation}
which is continuous, and every Borel myopic selection equals it.  If
\(b_t\ne0\) and \(-a_t/b_t,\bar x_t\in X\), the argument of \(\Pi_X\) is a
convex combination of these two points, the projection is inactive, and with
the predicted mean at the average action \(q_t\coloneqq a_t+b_t\bar x_t\),
\begin{equation}\label{S5-eq:gap}
 D_t=-\frac{b_t}{b_t^2+w_t}\,q_t,
 \qquad
 q_t=-\left(b_t+\frac{w_t}{b_t}\right)D_t .
\end{equation}
For every rule, \eqref{eq:design} and Lemma~\ref{lem:posterior-interface}\ref{lem:posterior-interface:a}
give, pathwise and hence under \(P^{\mu,\psi}\) and every
\(P^\psi_\vartheta\),
\begin{equation}\label{S5-eq:updates}
 \bar x_{t+1}=\bar x_t+\frac{D_t}{A_t+1},
 \qquad
 V_{t+1}=V_t+\kappa_tD_t^2,
 \qquad
 A_{t+1}\xi_{t+1}=A_t\xi_t-\varepsilon_{t+1},
\end{equation}
with \((\varepsilon_t)\) i.i.d.\ \(N(0,1)\) under both laws, so
\(\xi_t=(A_0\xi_0-\sum_{j\le t}\varepsilon_j)/(A_0+t)\).

\begin{lemma}[Nonsummability of the fitted-mean error]
\label{S5-lem:noise-nonsummable}
For every nondegenerate Gaussian prior \(\mu\) and every Borel decision rule
\(\psi\) from the feasible states into an arbitrary action set
\(X\subseteq\R\), \(\sum_t\xi_t^2=\infty\) almost surely, under
\(P^{\mu,\psi}\) and under every fixed-truth law \(P^\psi_\vartheta\).
\end{lemma}

\begin{proof}
Only the third identity in \eqref{S5-eq:updates}, which is pathwise algebra
and does not involve the action set, and the i.i.d.\ \(N(0,1)\) law of
\((\varepsilon_t)\) under both laws are used, so the argument applies to
every action set \(X\subseteq\R\).
Write \(\xi_t=\tilde\xi_t+A_0\xi_0/A_t\), with \(\tilde\xi_t\) the solution started
at zero; the second term is square summable on every path, so it suffices
to treat \(\tilde\xi\).  Squaring the recursion gives
\begin{equation}\label{S5-eq:energy}
 A_{t+1}\tilde\xi_{t+1}^2-A_t\tilde\xi_t^2
 =-\kappa_t\tilde\xi_t^2-2\kappa_t\tilde\xi_t\varepsilon_{t+1}
  +\frac{\varepsilon_{t+1}^2}{A_t+1}.
\end{equation}
On \(\{\sum_t\tilde\xi_t^2<\infty\}\) the martingale
\(\sum_{t<n}\kappa_t\tilde\xi_t\varepsilon_{t+1}\) converges, since its
accumulated conditional variance is finite \citep[Theorem~4.5.2]{d2019};
and \(\sum_t\varepsilon_{t+1}^2/(A_t+1)=\infty\) almost surely, because
\(\sum_t(\varepsilon_{t+1}^2-1)/(A_t+1)\) converges while
\(\sum_t(A_t+1)^{-1}=\infty\).  Summing \eqref{S5-eq:energy} then gives
\(A_n\tilde\xi_n^2\to\infty\), so \(\tilde\xi_n^2\ge A_n^{-1}\) eventually,
contradicting \(\sum_n\tilde\xi_n^2<\infty\).
\end{proof}

The same proof shows that Lemma~\ref{S5-lem:noise-nonsummable} holds for
every seeded policy, randomized or history-dependent as in
Remark~\ref{A-rem:predictable}, with values in an arbitrary set
\(X\subseteq\R\) and i.i.d.\ seeds of an arbitrary law, independent of
\(\vartheta\) and the noise. That is, \(\sum_t\xi_t^2=\infty\) almost surely under its
prior-predictive law and under each of its fixed-truth laws.

\begin{lemma}[Interior paths have infinite slope precision]
\label{S5-lem:interior-divergence}
Fix a nondegenerate Gaussian prior \(\mu\), the rule \(\psi_0\), an integer
\(T\ge0\) and constants \(c>0\) and \(C<\infty\), and let
\(\Omega_{T,c,C}\) be the event that \(c\le\lvert b_t\rvert\le C\),
\(-a_t/b_t\in X\) and \(\bar x_t\in X\) for every \(t\ge T\).  Then
\(\Omega_{T,c,C}\cap\{V_\infty<\infty\}\) is null under \(P^{\mu,\psi_0}\)
and under every \(P^{\psi_0}_\vartheta\).
\end{lemma}

\begin{proof}
On \(\Omega_{T,c,C}\cap\{V_\infty<\infty\}\),
\(\sum_tD_t^2<\infty\) by \eqref{S5-eq:updates}, and \eqref{S5-eq:gap} with
\(w_t\le w_0\) gives \(\lvert q_t\rvert\le(C+w_0/c)\lvert D_t\rvert\) for \(t\ge T\), so
\(\sum_tq_t^2<\infty\).  By Cauchy--Schwarz, \(\bar x_t\) converges to some
\(x_\infty\), with \(\lvert\bar x_t-x_\infty\rvert\le\sum_{j\ge t}\lvert D_j\rvert/(A_0+j+1)\);
Hardy's inequality in its tail form \citep[Theorem~331]{hlp1952} gives
\begin{equation}\label{S5-eq:tail-sum}
 \sum_{t\ge0}(\bar x_t-x_\infty)^2\le4\sum_{j\ge0}D_j^2 .
\end{equation}
Since \(\xi_t\to0\) by the strong law, \(q_t\to0\) and
\(\xi_t=\alpha+\beta\bar x_t-q_t\), we get \(\alpha+\beta x_\infty=0\) and
\(\xi_t=\beta(\bar x_t-x_\infty)-q_t\), which is square summable.  This
contradicts Lemma~\ref{S5-lem:noise-nonsummable} outside a null set.
\end{proof}

\begin{proposition}[No finite-precision trapping at an incorrect equilibrium]
\label{S5-prop:local-boundary}
Let the truth be \((0,1)\).  Then \((a,b,x)=(0,2,0)\) is a Berk--Nash
equilibrium, with \(a+bx=\alpha+\beta x\), \(x=0\) the unique maximizer of
\(-1-4z^2\) on \(X\), and \(G_m(0,0)=0\).  For every nondegenerate Gaussian
prior and every \(r\in(0,1/2)\),
\begin{equation}\label{S5-eq:no-trap}
 P^{\psi_0}_{(0,1)}\Bigl(
 \lvert a_t\rvert+\lvert b_t-2\rvert+\lvert\bar x_t\rvert<r\ \text{for every }t\ge0,\ \
 V_\infty<\infty\Bigr)=0 .
\end{equation}
\end{proposition}

\begin{proof}
On the event, \(3/2<b_t<5/2\), \(\lvert a_t\rvert,\lvert\bar x_t\rvert<1/2\) and
\(\lvert a_t/b_t\rvert<1/3\) at every date, so it lies in \(\Omega_{0,3/2,5/2}\), and
Lemma~\ref{S5-lem:interior-divergence} applies.
\end{proof}

\begin{proposition}[Complete learning with probability near one]
\label{S5-prop:global-boundary}
Fix \(s\in(0,1/32)\), the prior \(\mu_s=N((0,1)^\top,sI_2)\), and
\(B=\{\lvert\alpha\rvert\le1/4,\ 3/4\le\beta\le5/4\}\).  Every truth in \(B\) has the
unique interior complete-information maximizer \(-\alpha/\beta\), with
\(G_m=0\) there.  Nevertheless:
\begin{enumerate}
\item[\condtarget{S5-prop:global-boundary:i}{(i)}] \(P^{\mu_s,\psi_0}\bigl(V_\infty=\infty,\ (a_t,b_t)\to(\alpha,\beta),\
(\alpha,\beta)\in B\bigr)\ge1-32s\);
\item[\condtarget{S5-prop:global-boundary:ii}{(ii)}] the set of \(\vartheta\in B\) with
\(P^{\psi_0}_\vartheta\bigl(V_\infty=\infty,\ (a_t,b_t)\to\vartheta\bigr)>0\)
has positive Lebesgue measure.
\end{enumerate}
\end{proposition}

\begin{proof}
Here \(A_0=V_0=s^{-1}\), \(\bar x_0=a_0=0\), \(b_0=1\).  Doob's maximal
inequality \citep[Theorem~4.4.2]{d2019} for the martingales \(a_t\) and
\(b_t-1\), with second moments at most \(s\), shows that
\(E\coloneqq\{\lvert a_t\rvert<1/4,\ \lvert b_t-1\rvert<1/4\ \text{for every }t\}\) has
probability at least \(1-32s\).  On \(E\), \(\lvert a_t/b_t\rvert<1/3\), and by
induction \(x_{t+1},\bar x_t\in(-1/3,1/3)\), since each action is the convex
combination after \eqref{S5-eq:policy}; so \(E\subseteq\Omega_{0,3/4,5/4}\),
and Lemma~\ref{S5-lem:interior-divergence} gives \(V_\infty=\infty\)
almost surely on \(E\).  By Lemma~\ref{lem:complete-learning}, \((a_t,b_t)\)
then converges to \((\alpha,\beta)\), which lies in the closed rectangle
\(B\).  This is \ref{S5-prop:global-boundary:i}.  On the same event the action gaps vanish but are not
square summable. Indeed, \(w_t\to0\), and by \eqref{S5-eq:gap}
\(x_{t+1}+a_t/b_t=w_t(\bar x_t+a_t/b_t)/(b_t^2+w_t)\to0\), so
\(x_t\to-\alpha/\beta\), hence \(\bar x_t\to-\alpha/\beta\) and \(D_t\to0\);
while \(\kappa_t\le1\) and \eqref{S5-eq:updates} give
\(\sum_tD_t^2\ge\sum_t(V_{t+1}-V_t)=\infty\).  For \ref{S5-prop:global-boundary:ii},
Lemma~\ref{lem:disintegration} applied to the
joint event in \ref{S5-prop:global-boundary:i} gives
\(\int_BP^{\psi_0}_\vartheta(V_\infty=\infty,\ (a_t,b_t)\to\vartheta)\,
\mu_s(d\vartheta)\ge1-32s>0\), and \(\mu_s\) has a positive density.
\end{proof}

\end{document}